\documentclass{aa}

\usepackage{graphicx}
\usepackage{txfonts}
\usepackage{color}

\newcommand{\onvire}[1]{}

\newcommand{\beq}{\begin{equation}}
\newcommand{\eeq}{\end{equation}}
\newcommand{\excs}{\extracolsep{\fill}}

\usepackage{color}
\usepackage{epsfig}
\usepackage{amssymb}
\usepackage{graphicx,natbib}
\usepackage{longtable}
\usepackage{pdflscape}
\usepackage[figuresright]{rotating} 
\begin{document}
   \title{C2D Spitzer-IRS spectra of disks around T Tauri stars \\
 IV. Crystalline silicates}

   \author{J. Olofsson
          \inst{1}
          \and
          J.-C. Augereau
	  \inst{1}
	  \and
	  E. F. van Dishoeck
	  \inst{2,3}
          \and
          B. Mer\'{\i}n
          \inst{4}
          \and
          F. Lahuis
          \inst{5,2}
          \and
          J. Kessler-Silacci
          \inst{6}
          \and
          C. P. Dullemond
          \inst{7}
          \and
          I. Oliveira
          \inst{2,12}
          \and
          G. A. Blake
          \inst{8}
          \and
          A. C. A. Boogert
          \inst{9}
          \and
          J. M. Brown
          \inst{3}
          \and
          N. J. Evans II
          \inst{6}
          \and
          V. Geers
          \inst{10}
          \and
          C. Knez
          \inst{11}
          \and
          J.-L. Monin
          \inst{1}
          \and
          K. Pontoppidan
          \inst{12}
          }

   \offprints{olofsson@obs.ujf-grenoble.fr}
   
   \institute{Laboratoire d'Astrophysique de Grenoble, Universit\'e
     Joseph Fourier, CNRS, UMR 5571, Grenoble, France\\
     \email{olofsson@obs.ujf-grenoble.fr} \and
     Leiden Observatory, Leiden University, P.O. Box 9513, 2300 RA
     Leiden, The Netherlands \and
     Max Planck Institut f\"ur Extraterrestrische Physik,
     Giessenbachstrasse 1, 85748 Garching, Germany \and 
     Herschel Science Centre, SRE-SDH, ESA P.O. Box 78, 28691
     Villanueva de la Ca\~nada, Madrid, Spain \and
     SRON Netherlands Institute for Space Research, P.O.Box 800, 9700 AV
     Groningen, The Netherlands \and
     The University of Texas at Austin, Department of Astronomy, 1
     University Station C1400, Austin, Texas 78712–0259, USA \and
     Max Planck Institute for Astronomy, K\"onigstuhl 17, 69117
     Heidelberg, Germany \and
     Division of Geological and Planetary Sciences 150-21, California
     Institute of Technology, Pasadena, CA 91125, USA \and
     Infrared Processing and Analysis Center, MS 100-22, California
     Institute of Technology, Pasadena, CA 91125, USA \and
     Department of Astronomy and Astrophysics, University of Toronto,
     50 St. George Street, Toronto, ON, M5S 3H4, Canada \and
     Department of Astronomy, University of Maryland, College Park, MD
     20742, USA \and
     Division of Geological and Planetary Sciences 150-21, California
     Institute of Technology, Pasadena, CA 91125, USA}

   \date{Received \today; accepted }
 
   \abstract{}
   {Dust grains in the planet forming regions around young stars are
     expected to be heavily processed due to coagulation,
     fragmentation and crystallization. This paper focuses on the
     crystalline silicate dust grains in protoplanetary disks, for a
     statistically significant number of TTauri stars (96).}
   {As part of the Cores to Disks (c2d) Legacy Program, we obtained
     more than a hundred Spitzer/IRS spectra of TTauri stars, over a
     spectral range of 5--35\,$\mu$m where many silicate amorphous and
     crystalline solid-state features are present. At these
     wavelengths, observations probe the upper layers of accretion
     disks up to distances of a dozen AU from the central object.}
   {More than 3/4 of our objects show at least one crystalline
     silicate emission feature that can be essentially attributed to Mg-rich
     silicates. Fe-rich
     crystalline silicates are largely absent in the c2d IRS spectra.  The
     strength and detection frequency of the crystalline features seen
     at $\lambda > 20\,\mu$m correlate with each other, while they are
     largely uncorrelated with the observational properties of the 
     amorphous silicate $10\,\mu$m feature. This supports the idea that 
     the IRS spectra essentially probe two independent disk regions: a warm 
     zone ($\leq$1\,AU) emitting at $\lambda \sim 10\,\mu$m and a much
     colder region emitting at $\lambda > 20\,\mu$m ($\leq$10\,AU). We
     identify a {\it crystallinity paradox}, as the long-wavelength
     ($\lambda > 20\,\mu$m) crystalline silicate features are 3.5
     times more frequently detected ($\sim$55\% vs. $\sim$15\%) than
     the crystalline features arising from much warmer disk regions
     ($\lambda\sim10\,\mu$m).  This suggests that the disk has an
     inhomogeneous dust composition within $\sim$10\,AU. The analysis
     of the shape and strength of both the amorphous $10\,\mu$m
     feature and the crystalline feature around 23\,$\mu$m provides
     evidence for the prevalence of $\mu$m-sized (amorphous and
     crystalline) grains in upper layers of disks.}
   { The abundant crystalline silicates found far from their presumed
     formation regions suggests efficient outward radial transport
     mechanisms in the disks around TTauri disks. The presence of
     $\mu$m-sized grains in disk atmospheres, despite the short
     time-scales for settling to the midplane, suggests efficient
     (turbulent) vertical diffusion, likely accompanied by grain-grain
     fragmentation to balance the efficient growth expected. In this
     scenario, the depletion of submicron-sized grains in the upper layers of the disks points toward removal mechanisms such as stellar winds or
     radiation pressure.}

 \authorrunning{Olofsson et al.}  \titlerunning{Crystalline silicates
   in T\,Tauri disks}

   \maketitle
%
\section{Introduction}

The silicate dust grains that are originally incorporated into stellar
nebulae and eventually constitute planet forming disks are  thought to be of interstellar medium (ISM) size and composition,
namely sub-micron in diameter and extremely amorphous ($>$99\%) in
structure \citep[e.g.][]{Gail1998}. Solar System comets, on the other
hand, show high crystallinity fractions. The silicate grains in Oort comet Hale-Bopp are for instance composed of 40 to 60\% of Mg-rich crystalline grains \citep{Wooden1999,Wooden2007a,Crovisier1997}, while Jupiter Family comets have slightly lower crystalline fractions \citep[of about 25--35\%, e.g. comet 9P/Tempel 1,][]{Harker2007}. Although the actual crystalline fraction may depend somewhat on the methodology used to calculate the dust optical properties (e.g. \citealp{Min2008}), the clear detection of crystalline silicates features in comet spectra indicates their mass fraction to be high compared to the ISM.  The amorphous silicate grains that were incorporated into the Solar nebula must have been processed in the very early stages of planet formation before being incorporated into comets. Because of the high temperatures required to thermally anneal amorphous silicates and modify their lattice structure, it is generally accepted that the crystallization of amorphous silicates occured close to the young Sun. The crystalline silicates were then transported outwards due to processes such as turbulence or winds, leading to possible radial variations that may explain some of the mineralogic differences observed between small Solar System bodies today. Other processes could also explain the presence of crystals in the outer regions, such as shock waves triggered by gravitational instabilities in the solar nebula as described in \citet{Desch2002} and \citet{Harker2002}. Locally, such shocks could sufficiently heat up surrounding materials to anneal and crystallize amorphous grains.  In all these scenarii, silicate
crystalline grains can be considered as tracers of the history of the protoplanetary, circumsolar disk. 

The detection of silicates in the disks around young stars was
notoriously difficult until the {\it Infrared Space Observatory} (ISO)
and, later, the {\it Spitzer Space Telescope} were launched. Thanks to
ISO, mid- to far-IR spectroscopy of disks around young intermediate
mass stars (Herbig Ae/Be stars, hereafter, HAeBe) could be
performed. \citet{Acke2004} showed that 52\% of HAeBe stars exhibit
emission in the $10\,\mu$m (amorphous) silicate feature that arises
from the upper layers of the disks, while 23\% of these stars show features at
$11.3\,\mu$m, which is associated with forsterite, a Mg-rich
crystalline silicate. Using the mid-IR capabilities of MIDI on the VLT
Interferometer, \citet{2004} spatially resolved the 10\,$\mu$m
emission zone for three HAeBe stars. They found a gradient of
crystallinity, with a larger fraction of enstatite and forsterite
crystalline grains located close to the star (1--2\,AU) than further
away in the disk (2--20\,AU). This was among the first observations
where the disk dust content was proven to be radially inhomogeneous.

Although TTauri stars were too faint for observations with the ISO/SWS
and LWS, \citet{Natta2000} employed the lower spectral resolution
ISOPHOT instrument to detect the broad amorphous silicate emission
feature at around $10\,\mu$m for 9 stars in the Chameleon I cloud; but
the low resolution, the low signal-to-noise ratio and the limited
spectral range did not allow firm detection of any crystalline
features in the disks around young solar mass stars. Pioneering
ground-based observations of TTauri stars (hereafter TTs) nevertheless
showed evidence for forsterite emission features in the terrestrial
spectral window around $10\,\mu$m. \citet{Honda2003}, for example,
clearly identified the presence of crystalline silicate emission
features (forsterite at $10.1$, $10.5$ and $11.2\,\mu$m, and possibly
enstatite at $10.9\,\mu$m) in the N-band spectrum ($R\sim 250$) of
\object{Hen~3-600A}, and argued that the crystalline grains represent
about 50\% (30\% enstatite, 20\% forsterite) of the total mass of
grains emitting at $10\,\mu$m. Because of the difficulty of
disentangling between grain growth, the emission from PAH (Polycyclic Aromatic
Hydrocarbon) or from forsterite as the main cause of the
$11.3\,\mu$m ``feature'' in some disks (see e.g. \citealp{Sitko2000}
and the discussion in \citealp{Przygodda2003}), an unambiguous
assessment of the presence of crystalline silicates in large samples
of TTauri disks has awaited the launch of Spitzer.

The high sensitivity and large spectral range ($5$--$35\,\mu$m) of the
IRS spectroscopic instrument onboard Spitzer permits routine
mid-infrared spectroscopy of TTauri disks in nearby molecular clouds.
Surveys of young solar analogs in different star forming regions such
as those by \citet{Furlan2006} and \citet{Watson2009} for the Taurus
cloud, and of amorphous silicate features in other clouds by
\citet{Kessler-Silacci2006}, were performed. Detailed mineralogic fits
analyzing the amorphous and crystalline silicate fractions in TTauri
disks were performed by \citet[][7 TTs]{Bouwman2008} and \citet[][65
TTs]{Sargent2009a}, as well as for disks around much lower mass objects
like the borderline brown dwarf (hereafter, BD) \object{SST-Lup3-1}
\citep{Mer'in2007}, and the BD \object{2MASS J04442713+2512164} in the
Taurus cloud \citep{Bouy2008}, both of which show abundant crystalline
silicate grains despite their low (sub-)stellar
temperatures. According to \citet{Bouwman2008}, the crystalline
content of the inner regions of TTauri disks appear to be mostly
dominated by enstatite silicates while the crystalline content of the
outer region instead seems dominated by forsterite grains. Such
data point toward an inhomogeneous dust content within the disks which
could result from a radial dependence of crystallisation processes, or
a difference in the initial conditions under which these crystals
formed. The evolution of crystalline silicates has also been studied
by \citet[][84 TTs]{Watson2009}, who find that within the same
environmental conditions, the silicate crystalline mass fraction
varies greatly, from none to nearly 100\%. They find that 90\% of the
objects in their sample present crystalline features in the 10\,$\mu$m
region (either the 9.2\,$\mu$m enstatite feature or the 11.3\,$\mu$m
forsterite feature). Regarding other crystalline features at longer
wavelengths, they found that 50\% of the objects surveyed had
detectable crystalline features at $\sim$33\,$\mu$m.  Also, no strong
correlations between crystalline indices and stellar parameters have
been found, suggesting that some mechanisms (e.g. X-ray emission,
giant planets formation or migration) operate to erase the
correlations that are expected from the standard models of
crystallisation processes.

Here we present a comprehensive study of crystalline silicates in more
than a hundred disks around young stars based on 5--35$\,\mu$m IRS
spectroscopic observations obtained as part of the c2d Spitzer Legacy
Program “From Molecular Clouds to Planets” \citep{Evans2003}. This
work is the continuation of a series of c2d/IRS papers, studying
grains, PAHs and gas in inner disk regions around young stellar
objects (class II). Paper I by \citet{Kessler-Silacci2006} focuses on
amorphous silicate features as a proxy for grain growth in a subsample
of the c2d IRS Class II objects. They find that the grain size is a
function of spectral type. Paper II by \citet{Geers2006} shows
evidence for low fractions of PAHs around TTs, indicating low gas
phase PAHs abundance in disks as compared to the ISM. Paper III by
\citet{Lahuis2007} reports the first detections of [Ne II] and [Fe I]
gas-phase lines in TTauri disks.

In this paper, we present in Sec.\,\ref{secobs} IRS observations of
the entire c2d/IRS sample of class II objects, and we provide in
Sec.\,\ref{sec:overview} an overview of the detection statistics of
solid-state emission features, especially those from crystalline
silicates.  In Sec.\,\ref{sec:crystals} we study the properties of
crystalline silicate grains, including their size, and search for
correlations between amorphous and crystalline silicates and between
the SED shape and grain size. We further investigate the difference in
crystallinity between the inner and outer regions of disks. In
Sec.\,\ref{sec:discussion} we discuss the implications of our results
for the dynamics of disks at planet-forming radii, and we summarize
our results in Sec.\,\ref{sec:summary}. In a companion paper (Olofsson
et al. 2009b, in prep.), we analyse in more detail the dust mineralogy
based on compositional fitting of the complete spectra.

\section{Spitzer/IRS observations}\label{secobs}

We present in this study the infrared spectra of disks around 108
young stellar objects obtained as part of the c2d Spitzer Legacy
program. The spectra were obtained using the IRS instrument onboard
the Spitzer Space Telescope. It was used to expand the early
spectroscopic studies of HAeBe stars with ISO, and of a few TTs
observed from the ground in the $\lambda \sim$ 10\,$\mu$m atmospheric
window. Thanks to the high sensitivity of Spitzer, a large sample of
disks around low-mass and solar-mass stars could be efficiently
observed.

The IRS spectrograph is composed of four different modules that enable
a wavelength coverage from 5 to 37\,$\mu$m. Two modules, Short-Low and
Long-Low (SL and LL hereafter), provide a spectral resolution between
60-127 over ranges of 5.2\,-\,14.5\,$\mu$m and 14.0\,-\,38.0\,$\mu$m
for SL and LL, respectively.  The remaining two modules, Short-High
and Long-High (SH and LH hereafter), have a spectral resolution of
about 600. They cover more limited spectral ranges of
9.9\,-\,19.6\,$\mu$m and 18.7\,-\,37.2\,$\mu$m for SH and LH,
respectively.

\subsection{Stellar sample}

In this paper, we focus on Class II objects -- young stars no longer embedded in their protostellar envelope and surrounded by a circumstellar disk, where the dust may already be significantly processed. Our selection contains objects classified as Class II according to the literature, but also 10 objects with no known spectral type. In this sample, 30 objects are sources newly discovered with the instrument IRAC onboard Spitzer as part of the mapping phase of the c2d project. All such objects are named with the prefix SSTc2d, and are classified as Class II objects according to \citet{Evans2009} except for one source with no known classification (SSTc2d\,J182909.8+03446). The object \object{SSTc2d\,J162221.0-230403} is characterized as Class\,I, but shows amorphous 10\,$\mu$m emission without additional absorption features.  Two other SSTc2d objects have no classification. However, for the 10 objects with no classification in the literature, they were retained in the full list as they clearly show amorphous silicate features in emission.

The final sample contains 108 stars including 60 TTs, 9 HAeBe and 1 BD. The 38 remaining objects have no spectral classification in the literature, or are the new SSTc2d stars discovered with Spitzer. This sample is distributed toward several clouds: Perseus (16 objects), Taurus (9), Chamaeleon (23), Ophiuchus (25), Lupus (16), Serpens (15), and includes 4 isolated stars: 3 HAeBes (\object{BF\,Ori}, \object{HD\,132947} and \object{HD\,135344}) and one TT star (\object{IRAS\,08267-3336}). All six clouds are young star forming regions (1--5\,Myr), located at distances within 140 and 260\,pc from the Sun, with star densities (number of young stellar objects per square parsec) lying between 3 and 13. Some clouds are therefore quite extended, like Taurus or Lupus. The main cloud properties are given in \citet{Evans2009}. The complete list of IRS targets analysed in this paper, their classifications and spectral types, as well as details about the observations, are given in Table\,\ref{obs}.

Out of the 108 objects, four display spectra unusual for Class\,II objects. These objects are \object{DL\,Cha}, \object{HD\,132947}, \object{SSTc2d\,J182849.4+00604} and \object{SSTc2d\,J182859.5+03003}. The first three sources all display spectra reminiscent of those observed toward Oxygen-rich Asymptotic Giant Branch stars (see \citealp{Sloan2003} for example spectra) while the fourth object displays a spectrum similar to a C-rich AGB star. Both SSTc2d objects located in the Serpens cloud are objects newly discovered using Spitzer/IRAC surveys. DL\,Cha is a known variable star in Chamaeleon and HD\,132947 is an isolated Herbig star, classified in the literature as a Pre-Main Sequence star (see e.g. \citealp{Valenti2003}). We do not question here the classification of these four stars but as they display unusual spectra (see Fig.\,\ref{sp:agb}) we decided to remove them from the rest of the study.

\subsection{Data reduction}

The spectra presented in this paper were extracted from S13 pre-reduced (BCD) data using the c2d legacy team pipeline \citep{Lahuis2006} which makes available two different extraction methods: full aperture extraction and optimal Point Spread Function (PSF) extraction. The full aperture extraction method is applied to both low and high spectral resolution modules. For the low resolution modules, the c2d pipeline implements an extraction with a fixed-width aperture over the whole order. The source position in the slit is determined and the aperture is then centered on the source. The width is such that 99\% of the flux of a point-source falls within the window. For the high resolution modules, the full slit width is used for the extraction. One disadvantage of this method is that the reduced spectra present a number of spikes (mainly due to bad pixels response, or ``hot'' pixels on the array, see below) that cannot be easily removed by hands-off pipelines.

For the PSF extraction, the observed signal is assumed to be that of a
point source or slightly resolved source plus an uniform zero
level. This zero level, in most cases, represents the local extended
emission close to the source, and the typical PSF profile is
calculated from high signal-to-noise ratio calibrator data. As
  this zero-level may also contain residuals from, for example, raw
  pipeline dark current, we remove this contribution. PSF fitting is
less sensitive to bad data samples and unidentified bad pixels than
the aperture extraction method. The {\it Spitzer Science Center}
  provides masks with known bad pixels, but they are not all
  identified. Our pipeline detects such remaining
  pixels. The LH array is particularly affected by this issue, but
  other modules require attention. With the PSF extraction method, bad
  pixels are simply eliminated, whereas their values are simply
  interpolated for full aperture extraction method. The PSF extraction
  method, in addition, provides an estimate of the data zero level or
sky contribution to the observed spectrum.  An important disadvantage
of the PSF fitting method, though, is that for some modules (mainly
for the $2^{\mathrm{nd}}$ orders), the PSF is subpixel in size and the
extraction method can become unstable. Thus, for some sources, the
$2^{\mathrm{nd}}$ order of the SL module and the $2^{\mathrm{nd}}$
order of the LL module are unavailable, leading to incomplete spectra.

In this paper, we adopt the PSF extraction method in order to obtain
spectra with as few spikes as possible. Where the PSF extraction
  method is not available or turns out to be unstable, we use the full
  aperture extraction method to build spectra. We then opt for a
median filtering scheme to remove the remaining spikes: for the SL and
SH modules we use a median smoothing over three channels while we use
a median smoothing over five channels for the LL and LH modules. Some
problems were encountered with the data reduction for two objects in
our sample: EC69 and ISO-Cha237. For the latter source there are two
possible explanations: either the LL1 module spectrum poorly matches
the LL2 module spectrum, or the LL2 module pipeline reduction failed
and its slope is off. The second explanation seems the most likely,
given the shape of the red side of the 10\,$\mu$m feature. For EC69,
flux values for some orders were null. We decided to keep these two
objects in our sample, first because ISO-Cha237 shows a 10\,$\mu$m
feature that can be analyzed.  For EC69, the spectrum appears
featureless, and as it is classified as a TTauri star, we choose not
to bias our results by removed ``problematic'' spectra.

\subsection{Module merging and offset correction\label{offset_corr}}

The c2d pipeline corrects for telescope pointing errors that may lead to important flux offsets between the different modules. Some small offsets can nevertheless remain between modules even after pointing correction, and an additionnal post-processing offset correction is thus applied. This correction depends on the modules available for every spectrum. Table\,\ref{mod_cas} gives a summary of all the possible configurations we encountered. A tick-mark ($\surd$) means that data provided by this module are used to build the final spectrum. Priority was given to SL and LL, first because the spectral resolution is sufficient for detecting crystalline features, and second, because the high resolution modules provide noisier spectra than those from the low resolution modules. For each star, a final spectrum is then obtained by merging all the chosen modules and correcting manually for small remaining offsets. The spectra themselves are presented in Fig.\,\ref{spectre}, Fig.\,\ref{fig:C23} and Figs.\,\ref{sp:perseus}--\ref{sp:others}.

\begin{table}
\begin{center}
  \caption{\label{mod_cas}Overview of the modules used in all the
    different cases encountered to obtain the final spectra (see
    Sec.\,\ref{offset_corr} for details). The exact modules used for each
    individual star are given in Table\,\ref{obs}.}
\begin{tabular}{cccccccc}
\hline\hline
Case & SL1 & SL2 & SH & LL1 & LL2 & LH & Number of stars \\ \hline
0 & -   & -   & $\surd$ & - & - & $\surd$ & 9 \\
1 & - & - & $\surd$ & $\surd$ & - & $\surd$ & 2 \\
2 & $\surd$ & $\surd$ & - & $\surd$ & $\surd$ & - & 51 \\
3 & $\surd$ & $\surd$ & $\surd$ & - & - & $\surd$ & 18 \\
4 & $\surd$ & $\surd$ & $\surd$ & $\surd$ & - & $\surd$ & 22 \\
5 & $\surd$ & $\surd$ & $\surd$ & $\surd$ & $\surd$ & - & 6 \\
\hline
\end{tabular}
 \begin{list}{}{}
 \item Note: SL1: $7.4$--$14.5\,\mu$m , SL2: $5.2$ -- $7.7\,\mu$m ,
   LL1: $19.5$ -- $38.0\,\mu$m, LL2: $14.0$ -- $21.3\,\mu$m
 \end{list}
\end{center}
\end{table}

\subsection{Estimating the uncertainties}
\label{sec:uncertainties}

\begin{figure*}
\resizebox{\hsize}{!}{\includegraphics[angle=90,]{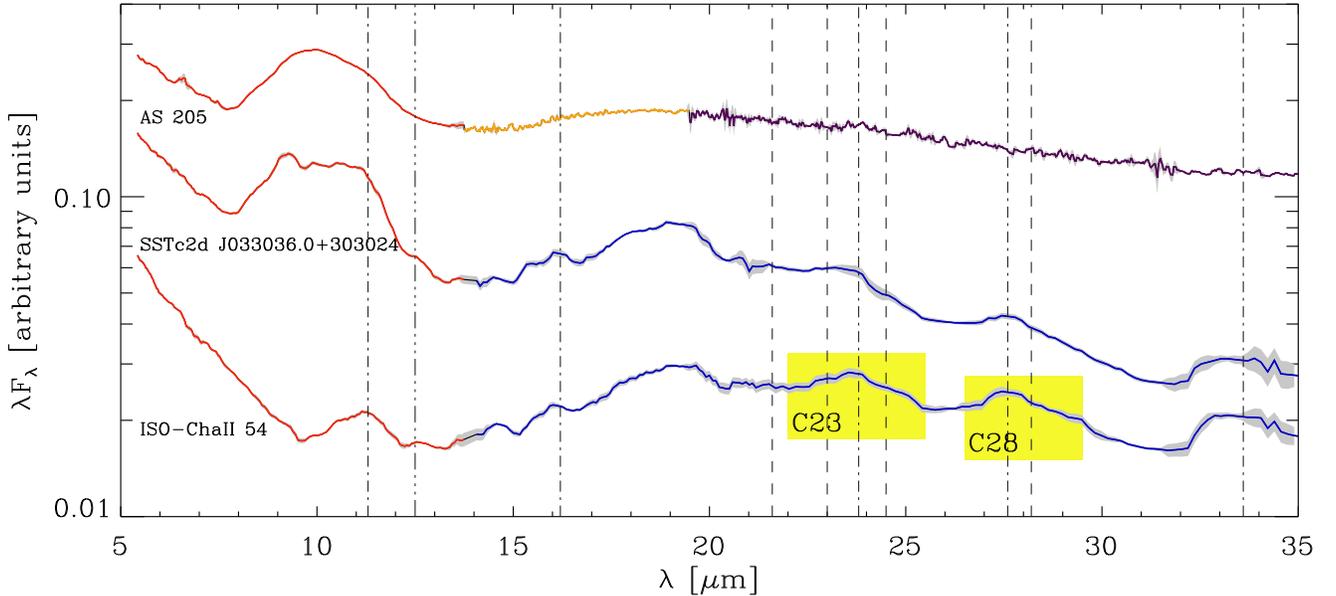}}
\caption{\label{spectre}Example Spitzer/IRS spectra of 3 T\,Tauri
  stars (top spectrum: SL+SH+LH modules, bottom spectrum: SL+LL). The
  grey envelopes correspond to 3-$\sigma$ uncertainties as estimated
  in Sec.\,\ref{sec:uncertainties}. The vertical lines show the peak
  positions of enstatite (dashed lines) and forsterite (dot-dashed
  lines) crystalline features that were searched for in every spectrum of 
  our 108 star sample (see also Table\,\ref{feat}). The 12.5\,$\mu$m
  feature (triple-dot dashed line) is attributed to silica, and the two
  yellow boxes correspond to the C23 and C28 crystalline complexes
  discussed in Sec.\,\ref{sec:forsens}.}
\end{figure*}

The uncertainties propagated through the c2d pipeline are dominated by
the flux calibration error estimates. As our study is oriented toward
the detection of features in emission  from silicates, we are only concerned
with relative uncertainties which are evaluated by dividing the
original spectrum by a smoothed version of itself obtained using a
Savitzky-Golay filter \citep{Savitzky1964}. This provides an estimate
of the signal-to-noise ratio (SNR) on the spectrum, and is used to
assess the presence of features as follows: The Savitzky-Golay
filtering process calculates a local 3$^{\rm{rd}}$-order polynomial
for every point of the spectrum, based on the four left and right
neighbors. The main advantage of this filter is that it maintains the
shape and contrast of the features, while ``pointy'' features tend to
be rounded out with classical average smoothing, thereby increasing
the uncertainties at the top of every feature and downgrading the
actual quality of the data. The Savitzky-Golay filter has only
  been used to calculate the uncertainties displayed as grey envelopes
  in Fig.\,\ref{spectre}, Fig.\,\ref{fig:C23} and
  Figs.\,\ref{sp:perseus}--\ref{sp:others}, not affecting the spectra
  themself presented on these figures and used for the scientific
  analysis.

\begin{table}
  \caption{\label{feat}Summary of the different crystalline silicate
    features examined in this study.}
\begin{tabular}{ccccc}
\hline\hline
Crystal & Peak  & Measured & Measured $\Delta \lambda$ \\ 
 & $\lambda$ [$\mu$m] & peak $\lambda$ [$\mu$m] & Eq. Width [$\mu$m] \\
\hline
\begin{tabular}{l} Enstatite \\ (MgSiO$_{3}$) \end{tabular}
& \begin{tabular}{l} 9.2$^{(1,2)}$ \\ 21.6$^{(1)}$ \\ 23.0$^{(1,2)}$ \\ 24.5$^{(1,2)}$ \\ 28.2$^{(1,2)}$ \end{tabular}
& \begin{tabular}{l} 9.24$\pm$0.08 \\ 21.47$\pm$0.16 \\ C23 \\ C23 \\ C28 \end{tabular}
& \begin{tabular}{l} 0.49$\pm$0.11 \\ 0.45$\pm$0.25 \\ C23 \\ C23 \\ C28 \end{tabular} \\
 \hline 
\begin{tabular}{l} Forsterite \\ (Mg$_{2}$SiO$_{4}$) \end{tabular}
& \begin{tabular}{l} 11.3$^{(2)}$ \\ 16.2$^{(2)}$ \\ 23.8$^{(2)}$ \\ 27.6$^{(2)}$ \\ 33.6$^{(2)}$ \end{tabular}
& \begin{tabular}{l} 11.26$\pm$0.12 \\ 16.08$\pm$0.16 \\ C23 \\ C28 \\ 33.1$\pm$0.63 \end{tabular}
& \begin{tabular}{l} 0.60$\pm$0.33 \\ 0.38$\pm$0.16 \\ C23 \\ C28 \\ 1.19$\pm$0.47 \end{tabular} \\
\hline
\begin{tabular}{l} Diopside \\ (CaMgSi$_{2}$O$_{6}$) \end{tabular} & 25.0$^{(2)}$ & 25.14$\pm$0.18 & 0.58$\pm$0.22 \\ \hline
\end{tabular}
\begin{list}{}{}
\item References: $^{(1)}$\citet{Koike2000}, $^{(2)}$\citet{Molster2002b}
\end{list}
\end{table}

\section{Overview of observed solid-state
  features\label{sec:overview}}

\subsection{Observed crystalline silicates}

\begin{figure}
\begin{center}
  \caption{\label{sp:perseus}Spitzer/IRS spectra of young stars in the
    Perseus cloud, in units of Jy. Gray areas correspond to 3-$\sigma$
    uncertainties as estimated in
    Sec.\,\ref{sec:uncertainties}. Sources are sorted by increasing
    Right Ascension. Red color is for SL module, orange for SH, blue
    for LL and dark purple for LH.}
  \resizebox{\hsize}{!}{\includegraphics{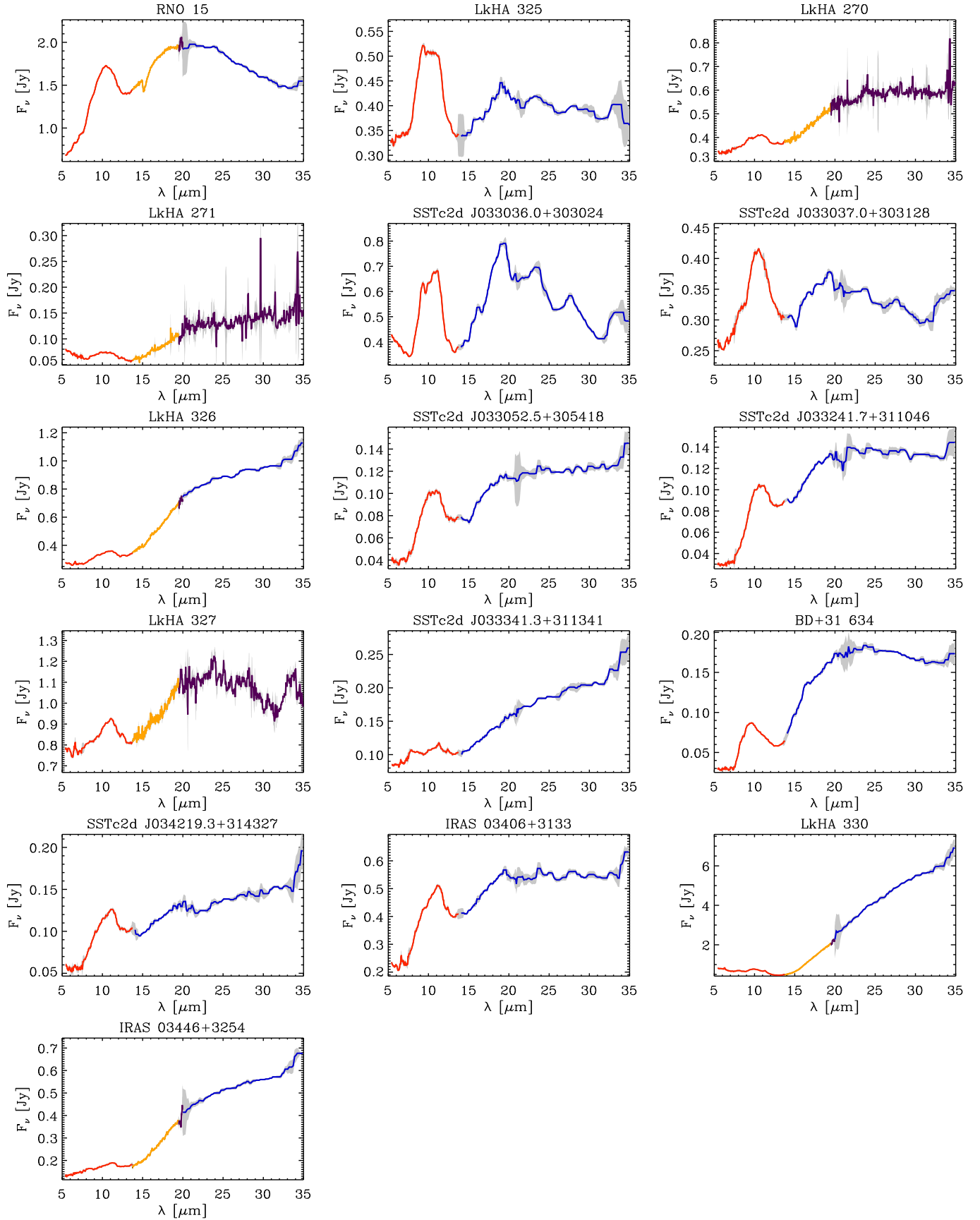}}
\end{center}
\end{figure}

\begin{figure}
\begin{center}
  \caption{\label{sp:lupus}Same as Fig.\,\ref{sp:perseus} but for the
    Lupus star sample}
\resizebox{\hsize}{!}{\includegraphics{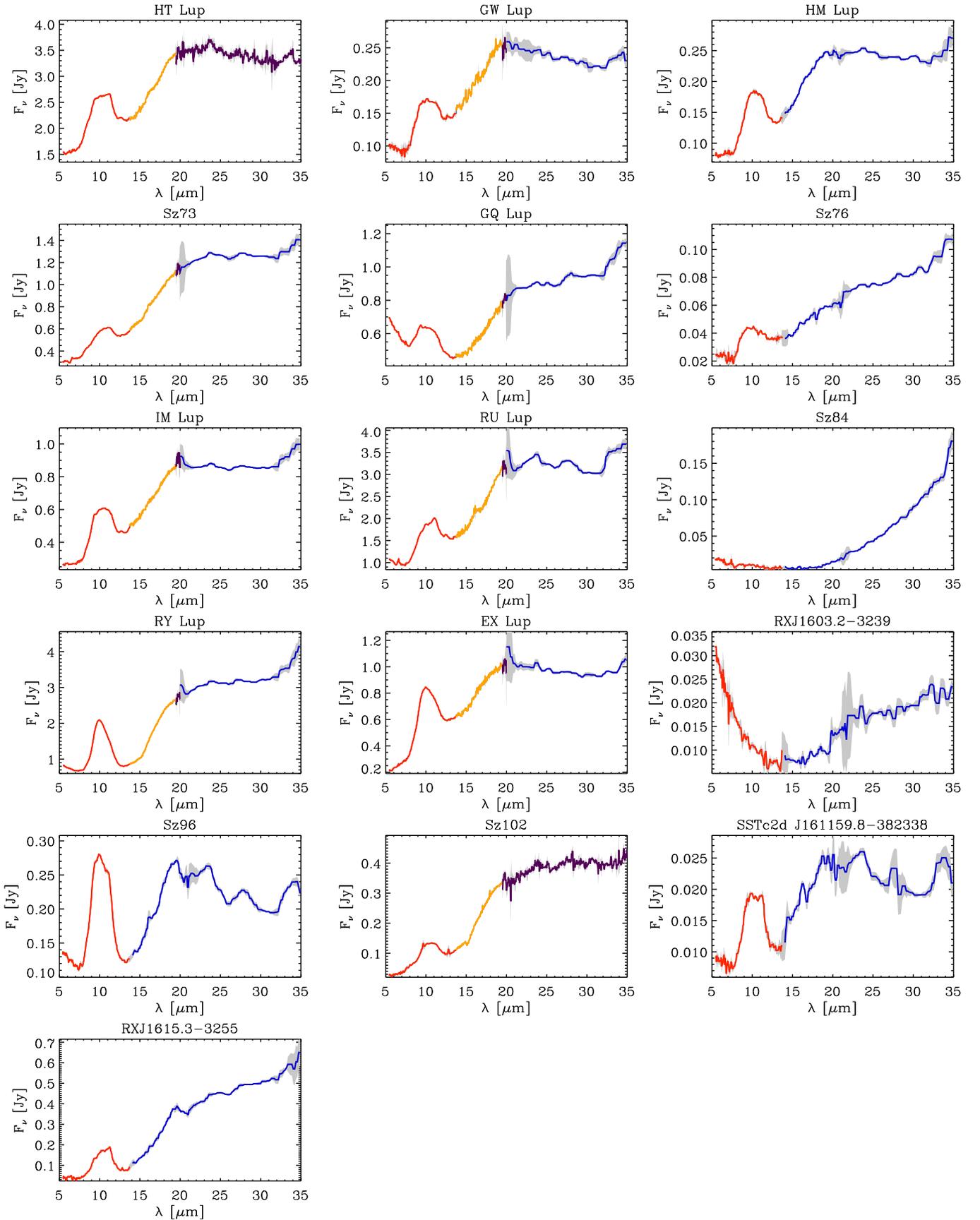}}
\end{center}
\end{figure}

\begin{figure}
\begin{center}
  \caption{\label{sp:oph_1}Same as Fig.\,\ref{sp:perseus} but for the
    Ophiuchus star sample.}
\resizebox{\hsize}{!}{\includegraphics{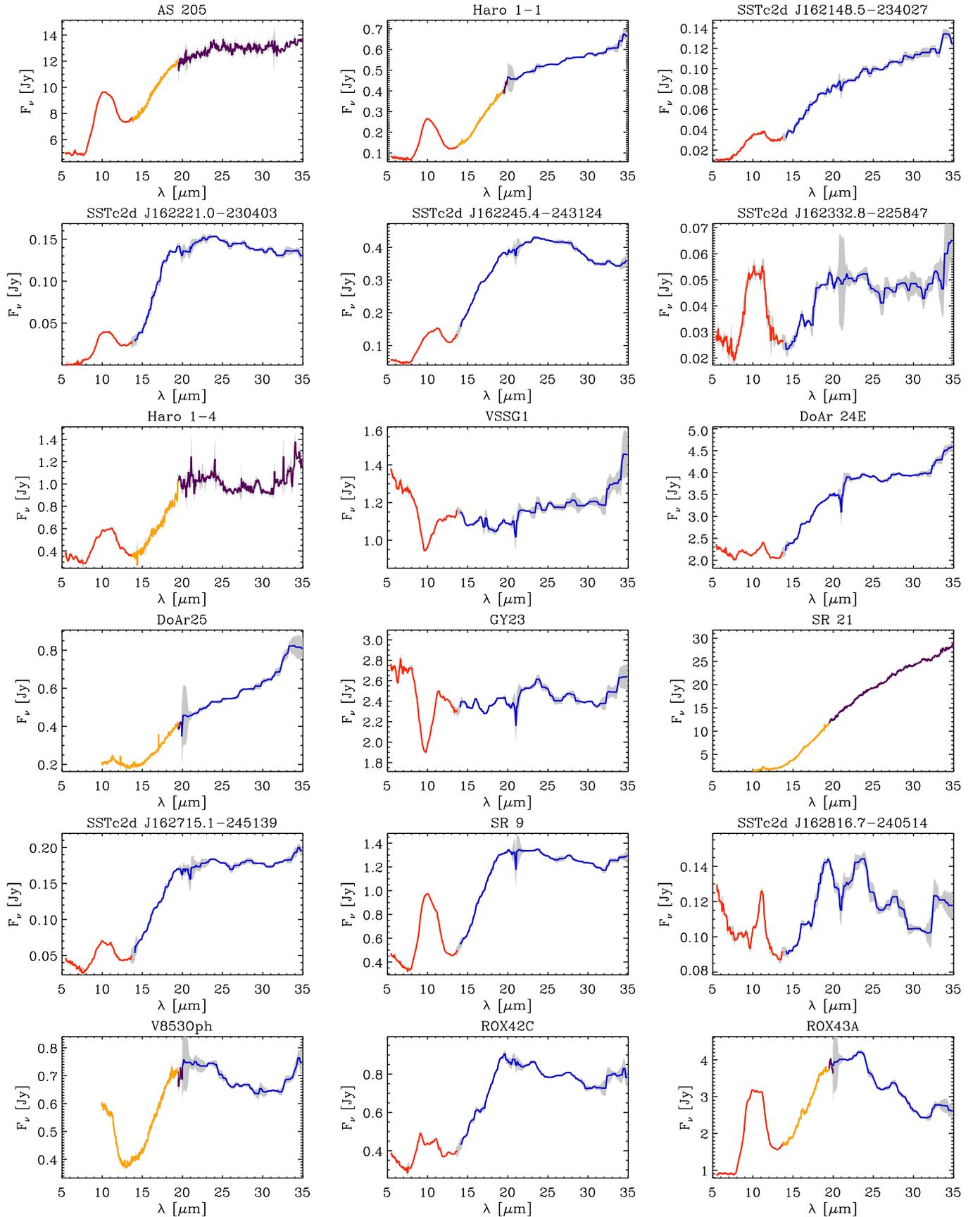}}
\end{center}
\end{figure}

\begin{figure}
\begin{center}
  \caption{\label{sp:oph_2}Spitzer/IRS spectra for the Ophiuchus
    sample: continued}
\resizebox{\hsize}{!}{\includegraphics{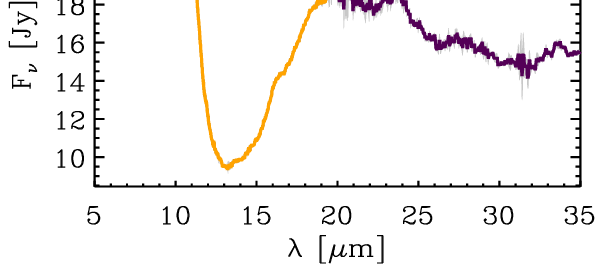}}
\end{center}
\end{figure}

\begin{figure}
\begin{center}
  \caption{\label{sp:serpens}Same as Fig.\,\ref{sp:perseus} but for
    the Serpens star sample.}
  \resizebox{\hsize}{!}{\includegraphics{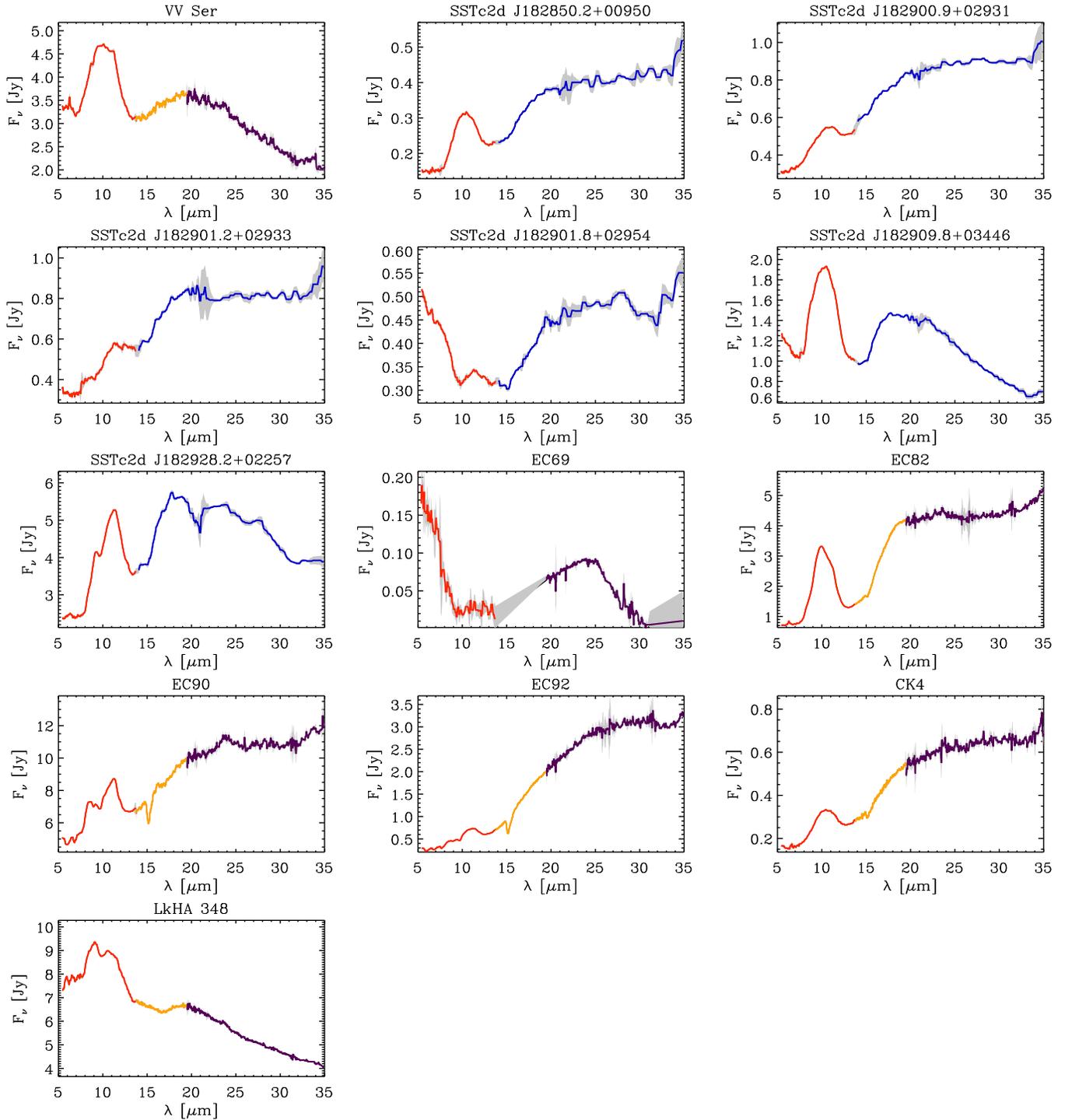}}
\end{center}
\end{figure}

\begin{figure}
\begin{center}
  \caption{\label{sp:others}Same as Fig.\,\ref{sp:perseus} but for
   the 3 isolated stars in our sample}
 \resizebox{\hsize}{!}{\includegraphics{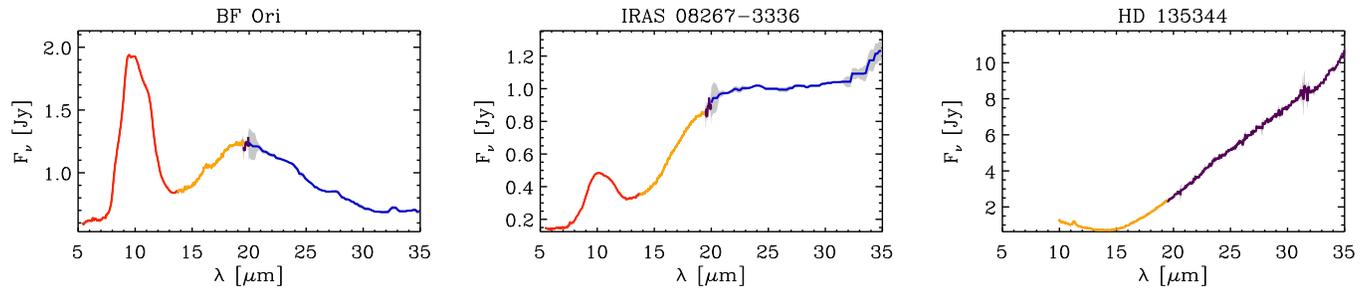}}
\end{center}
\end{figure}

\begin{figure}
\begin{center}
  \caption{\label{sp:agb}Same as Fig.\,\ref{sp:perseus} but for
   the 4 objects with spectra similar to C-rich and O-rich AGB stars}
 \resizebox{\hsize}{!}{\includegraphics{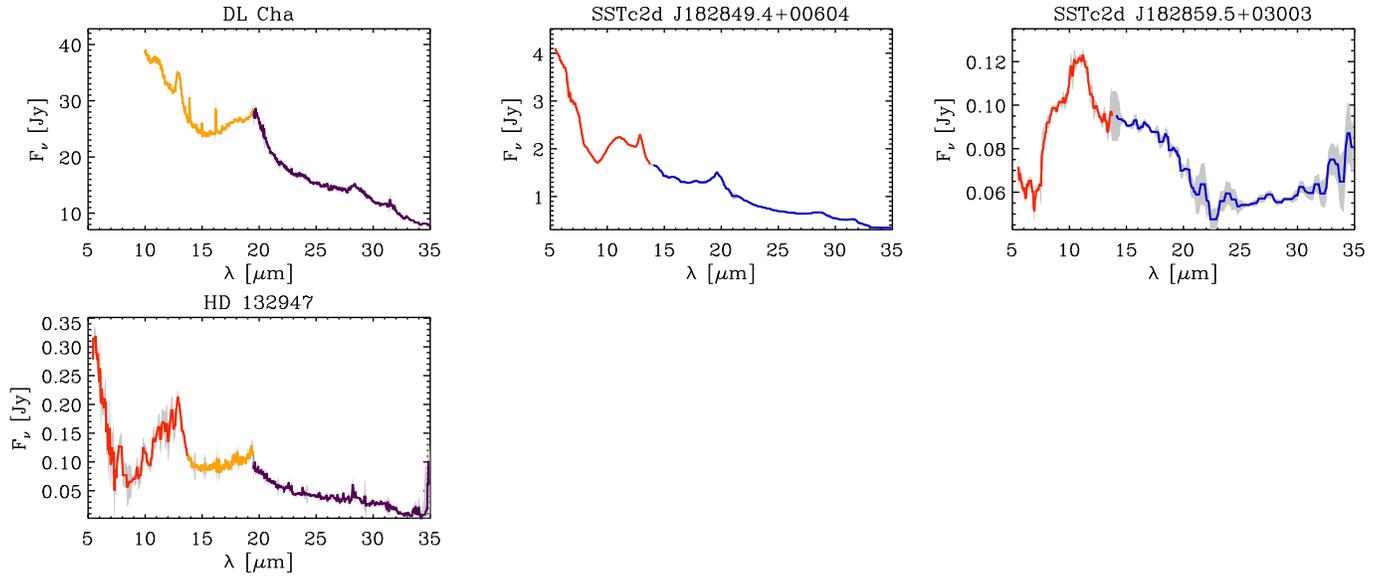}}
\end{center}
\end{figure}

The 104 spectra analyzed in this paper are displayed in
Figures~\ref{sp:perseus} to \ref{sp:others}, for wavelengths between 5
and 35\,$\mu$m and in units of Jy ($F_{\nu}$). They are sorted by
cloud and increasing Right Ascension. The spectra of the four isolated
stars in our sample are displayed in Fig.\,\ref{sp:others}.

Broad emission features at 10 and 20\,$\mu$m from amorphous silicates
can be identified in many spectra, as well as narrower emission
features. The latter are more easily identified beyond 20\,$\mu$m, and
are due to crystalline silicates. Figure\,\ref{spectre} illustrates
the diversity of the spectra seen in our
sample. \object{ISO-ChaII\,54} (bottom spectrum) shows many
crystalline features, with a weak 10\,$\mu$m band. In contrast,
\object{AS\,205} (top spectrum) is almost ``pristine'', with strong
amorphous bands and no evidence of crystalline
features\footnote{Extensive, narrow features not discussed in this
  paper, but seen between $10$ and $20\,\mu$m in Fig.\,\ref{spectre}
  are attributed to water emission lines \citep{Salyk2008}}. The third
spectrum, for \object{SSTc2d\,J033036.0+303024}, has both amorphous
and crystalline features. The crystalline features in disks can, in
most cases, be attributed to Mg-rich silicate minerals (with a
  possible contribution from Ca-rich silicates); in agreement with
some model predictions (e.g. \citealp{Gail2004} and ref. therein), and
as observed in solar system comets (e.g. \citealp{Wooden2007a}).

\begin{figure*}
\hbox to \textwidth
{
\parbox{0.5\textwidth}{
\includegraphics[angle=0,width=0.45\textwidth,origin=bl]{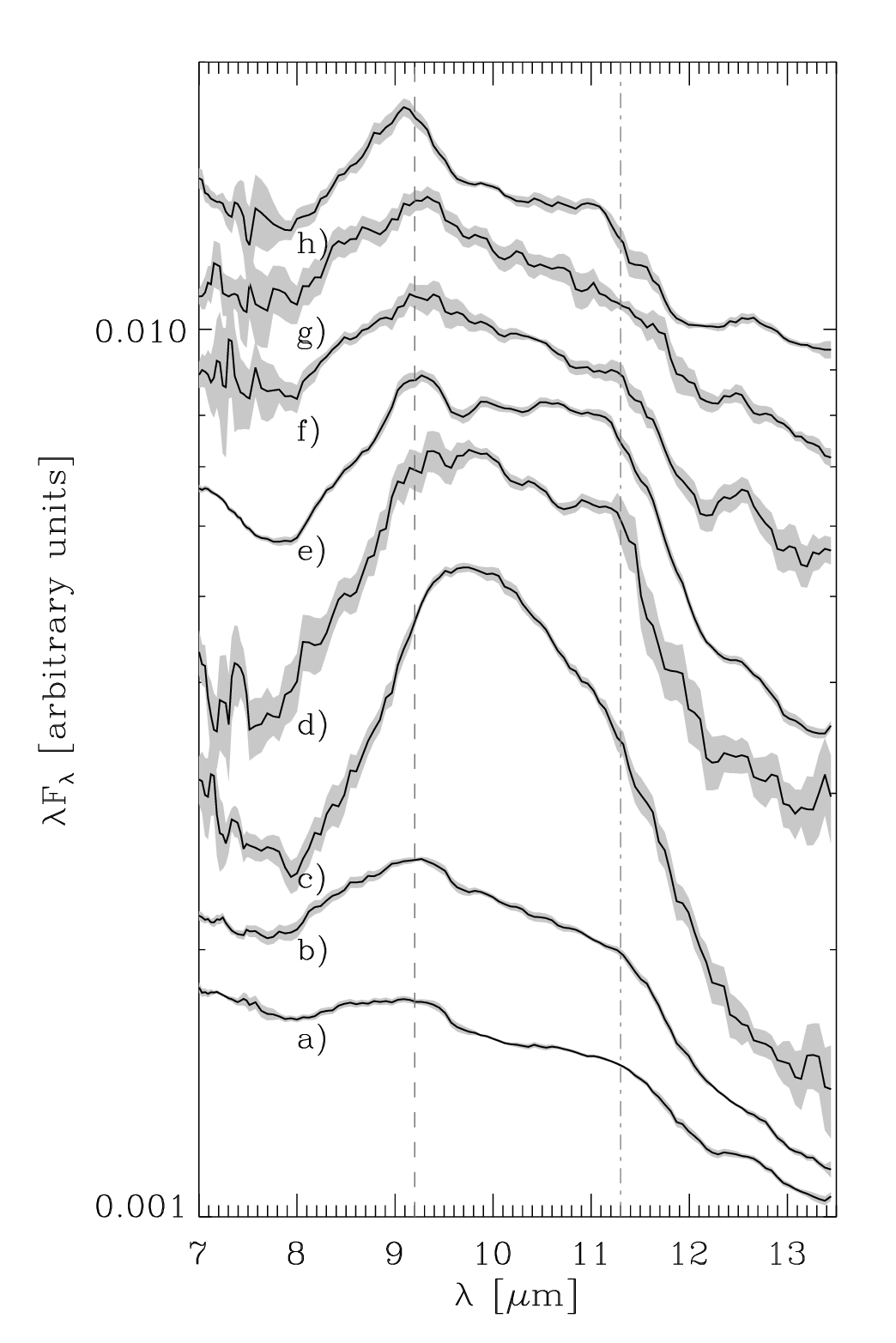}
}
\parbox{0.5\textwidth}{ 
\includegraphics[angle=0,width=0.45\textwidth,origin=bl]{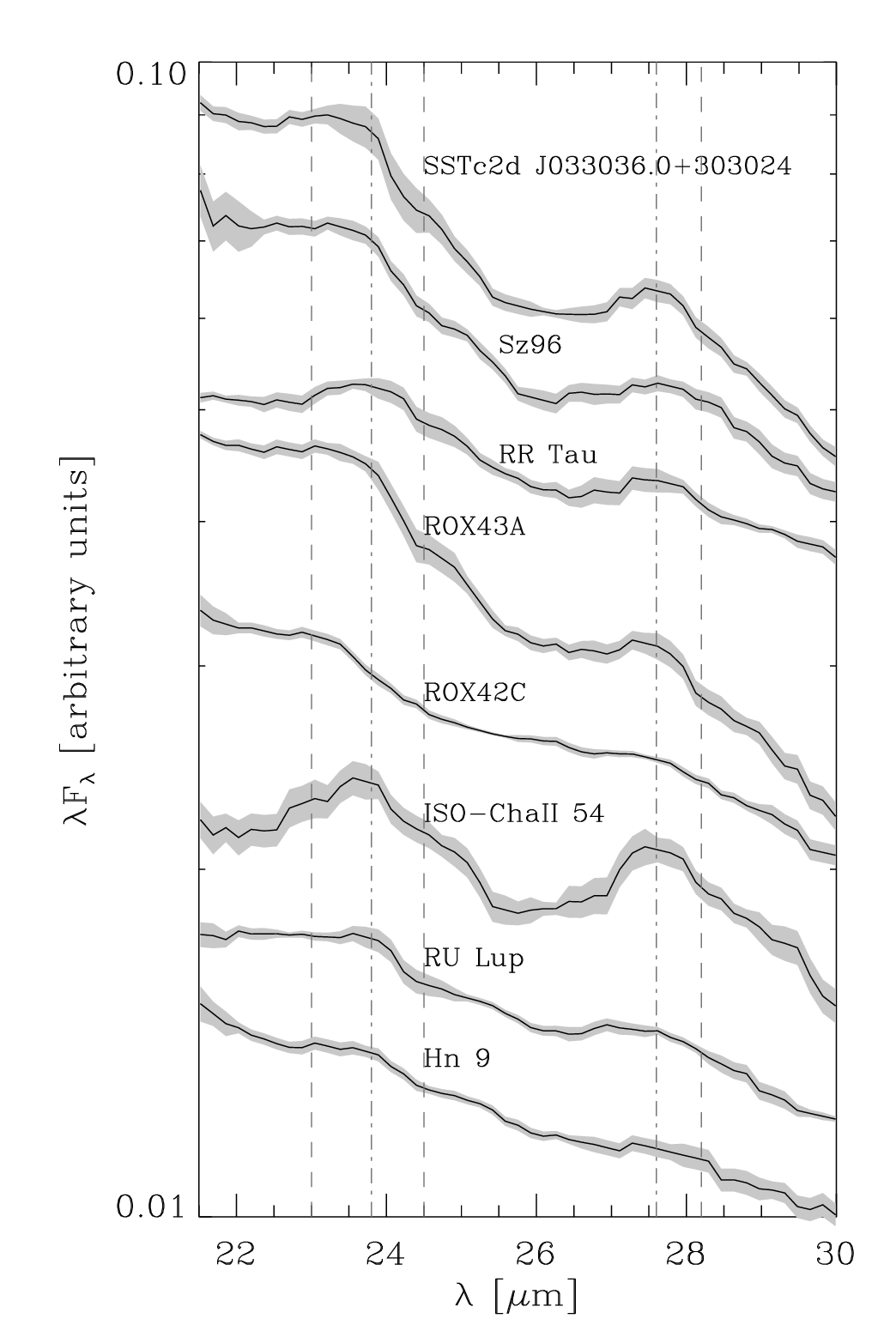}
}
}
\caption{\label{fig:C23}{\it Left panel}: Examples of the 9.2 (enstatite),
  11.3 (forsterite) and 12.5\,$\mu$m (silica) features superposed on
  the amorphous 10\,$\mu$m feature. {\it Right panel}: Examples of the C23
  and C28 crystalline silicate complexes. The grey areas correspond to
  3-$\sigma$ uncertainties. Dashed lines correspond to enstatite
  emission features, dot-dashed lines correspond to forsterite
  emission features. The sources in the left panel are
  a)~\object{Wa\,Oph\,6}, b)~\object{LkH$\alpha$\,325}, c)~\object{TW\,Cha},
  d)~\object{SSTc2d\,J161159.8-382338},
  e)~\object{SSTc2d\,J033036.0+303024}, f)~\object{Sz52},
  g)~\object{XX\,Cha} and h)~\object{ROX42C}.}
\end{figure*}

\subsubsection{Mg-rich crystalline silicates: Forsterite \& Enstatite}
\label{sec:forsens}
Table\,\ref{feat} presents the different crystalline silicate features
we searched for in every spectrum -- in addition to the amorphous
silicate feature at 10\,$\mu$m. In Table\,\ref{feat}, ``Peak
$\lambda$'' is taken from the literature, ``Measured $\lambda$'' and
``$\Delta \lambda$'' values are the peak positions and equivalent
widths measured by our observations. ``C23'' and ``C28'' refer to
silicate features blended in complexes. Uncertainties are standard
deviations. Here we consider features arising from Mg-rich silicates,
specifically enstatite (inosilicate belonging to the pyroxene group,
with chemical formula MgSiO$_{3}$) and forsterite, the nesosilicate
Mg-rich end-member of the olivine group (Mg$_{2}$SiO$_{4}$). Blowups
of representative spectral features are shown in
Fig.\,\ref{fig:C23}. Using IRS spectra, we cannot disentangle
differences between crystalline structures, for instance,
clino-enstatite or ortho-enstatite. Still, according to
\citet{Koike2000}, enstatite features are most likely attributed to
ortho-enstatite.

The example spectrum for ISO-ChaII\,54, displayed in
Fig.\,\ref{spectre} as well as the blowups in the right panel of
Fig.\,\ref{fig:C23}, shows that the 23.0, 24.5\,$\mu$m enstatite and
the 23.8\,$\mu$m forsterite features can be blended into one single
complex. Similarly, the 27.6\,$\mu$m forsterite feature can also be
blended with the 28.2\,$\mu$m enstatite feature. In the following, we
will treat these features as two complexes, at reference positions of
23 and 28\,$\mu$m, independent of the actual crystals responsible for
the observed emission.  We shall note these two complexes, C23 and
C28, which are by far the most frequently detected crystalline
features in disks (Sec.\,\ref{sec:measure}). Fig.\,\ref{fig:C23}
displays eight example spectra of the C23 and C28 complexes. Clearly,
the shape and peak positions may vary from one object to another. We
will show in Sec.\,\ref{sec:gsize} that some of these differences may
relate to the mean size of the emitting grains, although the
composition (more or less enstatite compared to forsterite), grain
shape, and strengh of the continuum might also contribute to the
diversity of the C23 and C28 features observed.

Beside the features listed in Table\,\ref{feat}, other enstatite bands
at 10.6, 11.7 and 19.6\,$\mu$m, as well as forsterite features at 10.1
and 19.7\,$\mu$m were also searched for with little success in a
subsample of 47 high SNR spectra. The detection rate of these
crystalline features is very low and not significant.  For the two
enstatite and forsterite features at $\sim$19.6 and 19.7\,$\mu$m, the
main difficulty lies in the added noise where SH and LH modules merge,
which prevents any firm detection of these two features.

\subsubsection{Fe-rich components:  Troilite \& Fayalite}
\label{sec:ferich}
\begin{figure}
\begin{center}
  \caption{\label{sp:chama_2}Spitzer/IRS spectra for the Chamaleon
    sample: continued}
\resizebox{\hsize}{!}{\includegraphics{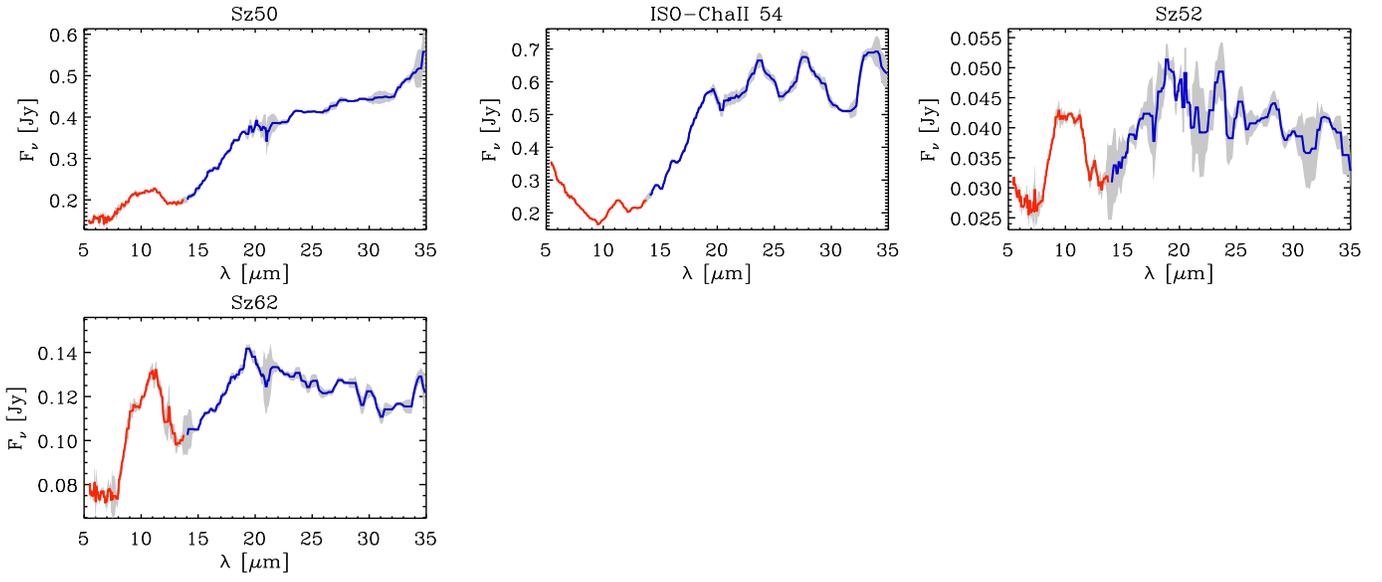}}
\end{center}
\end{figure}

The formation of Mg-rich crystalline silicates by direct condensation from a gas of solar composition (as opposed to formation by thermal annealing) is expected to additionally yield iron-bearing condensates, in particular FeS (or troilite, see \citealp{Wooden2007a}). This condensate is of great interest and, while a detailed study of the ISO spectra of two HAeBe stars (including \object{HD\,163296}, Fig.\,\ref{sp:chama_2}) carried out by \cite{Keller2002} indicates FeS in these systems, it remains difficult to detect, in general, because the main feature (between 17-26\,$\mu$m), centered at 23.5\,$\mu$m, is broad, and falls in regions where the IRS spectra are often noisy (due to module merging around 20\,$\mu$m). Further, the disk spectra are often dominated by the presence of the 18\,$\mu$m amorphous silicate feature as well as by frequent C23 emission features. Therefore, studying the presence or absence of FeS emission in the c2d sample is beyond the scope of this paper.

\begin{figure}
\begin{center}
  \caption{\label{sp:taurus}Same as Fig.\,\ref{sp:perseus} but for the
    Taurus star sample}
  \resizebox{\hsize}{!}{\includegraphics{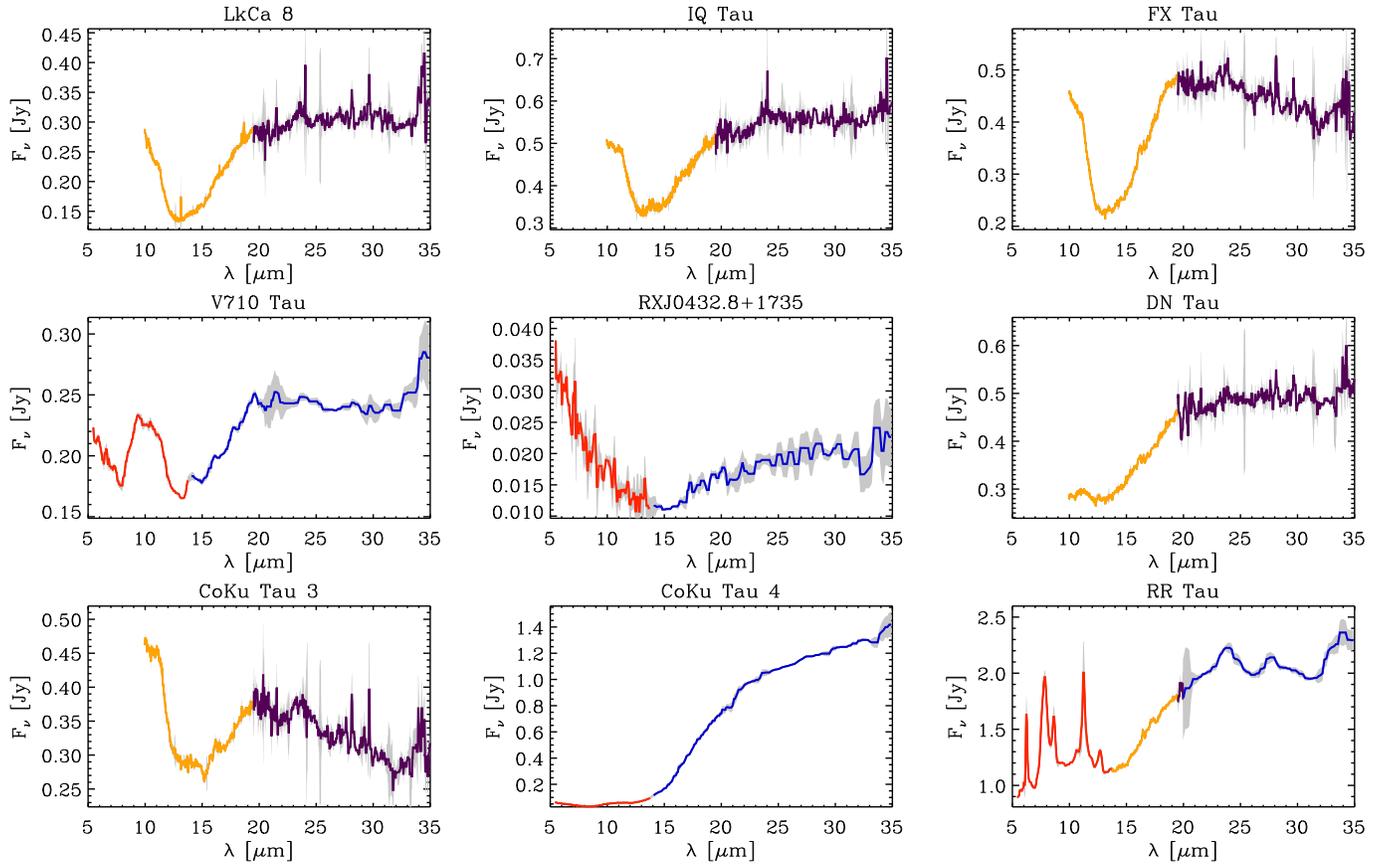}}
\end{center}
\end{figure}

\begin{figure}
\begin{center}
  \caption{\label{sp:chama_1}Same as Fig.\,\ref{sp:perseus} but for the
    Chamaleon star sample.}
\resizebox{\hsize}{!}{\includegraphics{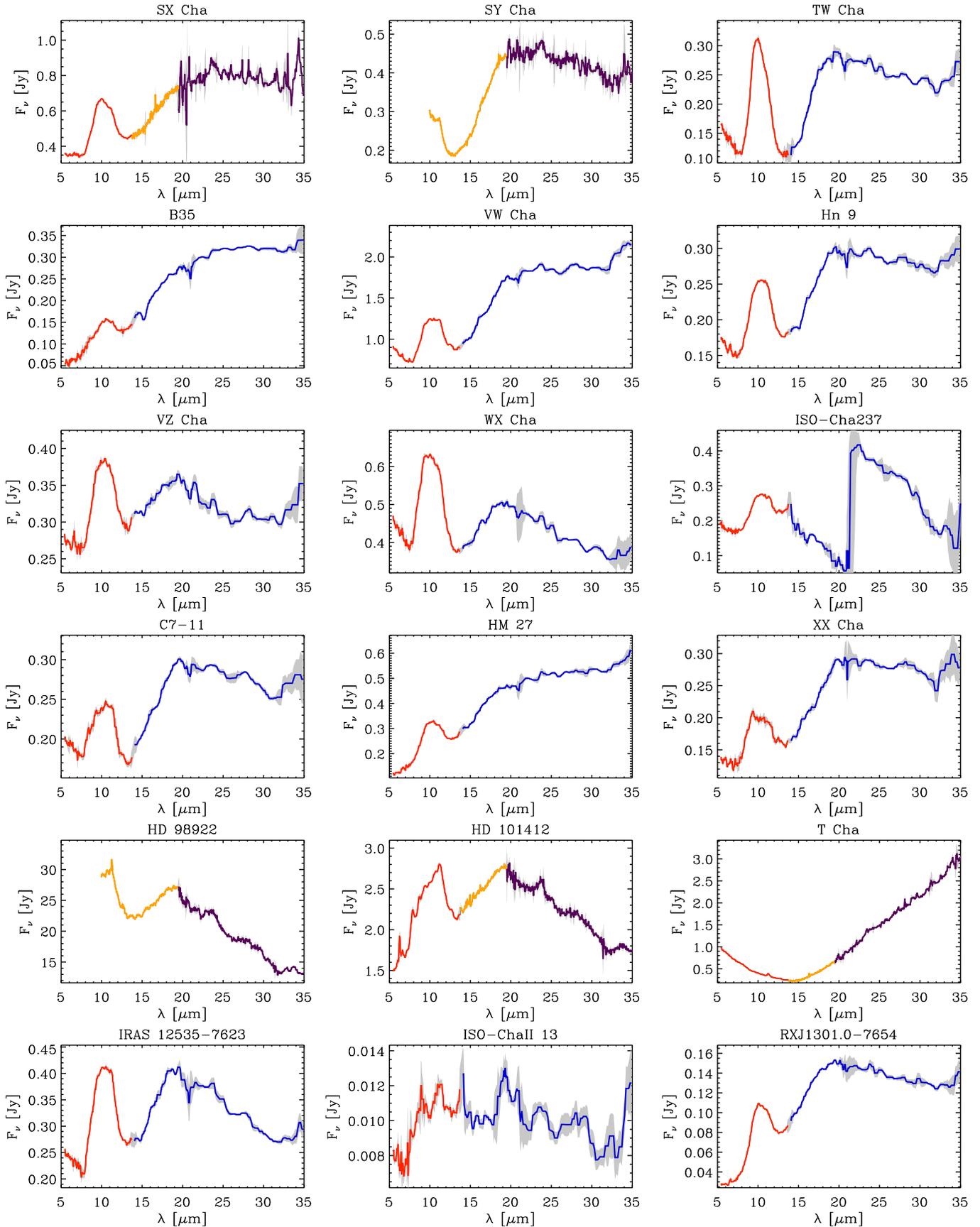}}
\end{center}
\end{figure}

Another Fe-rich component that can potentially be detected in IRS
spectra is fayalite, a crystal belonging to the olivine class
(Fe$_2$SiO$_4$). Theoretical opacities from \citet{Fabian2001} show
strong emission features around 30\,$\mu$m. For small grains, three
narrow features are expected at $\sim$27, 29.3 and 31.6\,$\mu$m.  We
find there are 4 objects where fayalite may tentatively be present,
all four in the Chameleon cloud (Fig.\,\ref{sp:chama_1}): TW\,Cha,
VZ\,Cha, WX\,Cha, and Sz62. Interestingly, the Sz62 spectrum also
shows other features that can be attributed to fayalite, at roughly
11\,$\mu$m and 19\,$\mu$m. All these objects notably display a
  feature centered at around 30.5\,$\mu$m, which is at shorter wavelength
  compared to the feature from pure fayalite which lies at
  31.6\,$\mu$m (e.g. \citealp{Fabian2001} or
  \citealp{Koike2003}). This may reflect a small departure from pure
  Fe-rich crystalline olivine (e.g. $\sim$10--15\% Mg fraction) as
  illustrated by the three top spectra in Fig.\,1 from
  \citet{Koike2003} where it is seen that the feature shifts to smaller
  wavelengths with increasing Mg-content.  We note that there are
other spectra that show some similar behaviours in the 30\,$\mu$m
region, but they are of lower signal-to-noise ratio and contain
possible data reduction artifacts (see e.g. RXJ0432.8+1735 spectrum,
Fig.\,\ref{sp:taurus}). These factors make it difficult to conclude
whether fayalite is present or not. Because of the small number of
positive detections and because of the ambiguity between possible
noise and real features, we will not investigate the presence of this
component further. We do note, however, that the compositional fits to
seven Spitzer/IRS spectra of TTs by \citet{Bouwman2008} show no
evidence of Fe-rich materials.

However, the previous conclusion means that we do not observe pure fayalite in our spectra. Considering the following formula for crystalline olivine (Mg$_x$Fe$_{1-x}$)$_2$SiO$_4$ (with $0 \leq x \leq 1$), features in the 20-30\,$\mu$m spectral range become weaker (especially around 23\,$\mu$m) for Mg fractions below 0.6. This means that we are actually probing crystalline olivine with at least more than 60\% of Mg compared to Fe. Still, as crystalline pyroxene is also contributing to C23 and C28 complexes, further interpretations remain difficult.

\subsubsection{Other silicates: Diopside \& Silica}
\label{sec:diotrosil}
Diopside, a calcium magnesium silicate, and a member of the pyroxene
group (CaMg(SiO$_{3}$)$_{2}$), can be searched for at
$\sim$$25\,\mu$m.  Because of the presence of the C23 complex,
however, the diopside feature can sometimes be blended with the
complex. Even worse, the shoulder of the complex can be
mis-interpreted as a 25\,$\mu$m feature. We therefore obtain only a
lower limit on the frequency at which this feature is present.  The
low detection rate of an additional (but weaker) $20.6\,\mu$m diopside
feature does not strengthen the confidence level of the detection of
the diopside in TTauri star disks.

Finally, silica (SiO$_{2}$) has also been identified in our spectra, with a feature arising at 12.5\,$\mu$m. Attribution of this single feature to amorphous or crystalline silica is not straightforward. \citet{Sargent2009a} showed that amorphous silica tends to produce a feature around 12.3--12.4\,$\mu$m, while various crystalline silica polymorphs produce a feature centered at or slightly to longer wavelengths of 12.5\,$\mu$m. Due to the presence of the broad amorphous 10\,$\mu$m feature, disentangling the different contributions is beyond the scope of our statistical study of the most prominent crystalline silicates. Example silica features are shown on Fig.\,\ref{spectre} and the left panel of Fig.\,\ref{fig:C23}.

\subsection{Silicate feature statistics for TTauri stars}

\subsubsection{The fraction of disks showing silicate(s) feature(s)}
\label{sec:measure}
We developed a routine that measures the characteristics of both
crystalline and amorphous (10\,$\mu$m) features. For each feature, it
assumes a local continuum which is built using a two point Lagrange
polynomial joining the feet of the feature, computed as follows:
\begin{eqnarray*}
  F_{\nu,\mathrm{c}} = F_{\nu_1} \times 
  \left( \frac{\lambda - \lambda_2}{\lambda_1 - \lambda_2} \right) 
  + F_{\nu_2} \times 
  \left( \frac{\lambda - \lambda_1}{\lambda_2 - \lambda_1}\right), 
\end{eqnarray*}
where $F_{\nu,\mathrm{c}}$ is the final continuum, $\lambda_1$ and $\lambda_2$ the blue and red feet of the feature, respectively, and $F_{\nu_1}$ and $F_{\nu_2}$, the observed fluxes at $\lambda_1$ and $\lambda_2$, respectively. This routine returns the band flux, its peak position, the mean wavelength and the peak flux. It also returns the 1-$\sigma$ uncertainty on the band flux. The results given by this procedure are presented in Table\,\ref{IDfeat}, in the form of the SNR on the measured feature fluxes using the above procedure. Results for only 101 objects out of total 104 are shown
since 3 of them (2.8\% of total sample) do not display any crystalline
or amorphous silicate features. Those objects are: \object{EC\,69},
\object{EC\,92} and \object{Sz\,84}.

Thus, the overwhelming majority ($\sim$97\%) of the disks do show at
least one feature that can be attributed to silicates. Since
$10\,\mu$m-sized or larger grains are essentially featureless in the
IRS spectral range (see, for example, the theoretical opacities in
Fig.\,\ref{blowopac}), our results qualitatively indicate that inner
disk atmospheres are populated by grains a few micrometers or less in
size, independent of the age of our stars, and of cloud
membership. Some differences between clouds are nevertheless addressed
in Sec.\,\ref{sec:prevalence}, and a discussion concerning the size of
the amorphous and crystalline silicate grains can be found in
Sec.\,\ref{sec:gsize}.

In the following, we remove the 8 HAeBe stars (plus HD\,132947
  that we already removed) in our survey from the statistical
analysis so that the sample is mainly composed of objects classified
as TTs.  The statistics in the paper are therefore made on the
remaining 96 objects.

\subsubsection{The prevalence of crystalline features}
\label{sec:prevalence}

\begin{figure}
  \hspace*{-0.5cm}\includegraphics[angle=0,width=1.1\columnwidth,origin=bl]{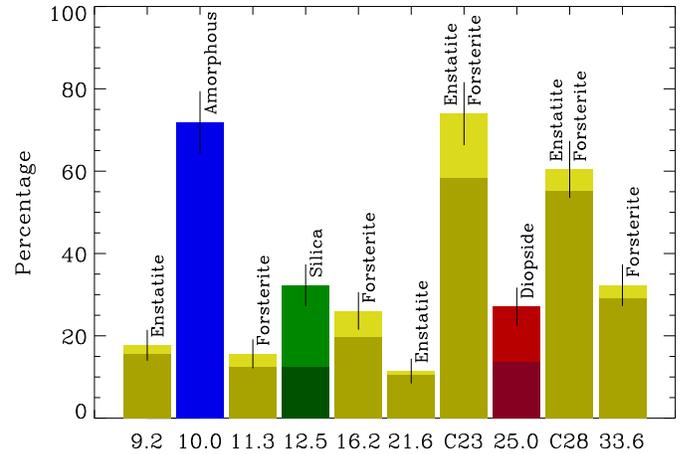}
  \caption{\label{fig:tt_feat}Detection statistics for crystalline
    silicate features (yellow and red bars) for the 10.0\,$\mu$m
    amorphous silicate feature (blue) and for silica at 12.5\,$\mu$m
    (green). Darker color bars are positive detections (SNR $> 20 $),
    and lighter color bars stand for tentative detections (SNR $\leq
    20 $). The uncertainties are those due to Poisson statistics.}
\end{figure}

Feature-by-feature statistics are displayed in Figure\,\ref{fig:tt_feat}, which shows detection fractions of crystalline features ranging between $\sim$10\% and $\sim$60\%. A visual inspection of the 28--29\,$\mu$m and 33--35\,$\mu$m features for a subset (40 TTs and 7 HAeBes) of the target sample analyzed in this paper led \citet{Kessler-Silacci2006} to conclude that $\sim$50\% of the spectra show crystalline silicate features. Considering now the full 96 star sample and only secure detections with SNR $> 20$ for the C28 and 33.6\,$\mu$m complexes, we obtain a fraction of $\sim$\,65\% (63 out of 96) for objects displaying at least one of the two features. Including all the other crystalline features (still with SNR $> 20$), this fraction rises to $\sim$78\% (75 out of 96 stars). We have checked that these results are not corrupted with possible PAH emission features, and that the 11.3\,$\mu$m can be associated with forsterite.  Indeed, \citet{Geers2006} identified only three objects in our sample that have potential PAHs emission at 11.3\,$\mu$m: \object{LkH$\alpha$\,330}, \object{T\,Cha} and \object{SR\,21}. All the other objects cited in \citet{Geers2006} are HAeBe stars and, therefore, not considered in our sample. With additional data, \citet{Geers2007} could confirm the presence of PAHs for \object{LkH$\alpha$\,330}, \object{T\,Cha} and \object{SR\,21}, and so these three objects have then been removed from our statistics for the analysis of the 11.3\,$\mu$m forsterite feature. One other object, \object{EC\,82} is identified by \citet{Geers2007} to have a 6.2\,$\mu$m PAH feature but with no 11.3\,$\mu$m counterpart, thereby avoiding any possible contamination in this particular case. Note that we have been as conservative as possible in identifying features, especially for the crystalline features around 10\,$\mu$m. Some spectra display a shoulder at around 11.3\,$\mu$m
  that could possibly be interpreted as a forsterite feature (see
  e.g. \object{Sz\,52} spectrum, Fig.\,\ref{fig:C23}, label {\it
    f}). We choose not to attribute such kind of shoulder to
  forsterite.  Therefore, we conclude from our analysis that
about 3/4 of the protoplanetary disks in our sample of young
solar-analogs show at least one crystalline feature in their
5--35\,$\mu$m spectrum, meaning that dust crystallization is not a
marginal process but is instead widespread at this stage of disk
evolution.

It is interesting to compare these results to those of \citet{Watson2009} who analyzed 84 TTs in the Taurus-Auriga association, also using Spitzer/IRS. Developing indices measuring the departure from an ISM-like pristine 10\,$\mu$m feature, they found that 90\% of the Taurus-Auriga disks show either the 9.2\,$\mu$m enstatite feature or the 11.3\,$\mu$m forsterite feature, while we obtain about 20\% for both features (Fig.\,\ref{fig:tt_feat}). In \citet{Watson2009}, any departure from the 10\,$\mu$m pristine reference spectrum is considered as being evidence for crystalline features, while we argue that this frequent departure that we also observe has rather more to do with a grain size effect (see Sec.\,\ref{sec:gsize} and Sec.\,\ref{sec:depl} for further discussions). There is therefore a difference in the interpretation of the measurements that leads to inconsistent conclusions. \citet{Watson2009} used the \citet{Dalessio2006} model to investigate the impact of dust sedimentation on the observed dispersion in equivalent width for the 10\,$\mu$m amorphous feature. They could reproduce a broad range of observed equivalent widths by depleting the dust fraction (that consists of grains $\leq 0.25\,\mu$m in their model) in the upper layers of the disks, thereby mimicking dust settling. In a more recent paper, \citet{Dullemond2008} find that sedimentation has, nevertheless, a very limited effect on the shape of the 10\,$\mu$m feature as long as a continuous grain size distribution is considered (as opposed to fixed grain sizes). For longer wavelength crystalline features, comparison with the work of \citet{Watson2009} is possible for
the 33.6\,$\mu$m forsterite feature, but not for the features around
23 and 28\,$\mu$m as they did not study these two
complexes. They find that 50\% of the Taurus-Auriga objects show the
33.6\,$\mu$m crystalline feature, compared to 28\% for our sample, but
this detection statistic is consistent with the statistics for the C23
(55\%) and C28 (54\%) crystalline complexes.

\begin{figure}
\hspace*{-0.5cm}\includegraphics[angle=0,width=1.1\columnwidth,origin=bl]{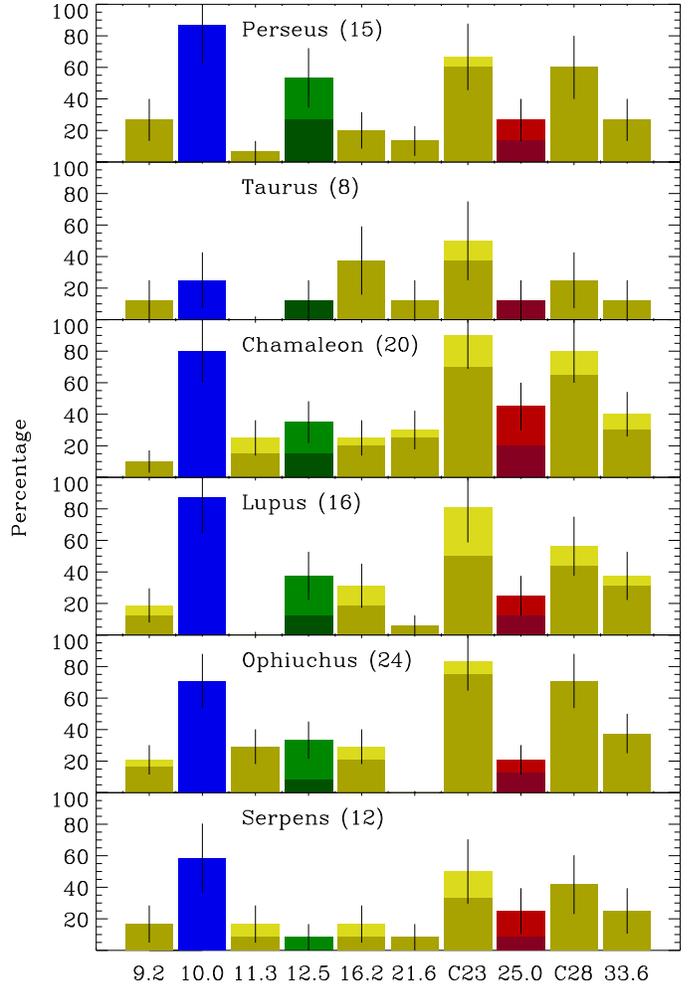}
\caption{\label{fig:clouds}Same as Fig.\,\ref{fig:tt_feat} but now
  sorted by clouds. Uncertainties are again Poisson noise, derived from the
  number of objects per cloud: 15 for Perseus, 8 for Taurus, 20 for
  Chamaeleon, 16 for Lupus, 24 for Ophiuchus and 12 for Serpens.}
\end{figure}

Figure\,\ref{fig:clouds} shows feature-by-feature statistics sorted by
clouds. The largest occurence of the C23 complex is found in the
Ophiuchus cloud with about 80\% of the objects showing this
feature. Disks in Serpens have a prevalence of crystalline features
significantly smaller than in the Perseus, Chamaeleon, Ophiuchus and
Lupus clouds which show comparable fractions of amorphous and
crystalline features. The Taurus cloud shows a noticeably low
frequency of the 10\,$\mu$m amorphous feature compared to the other
clouds and does not appear to be balanced by a remarkably high
fraction of crystalline grains when compared to the other regions. Our
sample contains only eight objects in the Taurus cloud, however, which
introduces a strong bias. \citet{Watson2009} found that the amorphous
10\,$\mu$m feature is present for a large majority of the objects in
their survey of 84 TTs. Therefore, discussing the differences is not
statistically significant for our Taurus sample. One can notice that,
except for Taurus, the five remaining clouds show very similar
behavior.  This points toward a rather homogeneous distribution of
amorphous and crystalline silicate dust grains in disks throughout
clouds. Detection statistics in the Serpens cloud are overall smaller,
but the trends are the same as that in the other clouds. Our current
result on the Serpens cloud will be revisited in a forthcoming paper
by Oliveira et al. (in prep) who analyze a much larger sample of more
than a hundred Serpens objects observed with Spitzer/IRS. 

In the following, we will treat all the objects as a single sample. The first reason for this is that cloud-to-cloud differences suffer from low statistics and merging the entire sample will increase the significance of our inferences (96 objects in total). The second reason is that the trends of the different clouds seem to be similar (except for our Taurus cloud smaple which has a low number of stars). Merging the sample will strenghten our conclusions thanks to the large number of Class\,II objects.

\section{Properties of crystalline silicates in the disks around young stars}
\label{sec:crystals}

One may expect that the relatively large spectral range of the
IRS instrument (5--35$\,\mu$m) implies that the detected short- and
long-wavelength emission features could arise from physically distinct
regions in the disk. \citet{Kessler-Silacci2006} and
\citet{Pinte2008}, among others, have shown that for a typical TTs,
the $10\,\mu$m emission comes from a region about 10 times smaller
than the $20\,\mu$m emitting zone. We examine in this section the
correlations between the detected features. In particular, we inspect
the relation between the properties of the $10\,\mu$m silicate
amorphous feature (its energy, the size of its carriers, its
apparition frequency) and those of the crystalline silicate features
at $\lambda > 20\,\mu$m.

\subsection{Growth of warm amorphous silicates}
\label{sec:gsize}

The amorphous 10\,$\mu$m feature is present at a rate of about 70\% in
our sample. \citet{Bouwman2001}, \citet{2003} and
\citet{Kessler-Silacci2006} have shown that detailed studies of this
feature reveal much information on dust characteristics. Especially
noteworthy is that the observed shape versus strength relation for the
10\,$\mu$m feature can be reproduced by varying the characteristic
grain size over the range of $0.1$--$3\,\mu$m.  Here we analyse the
10\,$\mu$m feature by adopting the same continuum normalization and
the same computational method for the shape (flux ratio
$S_{11.3}/S_{9.8}$) and strength ($S_{\mathrm{Peak}}^{10 \mu
  \mathrm{m}}$) of the 10\,$\mu$m feature as in
\citet{Kessler-Silacci2006}. The different values $S_{11.3}$,
$S_{9.8}$ and $S_{\mathrm{Peak}}$ are obtained by normalizing the
observed flux $F_{\nu}$ as follows: $ S_{\mathrm{\nu}} = 1+
(F_{\mathrm{\nu}} - F_{\mathrm{\nu,c}})/\langle F_{\mathrm{\nu,c}}
\rangle$, where $F_{\mathrm{\nu,c}}$ is the estimated linear 
continuum between the red and blue feet of the feature and
$\langle F_{\mathrm{\nu,c}} \rangle$ its mean value between
the two feet (all fluxes expressed in Jy).  We choose a wavelength range of $\pm 0.1$\,$\mu$m around 9.8\,$\mu$m and 11.3\,$\mu$m to calculate $S_{11.3}$ and $S_{9.8}$.

\begin{figure*}
\centering
\hbox to \textwidth
{
\parbox{0.5\textwidth}{
\includegraphics[angle=0,width=0.5\textwidth,origin=bl]{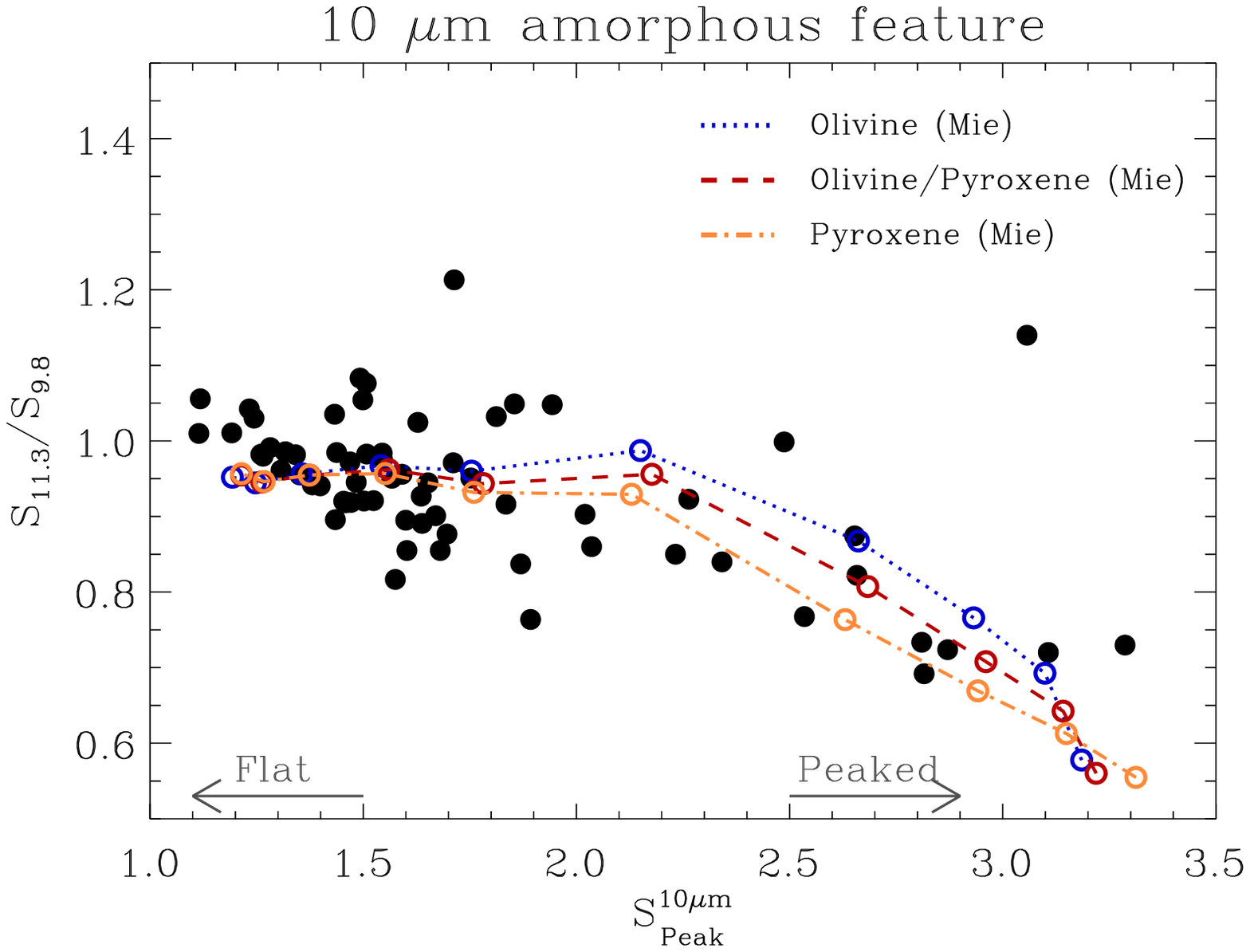}
}
\parbox{0.5\textwidth}{ 
\includegraphics[angle=0,width=0.5\textwidth,origin=bl]{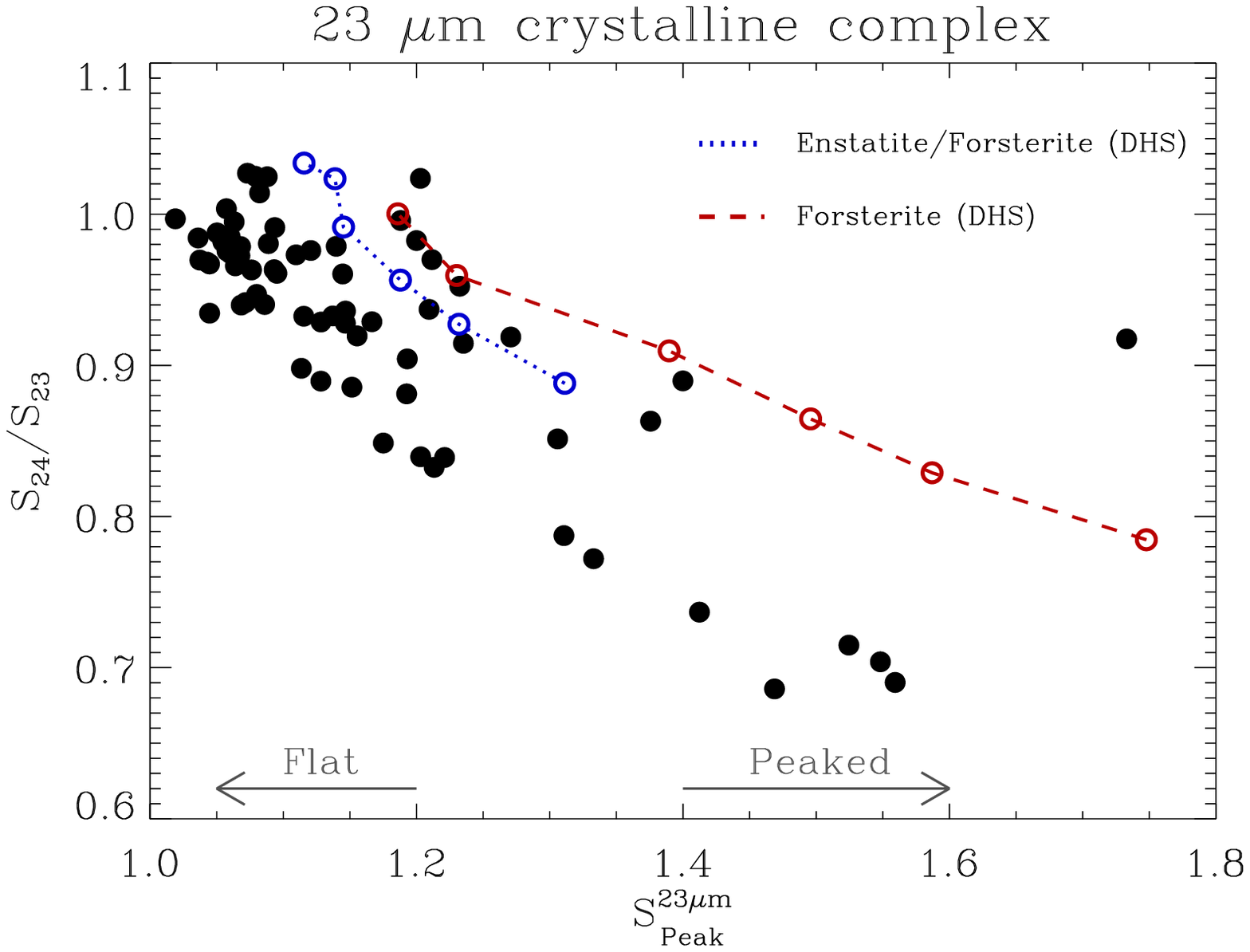}
}
}
\caption{\label{kurt_sp} {\it Left panel:} Correlations between shape
  ($S_{11.3}/S_{9.8}$) and strength ($S_{Peak}$) of the amorphous
  10\,$\mu$m feature. {\it Right panel}: Shape ($S_{24}/S_{23}$)
  versus strength ($S_{\mathrm{Peak}}$) for the C23\,$\mu$m
  crystalline complex. Colored points and curves are derived from
  theoretical opacities for different mixtures. For every curve on
  left panel, the open circles represent different grain sizes, from
  bottom to top, 0.1, 1.0, 1.25, 1.5, 2.0, 2.7, 3.25, 4.3, 5.2, and
  6.25\,$\mu$m. For every curve on right panel, the grain sizes are,
  from bottom to top, 0.1, 1.0, 1.25, 1.5, 2.0 and 2.7\,$\mu$m.}
\end{figure*}

Figure\,\ref{kurt_sp} (left panel) shows the known correlation between the flux ratio $S_{11.3}/S_{9.8}$ ($10\,\mu$m feature shape) and $S_{\mathrm{Peak}}$ ($10\,\mu$m feature strength). To quantify this correlation we use a Kendall $\tau$ test. The Kendall $\tau$ rank correlation coefficient measures the degree of correspondance between two datasets. If the agreement between the two sets of values is perfect then $\tau = 1$, if the disagreement is perfect then $\tau = -1$. The Kendall $\tau$ procedure also returns a probability $P$, computed using the complementary error function of $| \tau |$ (see \citealp{Press1992} for further informations on the Kendall $\tau$ and $P$ calculations). The $P$ probability is the two-sided significance; the smaller the $P$ value, the more significant the correlation.

For the shape versus strength of the amorphous 10\,$\mu$m feature, we find an anti-correlation ($\tau = -0.42$) with a significance probability $P = 3.58 \times 10^{-7}$. This correlation can be interpreted as larger grains producing a flatter 10\,$\mu$m amorphous feature, while smaller grains produce narrower features (e.g. \citealp{2003}).  A thorough analysis by \citet{Kessler-Silacci2006} and \citet{Kessler-Silacci2007} of the shape-strength trend showed that more massive stars (HAeBe) tend to have sharper $10\,\mu$m features than do TTs and BDs, which cluster in a region more consistent with micron-diameter grains as the typical emitting grain size. The comparison of our observations to synthetic $10\,\mu$m amorphous features calculated for different grain sizes (see Sec.\,\ref{sec:contrast} for details on the computation method of grain opacities) confirms this trend as the bulk of our sample clusters in a region consistent with grains that have grown to sizes larger than about $2\,\mu$m (left panel of Fig.\,\ref{kurt_sp}).

\subsection{Growth of cold crystalline silicates}
\label{sec:gsize2}
Although a similar analysis of the $20\,\mu$m amorphous emission
features is rendered more difficult because of the broadness of the
feature, the narrower longer-wavelength crystalline silicate features,
especially the C23 complex, show a grain size-dependent behaviour that
can be tested with our disk sample (see the theoretical opacities in
Fig.\,\ref{blowopac}). Similar trends could not be exploited for the
C28 complex as its shape is not much sensitive to grain size (over the
0.1-3\,$\mu$m range).
For the C23 complex, we build a feature shape index, the $S_{24}/S_{23}$ flux ratio, computed using the same approach as for the amorphous 10\,$\mu$m feature. The $S_{23}$ normalized flux is obtained by averaging the normalized (local) continuum-subtracted flux between $23.5$ and $23.8\,\mu$m, and between $24.1$ and $24.4\,\mu$m for $S_{24}$.  Figure\,\ref{kurt_sp} (right panel) shows the $S_{24}/S_{23}$ ratio as a function of the strength of the C23 complex (C23 $S_{\mathrm{Peak}}$).  The Kendall test gives $\tau = -0.54$ with a significance probability below $10^{-38}$, indicative of a clear anticorrelation between these two quantities.

To better evaluate the relationship between grain size and the C23
shape, we calculate theoretical $S_{24}/S_{23}$ ratios for synthetic
crystalline silicates opacity curves obtained using the Distribution
of Hollow Spheres (DHS) method \citep{Min2005}. We consider grains
between $0.1$ and $2.7\,\mu$m in radius, and two different
compositions: 100\% forsterite (Mg$_{2}$SiO$_{4}$, using optical
constants from \citealp{Servoin1973}) and a mixture composed of 50\%
forsterite plus 50\% enstatite (MgSiO$_{3}$, optical constants taken
from \citealp{Jaeger1998}). Grains larger than $\sim3.0\,\mu$m, for
either enstatite or forsterite, do not give striking results, as the
feature disappears into the continuum. Also, a pure enstatite
composition turns out to be incompatible with the observed shape for
the C23 complex as it shows two unblended features at 23.8 and
24.5\,$\mu$m. This indicates that we are not observing pure enstatite
grains in our sample, a result which seems to first order consistent
with \citet{Bouwman2008} who found that the very inner ($\sim$ 1\,AU)
warm dust population is dominated by enstatite while the dust
population at large radii is dominated by forsterite. The overall
shape-strength trend for the C23 crystalline complex is better
reproduced for a forsterite/enstatite mixture, with characteristic
grain sizes larger than 1.0\,$\mu$m in order to to match the bulk of
the measured values. This suggests that not only have the warm
amorphous silicate grains grown in our sample of TTs (left panel of
Fig.\,\ref{kurt_sp}), but also that the colder crystalline silicate
grains have coagulated into $\mu$m-sized particles and remain
suspended in the upper layers of the disk atmospheres. 

The offset in the right panel of Fig.\,\ref{kurt_sp} between the data points distribution and the curves obtained using theoretical opacities may come from various effects. First, we consider pure crystalline dust grains, which may be a too restrictive assumption to reproduce the observations. It allows us to see the impact of grain size, however, and to show that it matches reasonably well the observed trend. Recently, \citet{Min2008} studied the impact of inclusions of small crystals into larger amorphous grains, but this is too much of a refinement for our purposes. Second, we only consider pure Mg-rich crystals (no iron). According to \citet{Chihara2002}, the 24.5\,$\mu$m feature of crystalline pyroxenes shifts towards slightly larger wavelengths with increasing Fe-content. Also, according to \citet{Koike2003}, the peak position of the 23.8\,$\mu$m crystalline olivine feature does not shift unless the Fe-content is larger than 20\%. As the $S_{24}$ index is computed integrating between 24.1 and 24.4\,$\mu$m, raising the Fe-content will produce a smaller $S_{24}$ and therefore a smaller $S_{24}/S_{23}$, while the strength of the complex will slightly decrease as the 24.5\,$\mu$m feature shifts to larger wavelengths. To conclude, even if there is a small shift when comparing the observations to theoretical points, the trend is well reproduced by varying the grain size.

Interestingly, we find no correlation between the grain size proxy for the amorphous 10\,$\mu$m feature and the grain size proxy for the C23 complex. Considering the sets of values $(S_{11.3}/S_{9.8})/S_{\mathrm{Peak}}^{10}$ as a function of $(S_{24}/S_{23})/S_{\mathrm{Peak}}^{23}$, we indeed obtain a Kendall's $\tau$ of 0.07, with a significance probability $P$ of 0.44. This means that even if grain growth took place in both components, the growth of the amorphous and crystalline grains, respectively, does not seem to be correlated.

\subsection{Relationship between the 10\,$\mu$m and C23 features with
  disk properties}

\begin{figure}
  \centering
  \hspace*{-0.5cm}\includegraphics[angle=0,width=1.\columnwidth,origin=bl]{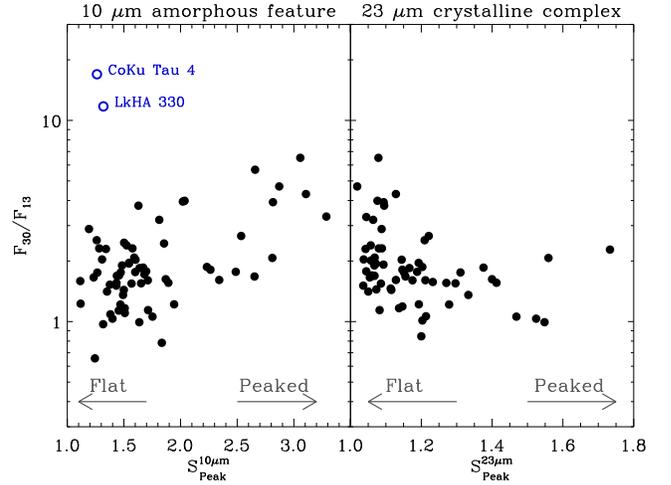}
  \caption{\label{fig:slope} {\it Left panel:} Correlations between
    the slope of spectra ($F_{30}/F_{13}$) as a function of the
    strength ($S_{Peak}$) of the amorphous 10\,$\mu$m feature. The two
    outliers (blue open circles) are cold disks:
    \object{LkH$\alpha$\,330} and \object{CoKu\,Tau/4}.  {\it Right
      panel}: Spectral slope ($F_{30}/F_{13}$) versus strength
    ($S_{\mathrm{Peak}}$) for the C23\,$\mu$m crystalline complex.}
\end{figure}

Young flared disks are expected to show rising spectra in the mid-IR, as opposed to more evolved, settled systems having self-shadowed disks with much flatter mid-IR spectra \citep[e.g.][]{Dominik2003}. Here we measure the slope of the spectra by the ratio of the $F_{30}$ and $F_{13}$ indexes, where $F_{13}$ is the mean flux value (in Jy) in the range 13\,$\mu$m$\pm 0.5\,\mu$m, and between 29\,$\mu$m and 31$\,\mu$m for $F_{30}$. The left panel of Figure~\ref{fig:slope} shows the correlation between the strength of the amorphous 10\,$\mu$m feature and the slope of the spectra ($F_{30}/F_{13}$ ratio). Taking out the two outliers (blue open circles), which are known cold disks \citep[\object{LkH$\alpha$\,330} and \object{CoKu\,Tau/4},][]{Brown2007}, we find a correlation coefficient $\tau = 0.29$ with a significance probability $P$ smaller than $5.2 \times 10^{-4}$ suggestive of a correlation. A similar trend is found by \citet{Bouwman2008} within their FEPS sample of seven stars, and by \citet{Watson2009} with the Taurus-Auriga sample of 84 TTs. This can be interpreted as smaller grains (large $S_{\mathrm{Peak}}^{10 \mu
  \mathrm{m}}$ values) being probed in flared disks (high values of
$F_{30}/F_{13}$), while flattened disk (low $F_{30}/F_{13}$ values)
atmospheres are characterized by larger grains.

If dust sedimentation is the main process responsible for disk
flattening, one would expect the upper layers of the disks to be depleted in
large grains, hence the mid-infrared spectroscopic signature to be
that of very small grains (peaked 10$\,\mu$m features). The observed
trend (Fig.\,\ref{fig:slope}) very clearly refutes such a possible
trend in the relation between the SED shape and that of the 10$\,\mu$m
feature. The effect of dust sedimentation on the shape of the
10$\,\mu$m amorphous silicate feature was recently discussed by
\citet{Dullemond2008} based on models. They showed that the
sedimentation of the larger grains toward the disk midplane is
generally not enough to transform a flat 10$\,\mu$m feature into a
peaked feature. Our observations tend to support these results as
sedimentation, if assumed to be revealed by disk flattening, is
generally associated with flat 10$\,\mu$m features. This trend is in
fact valid for the majority of the TTs in our sample as the objects
cluster in the region with flat SEDs and flat silicate features in
Figure~\ref{fig:slope}.

For the C23 crystalline complex (right panel of Fig.\,\ref{fig:slope}), we find a rather weak anticorrelation, with $\tau = -0.21$ and a significance probability $P <$0.01. The small crystalline grains with $S_{\mathrm{Peak}}$ larger than about 1.2 are only found in flat or slowly rising disks ($F_{30}/F_{13}$ smaller than about 2), while the larger grains can be found in all kinds of disks. Interestingly, the very flared disks which have large crystalline grains emitting at 23$\,\mu$m also show small amorphous grains in their warm component (that is, the region probed at 10$\,\mu$m). Seven out of the eight objects (\object{RY Lup}, \object{RXJ1615.3-3255}, \object{Haro 1-1}, \object{SSTc2d
  J162148.5-234027}, \object{SSTc2d J162245.4-243124}, \object{SSTc2d
  J162715.1-245139}, \object{Haro 1-16}) whith $F_{30}/F_{13} > 3$ on
the right panel of Fig.\,\ref{fig:slope} indeed have their counterpart
in the left panel with $S_{\rm Peak}^{10}$ larger than about 2 (the
last object being \object{DoAr 25}).
\begin{figure*}
\hbox to \textwidth
{
\parbox{0.5\textwidth}{
\includegraphics[angle=0,width=0.5\textwidth,origin=bl]{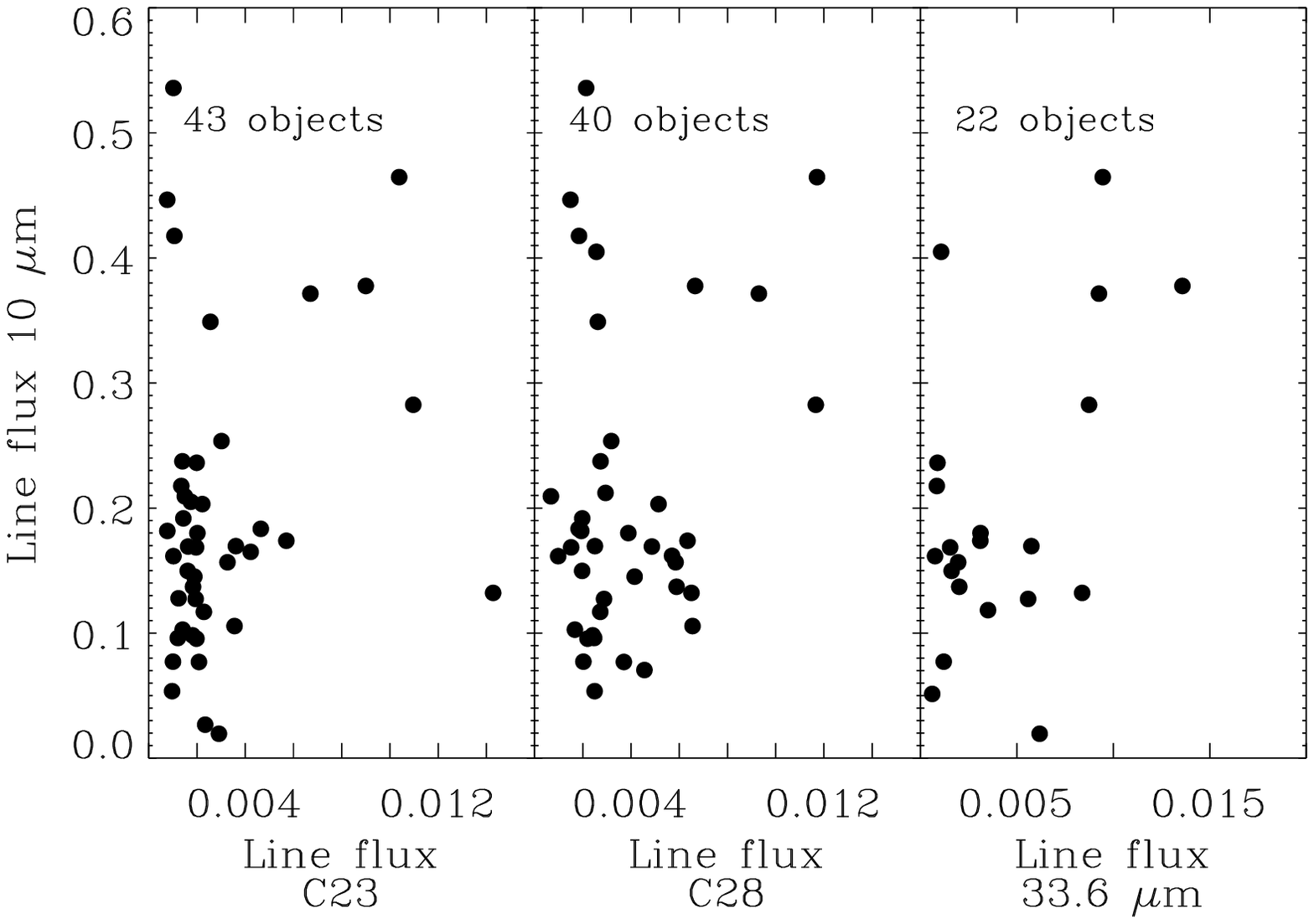}
}
\parbox{0.5\textwidth}{ 
\includegraphics[angle=0,width=0.5\textwidth,origin=bl]{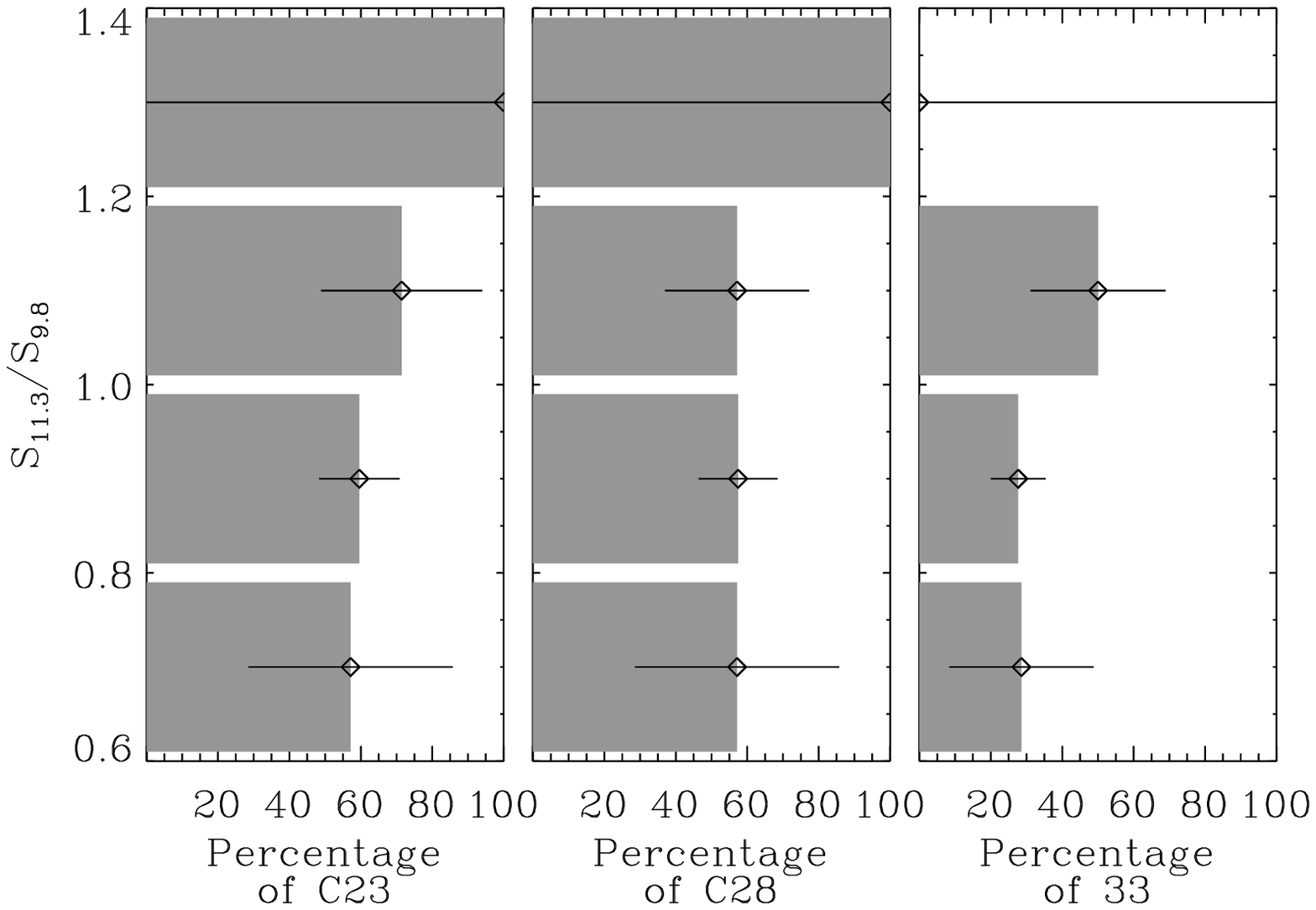}
}
}
\caption{\label{LF10}{\it Left panel: }From left to right:
  correlations between the unitless line fluxes of the 10\,$\mu$m
  feature and the C23, C28 complexes plus 33.6\,$\mu$m
  forsterite feature. {\it Right panel: }From left to right:
  percentages of objects showing the C23/C28 complexes and
  33.6\,$\mu$m forsterite feature as a function of the shape of the
  10\,$\mu$m feature. Larger, amorphous grains are located at the
  top of the plot.}
\end{figure*}

\subsection{Relationship between the 10\,$\mu$m feature and
  crystallinity\label{sec_cryst_10}}

We find in Sec.\,\ref{sec:gsize2} that the growth of the warm
amorphous silicates (probed by the $10\,\mu$m feature) is
statistically disconnected from the growth of the cold crystalline
silicates (probed by the C23 complex). We further investigate the
relationship between the amorphous and crystalline silicates features
by searching for a correlation between the energy contained in the
amorphous $10\,\mu$m feature and the energy contained in the C23, C28
and the $33.6\,\mu$m features. To get rid of possible distance or
brightness effects we normalized the line fluxes as follows:
$\mathrm{Line\,\,flux\,\,[unitless]} = \mathrm{Line\,\,flux}
[\rm{W.m}^{-2}] / (\lambda \langle F_{\mathrm{\lambda, c}} \rangle)$,
where $\langle F_{\mathrm{\lambda, c}} \rangle$ is the mean value of
the local feature continuum and $\lambda$ is the measured central
wavelength of the considered feature. 

The unitless line fluxes are displayed in the left panel of Figure\,\ref{LF10}, which shows that when both the C23 and the 10\,$\mu$m feature are present with SNR $> 20$, their emission energies are not correlated ($\tau = 0.06$, significance probability $P = 0.55$), although the plot would suggest that objects with low 10\,$\mu$m line flux usually also have a low C23 line flux. The middle panel of Figure\,\ref{LF10} shows a similar result when both the C28 and the 10\,$\mu$m feature are present with SNR $> 20$: $\tau = 0.023$ (no correlation between these features) with a significance probability $P = 0.83$, reflecting the dispersion. Finally, the right panel of Fig.\,\ref{LF10} shows a similar trend for the 33.6\,$\mu$m and the amorphous 10\,$\mu$m feature: $\tau = 0.22 $ and a significance probability $P = 0.15$. Therefore, the flux radiated by the 10$\,\mu$m amorphous feature is largely unrelated to the energy contained in the crystalline silicates features that appear at wavelengths longer than 20$\,\mu$m.

Similarly, the shape of the amorphous 10\,$\mu$m feature (that is, the
$S_{11.3}/S_{9.8}$ ratio, which is a proxy for grain size,
Sec.\,\ref{sec:gsize}) seems uncorrelated with the emergence of the
23\,$\mu$m complex, the 28\,$\mu$m complex, or the 33.6\,$\mu$m
forsterite feature. The right panel of Fig.\,\ref{LF10}, for example,
shows the number of positive detections of the C23, C28 and
33.6\,$\mu$m features per $S_{11.3}/S_{9.8}$ bin, where the errors
bars correspond to Poisson noise (square root of the number of
detections). The fraction of disks with C23 and C28 features is found
to be independent of the mean grain size probed by the 10\,$\mu$m
amorphous feature, further suggesting that the 10\,$\mu$m feature and
the C23, C28 features essentially probe disconnected populations. For
the 33.6\,$\mu$m feature, the fraction of objects per bin seems less
independent because of the smaller numbers of detections (23 objects
with both the 10 and 33.6\,$\mu$m features).

\begin{figure}
\resizebox{\hsize}{!}{\includegraphics{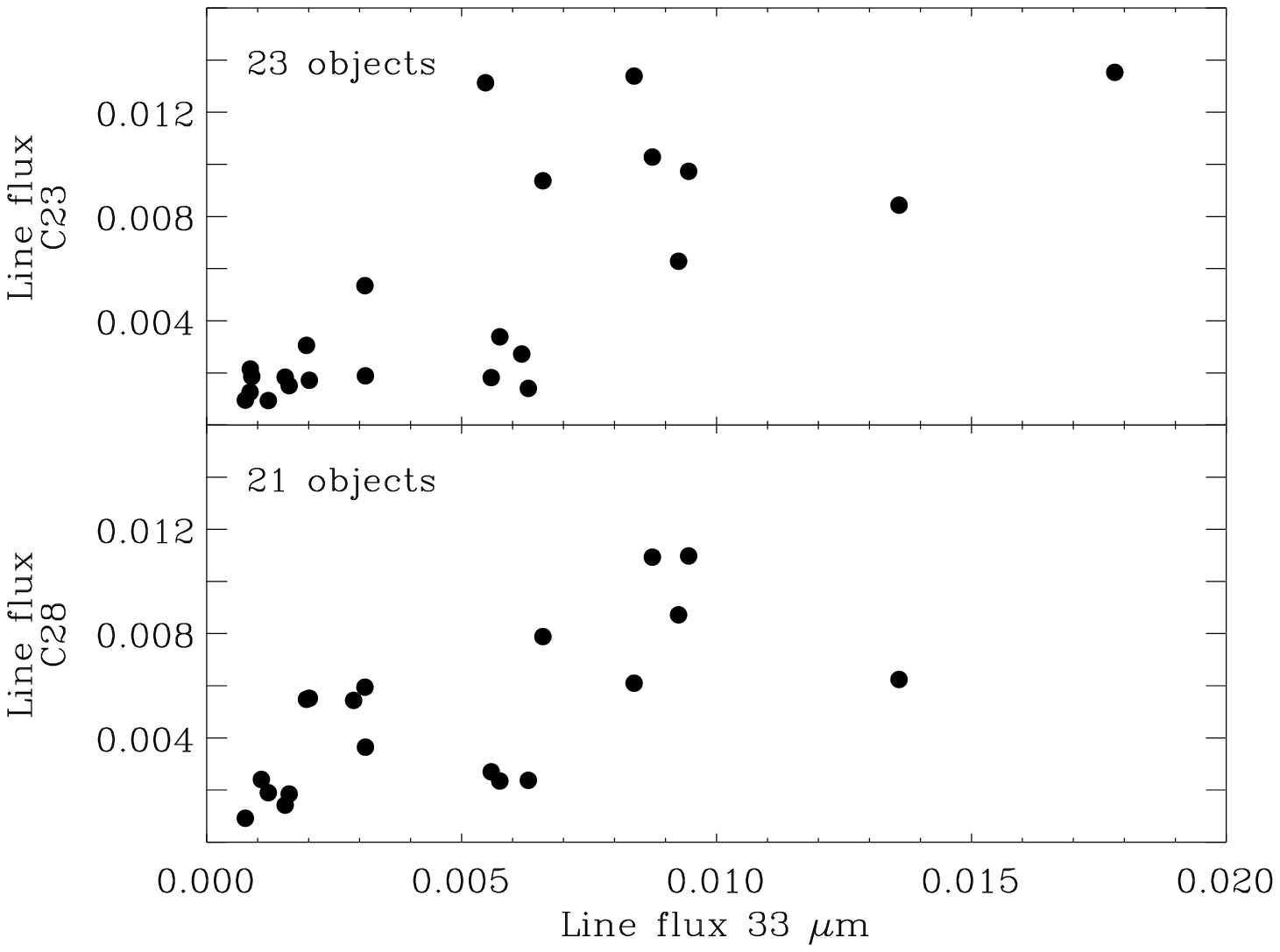}}
\resizebox{\hsize}{!}{\includegraphics{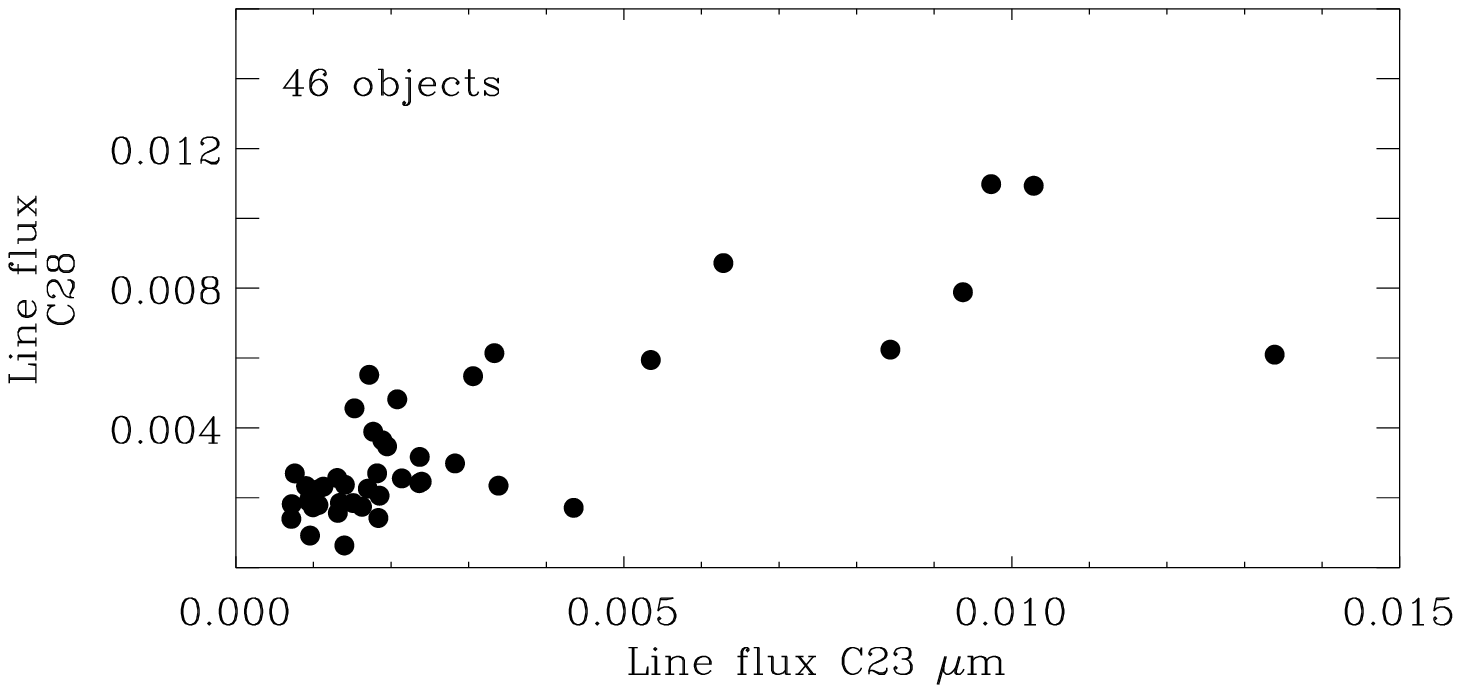}}
\caption{\label{LF}{\it Top panel: }Normalized line fluxes for the
  33.6\,$\mu$m feature and the C23 complex (Kendal $\tau = 0.55$ and
  significance probability $P = 2.41 \times 10^{-4}$). {\it Middle panel: }Same but with the C28 complex
  ($\tau = 0.56$, $P = 3.66 \times 10 ^{-4}$). {\it Bottom panel: }
  Normalized line fluxes between the C23 and C28 complexes ($\tau =
  0.54$, $P = 1.19 \times 10^{-7}$).}
\end{figure}

Overall, these results suggest that the crystalline features appearing at wavelengths longer than 20\,$\mu$m and the 10\,$\mu$m amorphous feature arise from unrelated populations. On the other hand, Figure\,\ref{LF} shows the correlation between the normalized line fluxes of the C23 and C28 complexes where both are present with SNR $> 20$, and also their correlation with the 33.6\,$\mu$m feature. The Kendall $\tau$ tests and associated significance probabilities confirm the expected result that the crystalline features at wavelengths longer than 20\,$\mu$m have emission fluxes that correlate to each other.

\subsection{Feature correlation coefficients}

\begin{table*}
\begin{center}
  \caption{\label{correl_f}Correlation coefficients between the features detected with SNR$> 20$. Entries below the diagonal give the Kendall $\tau$ values and the number of spectra with common detections of considered features between parenthesis. The diagonal displays the total number of detections for a given feature, while the entries above the diagnoal report the Kendall $P$ significance probabilities.}
\begin{tabular}{c|cccccccccc}
\hline\hline
&    9.2\,$\mu$m & 10.0\,$\mu$m   &  11.3\,$\mu$m  &  12.5\,$\mu$m  &  16.2\,$\mu$m  &  21.6\,$\mu$m  &  C23   &  25.0\,$\mu$m &  C28   &  33.6\,$\mu$m  \\ \hline
9.2\,$\mu$m    &      {\bf 15} &       0.008 &       0.238 &      0.012 &      0.003 &      0.622 &      0.000 &     0.887 &       0.000 &  0.195 \\
10.0\,$\mu$m  &  0.187 (14)  &        {\bf 69} &       0.000 &      0.293 &      0.049 &      0.088 &      0.468 &     0.203 &       0.858 &  0.441 \\
11.3\,$\mu$m  &  -0.084 (1) &   -0.447 (3) &         {\bf 12} &      0.556 &      0.557 &      0.056 &      0.001 &     0.363 &       0.000 &  0.182 \\
12.5\,$\mu$m  &   0.179 (4)  &  0.075 (10)  &  0.042 (2)  &       {\bf 12} &      0.087 &      0.308 &      0.113 &     0.003 &       0.004 &  0.182 \\
16.2\,$\mu$m  &   0.211 (6)  & -0.140 (12) &  0.042 (3)  &  0.121 (4) &       {\bf 19} &      0.000 &      0.000 &     0.732 &       0.115 &  0.068 \\
21.6\,$\mu$m  &   0.035 (2)  &    0.121 (9)  & -0.135 (1) &  0.072 (2) &  0.253 (5) &      {\bf 10} &      0.053 &     0.025 &       0.025 &  0.963 \\
C23                   &  0.354 (15)  &  0.051 (43)  & 0.244 (11)  &  0.112 (9) & 0.299 (17) & 0.137 (8)  &         {\bf 56} &     0.000 &       0.000 &  0.000 \\
25.0\,$\mu$m  &  -0.010 (2) &   0.090 (11)  &  -0.064 (1) &  0.214 (4) &  0.024 (3) & 0.159 (3)  & 0.325 (13) &       {\bf 13} &       0.178 &  0.318 \\
C28                   & 0.260 (13)  &   0.013 (40)  & 0.332 (12)  &  0.202 (10) & 0.112 (13) & 0.158 (8)  & 0.619 (46) & 0.095 (9) &         {\bf 53} &  0.001 \\
33.6\,$\mu$m  &  0.092 (6)  &    0.055 (22)  &  0.095 (5)  &   0.095 (5) &  0.129 (8) & -0.003 (3) & 0.288 (23) & 0.071 (5) &  0.233 (21) &    {\bf 28} \\
\hline
\end{tabular}
\end{center}
\end{table*}
The one-to-one correlation probabilities between the detection frequencies of all the features examined in this study with SNR $> 20$ are displayed in Table\,\ref{correl_f}. The diagonal shows the total number of positive detections for any given feature (these numbers being close to percentages of detections as we have 96 objects). The lower part of the Table displays the Kendall $\tau$ values as well as the number of spectra with common detections of the considered features between parenthesis. The Kendall significance $P$ probabilities are displayed in the upper part.

At this point, it is interesting to understand the meaning of the inferred Kendall $\tau$ values in this Table. Some correlations can possibly be found according to the Kendall $\tau$ values but only because both features are absent from spectra. Consider as an example the 11.3\,$\mu$m forsterite feature and the C23 complex, for which we obtain $\tau = 0.244$ and $P = 0.001$, for 12 detections of the 11.3\,$\mu$m feature and 56 detections of the C23 complex, indicating a weak correlation. Out of the 96 stars, 11 spectra show both features, one spectrum shows the 11.3\,$\mu$m feature only, and 45 objects show the C23 complex only. The other 39 ($96-45-1-11$) objects show neither the 11.3\,$\mu$m feature nor the C23 complex, and these objects account for concordant pairs. This explains why we find a weak correlation (non-zero $\tau$ value), as the number of concordant pairs (50) is of the order of the number of discordant pairs (46) for this example. In case we would eliminate from the statistics objects that do not show any of these two features, we would obtain an anti-correlation with $\tau = -0.25$ and $P = 0.004$. Therefore, in order to have comparable numbers for all the features, irrespective of the number of detections, we decide to consider the entire sample for the calculation of the correlation coefficients, even when spectra do not show any of the considered features.

The detection frequencies of the 10\,$\mu$m amorphous feature and the C23 complex are found to be essentially uncorrelated ($\tau = 0.051$ with $P = 0.47$). Similar results are obtained when considering the 10\,$\mu$m feature and the C28 complex ($\tau = 0.013$ with $P = 0.86$), and also the 10\,$\mu$m and 33.6\,$\mu$m features ($\tau = 0.055$ with $P = 0.44$). For the C23, C28 complexes and the 33.6\,$\mu$m feature, we have, respectively, 43, 40 and 22 common detections with the amorphous 10\,$\mu$m feature (which is observed in 69 spectra). The C23 and C28 complexes, as well as the 33.6\,$\mu$m forsterite feature, do therefore appear in spectra independently from the presence of the 10\,$\mu$m feature. On the other hand, the C23 complex appears to be coincident with the C28 complex in our spectra ($\tau = 0.619$ with $P < 10^{-38}$), which concurs with our results based on line fluxes (Sec.\,\ref{sec_cryst_10}). The 33.6\,$\mu$m feature also appears to be coincident, to a slightly lesser extent with the C23 complex ($\tau = 0.288$ with $P < 10^{-38}$), and the C28 features ($\tau = 0.233$ with $P = 3 \times 10^{-3}$). The number of common detections also confirm these trends.

Table~\ref{correl_f} further shows that the forsterite 11.3\,$\mu$m
feature is anti-correlated with the amorphous 10\,$\mu$m feature
($\tau = -0.44$, only 3 objects show both the 10\,$\mu$m and the
11.3\,$\mu$m features). The anti-correlation between the forsterite
11.3\,$\mu$m feature and the amorphous 10\,$\mu$m feature could be
explained by a contrast effect. The amorphous feature is indeed rather
stronger than the 11.3\,$\mu$m feature which could therefore
potentially be present but hidden in the amorphous feature (see
discussion in Sec.\,\ref{sec:contrast} for a more detailed analysis of
this effect).

Interestingly, the crystalline features in the 10\,$\mu$m range (9.2 and 11.3\,$\mu$m) seem to be correlated with both the C23 and C28 complexes: if crystalline features are present in the 10\,$\mu$m spectral range then the presence of the two complexes is highly probable. Still the reverse consideration is not true, if one of the two complexes (C23 or C28) is present, it is not guaranteed that we will find the 9.2 or 11.3\,$\mu$m features. This reflects that the correlation is dominated by objects where none of the considered features is present. Furthermore, as we find no strong correlation for the line fluxes regarding these features, this would mean that there is a general increase of crystallinity for several objects. Also, the two features at 9.2 and 11.3\,$\mu$m do not seem to be correlated with the 33.6\,$\mu$m forsterite feature. The declining quality of the data (lower SNR) at the very end of the IRS spectra may explain this trend in part, as some (lower contrast) 33.6\,$\mu$m features may not have been detected. Another possible explanation would be that the 33.6\,$\mu$m forsterite feature may probe an even colder disk region than the population probed by C23 and C28 complexes.

To summarize, our results show that the C23, C28 and 33.6\,$\mu$m
crystalline features probably arise from the same population of
grains. Also, the detections of the 9.2 and 11.3\,$\mu$m crystalline
features are correlated with the detections of C23 and C28 (but
  the presence of one of the longer wavelengths complexes does not involve the detection of 9.2 or 11.3\,$\mu$m features), which can be interpreted as a general increase of crystallinity in the inner and outer regions of disks. However, this crystalline population seems to be disjoint from the population of amorphous silicate grains that produce the 10\,$\mu$m feature.

\subsection{Degree of crystallinity of the warm versus cold disk regions}
\label{sec:contrast}

\begin{figure*}
  \hspace*{-0.85cm} \includegraphics[angle=0,width=1.05\textwidth,origin=bl]{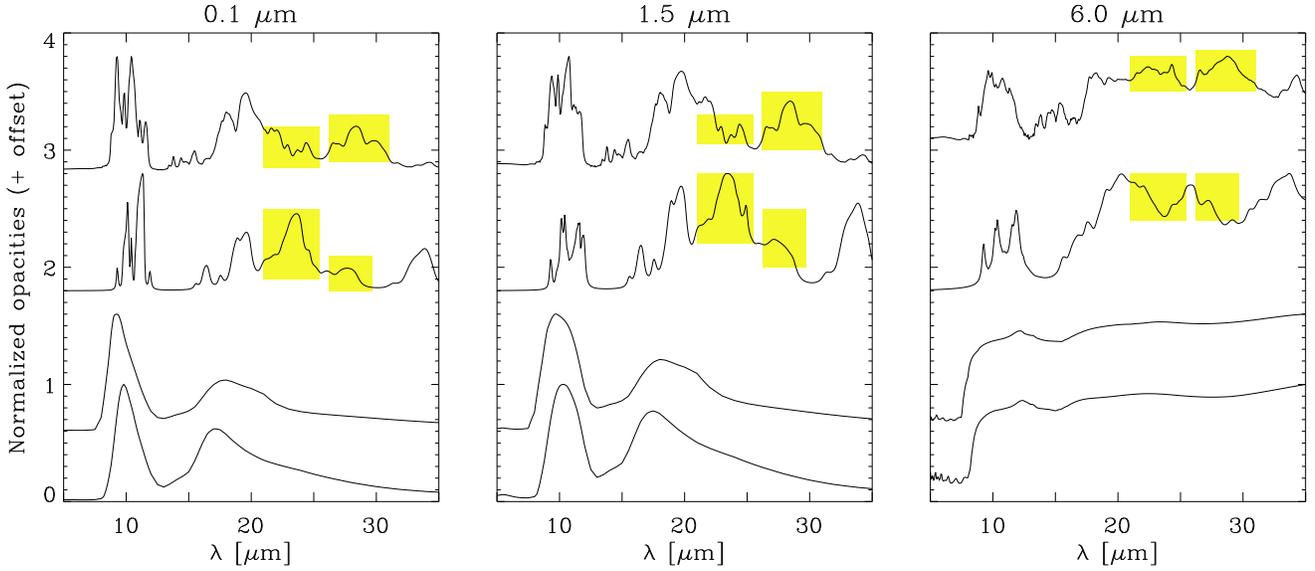}
  \caption{\label{blowopac}Example mass opacities (in arbitrary units)
    used to model the IRS observations. From bottom to top: amorphous
    olivine, amorphous pyroxene, crystalline forsterite and
    crystalline enstatite. The left panel shows opacities for a grain size
    $a=0.1\,\mu$m, $a=1.5\,\mu$m for the middle panel and
    $a=6.0\,\mu$m for the right panel. Some of the features discussed
    in Secs.\,\ref{sec:forsens} and \ref{sec:diotrosil} can be
    recognized. Yellow boxes are places of both C23 and C28 crystalline complexes.}
\end{figure*}
\begin{figure*}
\centering
  \hspace*{-0.85cm} \includegraphics[angle=0,width=.525\textwidth,origin=bl]{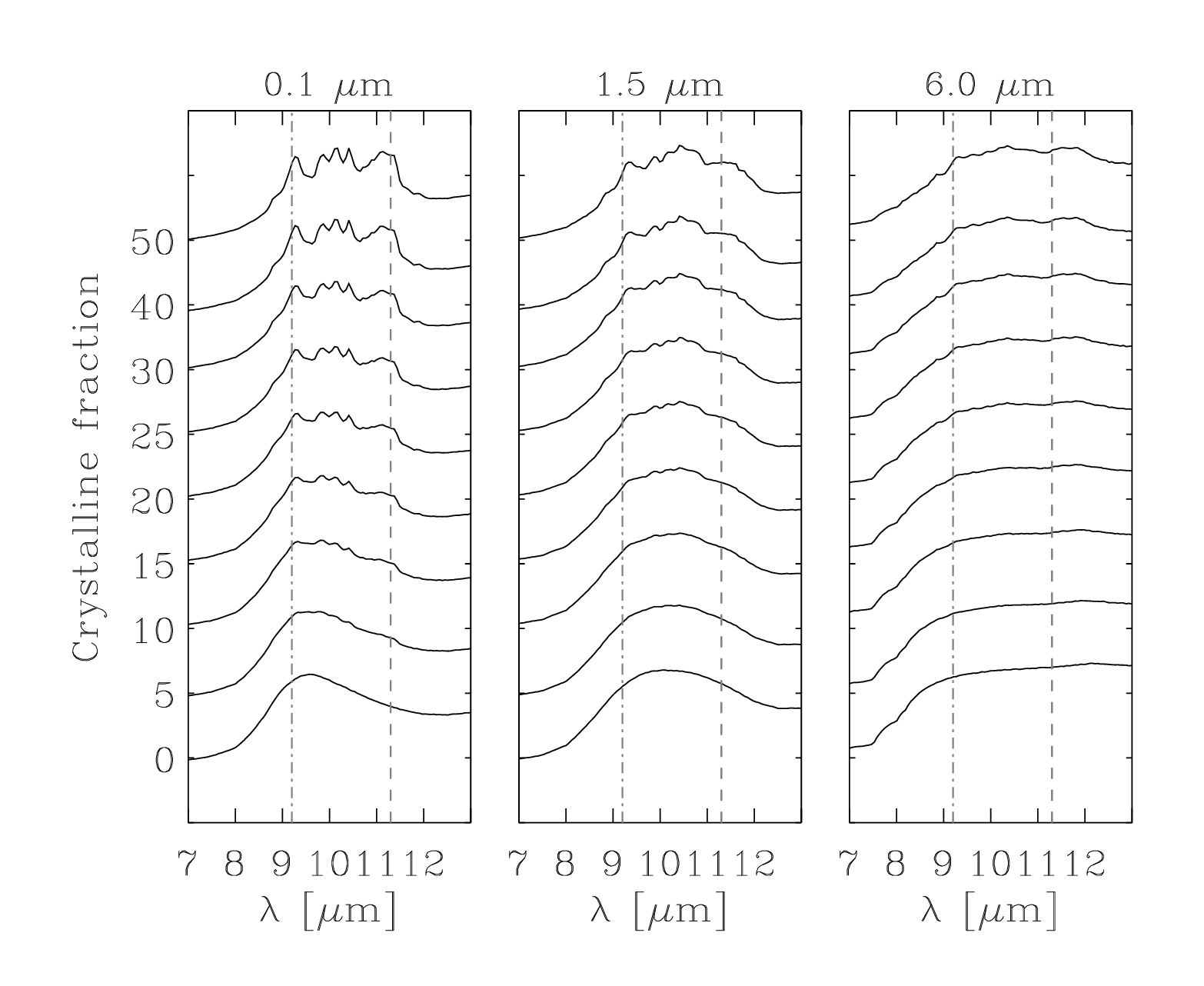}
  \hspace*{-0.85cm} \includegraphics[angle=0,width=.525\textwidth,origin=bl]{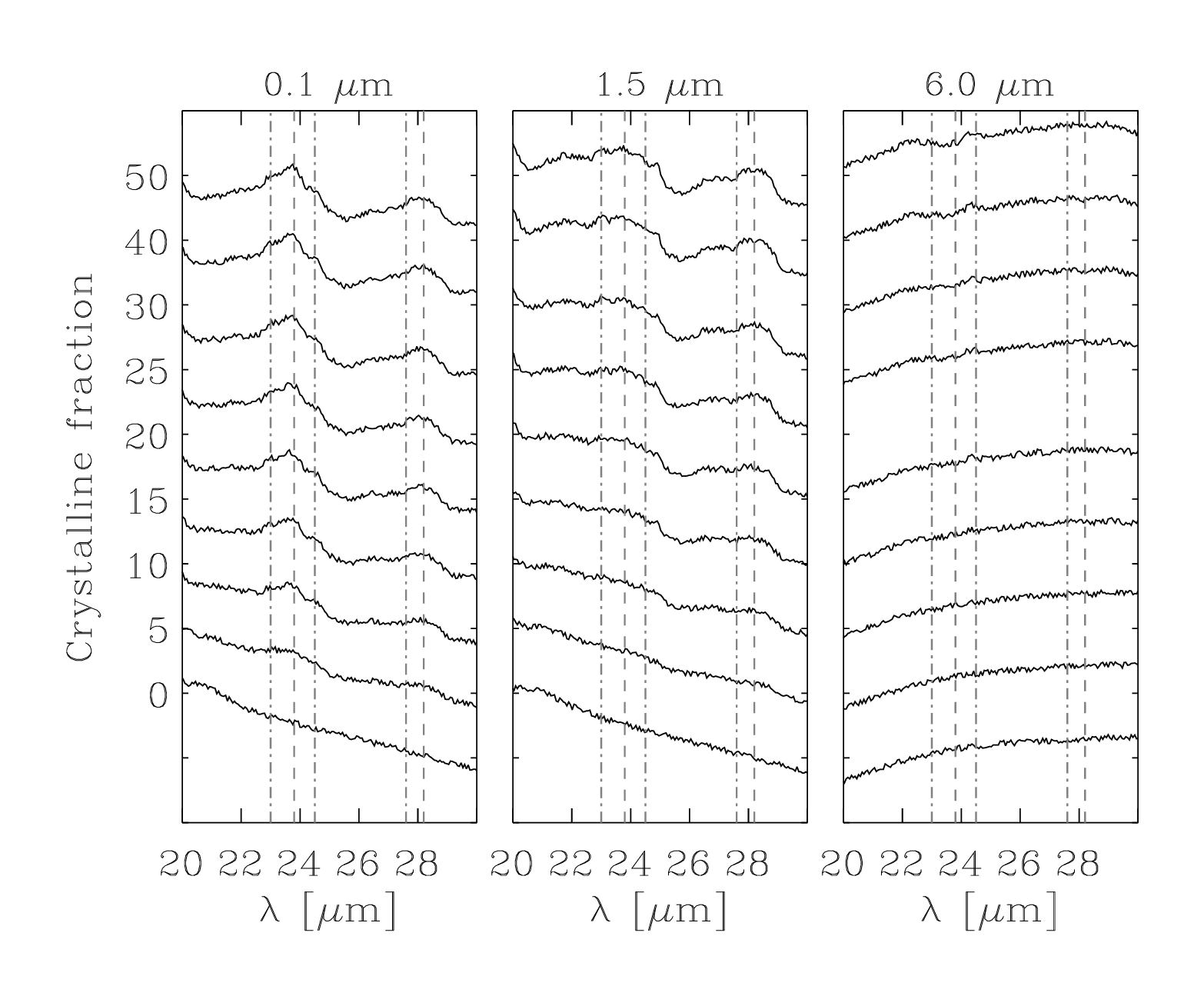}
 \caption{\label{fig:contrast}Synthetic spectra, in
  arbitrary units, for mass fractions of crystalline silicates
  from 0\% to 50\% ( steps are: 0, 5, 10, 15, 20, 25, 30, 40, 50\%, see Sec.\,\ref{sec:contrast} for more details). The impact of the grain size on the detectability of the
  11.3\,$\mu$m, C23 and C28 features is considered for 0.1\,$\mu$m,
  1.5\,$\mu$m and 6.0\,$\mu$m grains.}
\end{figure*}
Both Figures \ref{fig:tt_feat} and \ref{fig:clouds} show larger
occurrences for the C23 and C28 emission features than for smaller
wavelength crystalline silicate features, especially the 9.2
(enstatite) and 11.3\,$\mu$m (forsterite) features. As silicates are
expected to thermally anneal close to the star, this result may
reflect a counterintuitive positive gradient of the crystalline
silicate fraction toward colder temperatures. The presence of the
broad, and generally strong, amorphous $10\,\mu$m feature may lower
the contrast of (or even hide) the shortest wavelength crystalline
features, however, preventing direct conclusions. We examine in this
section this apparent {\it crystallinity paradox} by comparing our
Spitzer/IRS observations to synthetic spectra.

To better gauge the actual differences between the degrees of
crystallinity of the warm ($\lambda\sim10\,\mu$m) versus cold
($\lambda >20\,\mu$m) disk regions, we generate synthetic spectra
representative of the observations, and from which we extract line
fluxes as was performed for the IRS spectra in
Sec.\,\ref{sec:measure}. We consider amorphous species that include
silicates of olivine stoichiometry (glassy MgFeSiO$_{4}$, optical
constants from \citealp{Dorschner1995}), and silicates of pyroxene
stoichiometry (glassy MgFeSi$_{2}$O$_{6}$, optical constants from
\citealp{Dorschner1995}). For the crystalline species, we use
enstatite (MgSiO$_{3}$) optical constants from \citet{Jaeger1998} and
forsterite (Mg$_{2}$SiO$_{4}$) optical constants from
\citet{Servoin1973}. Theoretical mass opacities $\kappa_\lambda$ are
computed using Mie theory (valid for hard spheres) for the amorphous
species and DHS theory (Distribution of Hollow Spheres,
\citealp{Min2005}) for the crystalline silicates.  Twelve example
opacity curves for the four compositions and three typical (in terms
of spectroscopic signature) grain sizes ($0.1\,\mu$m, $1.5\,\mu$m and
$6.0\,\mu$m) are displayed in Fig.\,\ref{blowopac}.

The synthetic spectra of Fig.\,\ref{fig:contrast} have been generated by adding to a representative continuum ($C_{\nu}$), the mass opacities of amorphous and crystalline silicates, multiplied by relative masses and by a blackbody at reference temperatures (350\,K for the warm region, and 100\,K for cold region).  The amorphous content is a 50:50 mixture of olivine and pyroxene and the crystalline content is a 50:50 mixture of enstatite plus forsterite. The adopted $10\,\mu$m representative continuum is given by the median values (mean slopes and mean offsets: $C_{\nu} = \bar{a} \times \lambda
  + \bar{b}$) of all the local continua linearly ($C_{\nu}^{i} = a
^{i} \times \lambda + b ^{i}$) estimated in Sec.\,\ref{sec:measure}
for the observed amorphous 10\,$\mu$m features: $C_{\nu} = 0.01 \times
\lambda + 0.097$, with $C_{\nu}$ in Jy and $\lambda$ in
$\mu$m. Similarly, a representative C23/C28 continuum has been derived
for cold synthetic spectra: $C_{\nu} =-0.001 \times \lambda + 0.43$
($C_{\nu}$ in Jy and $\lambda$ in $\mu$m).

To obtain as representative synthetic spectra as possible, we compute
the median observed continuum over feature ratio (i.e., the continuum
flux divided by the height of the feature) for the 10\,$\mu$m feature
and the C23 complex. For the amorphous 10\,$\mu$m feature, we obtain a
ratio of 2 and we therefore normalize the opacities, before
  adding them to the continuum ($C_{\nu}$), so that the synthetic
spectra show similar ratios. For the C23, we measure a median
continuum over feature ratio of 20$\pm$10, and we normalize the
opacities in order to get a ratio between 10 and 30 in the synthetic
spectra. This exercise could not be reproduced for the large
6.0\,$\mu$m grains because the shapes of their C23/C28 features are
not representative of the observed features (especially for the
forsterite grains, see right panel of Fig.\,\ref{blowopac}),
suggesting such large grains are not the main carriers of these
features. As a final step, the synthetic spectra are degraded by
adding random noise, with a maximum amplitude of $1.8 \times 10^{-3}$
Jy for the warm spectra and $7.2 \times 10^{-3}$ for the cold spectra,
representative of the quadratic sum of the measured uncertainties on
the Spitzer data over in the 10\,$\mu$m region and in the C23/C28
spectral range. The synthetic observations displayed in
Fig.\,\ref{fig:contrast} are then processed as in
Sec.\,\ref{sec:measure} to derive the band fluxes of the crystalline
features.
\begin{figure*}
\centering
\hbox to \textwidth
{
\includegraphics[angle=0,width=0.9\textwidth,origin=bl]{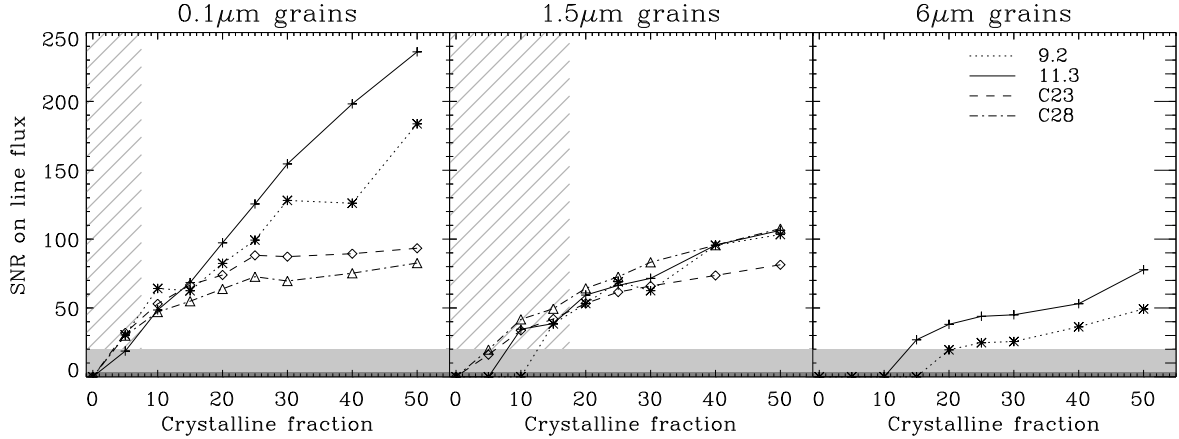}
}
\caption{\label{fig:snr_vs_crys} Signal-to-noise ratio of the 9.2\,$\mu$m, 
  11.3$\,\mu$m, C23 and C28 integrated line fluxes measured as in
  Sec.\,\ref{sec:measure} but on the synthetic spectra displayed in
  Fig.\,\ref{fig:contrast} (see text for details about the
  models). The light gray area represents the conservative detection
  threshold used in this paper (SNR=20), while the dark gray area
  shows the SNR=3 limit. Grey diagonal stripes are the loci where
  theoretical opacities are not consistent with typical observations.}
\end{figure*}

The theoretical SNRs for the $9.2$, $11.3\,\mu$m, C23 and C28 crystalline silicate features, assuming crystalline mass fractions between 0\% and 50\% and characteristic grain sizes, are displayed in Figure\,\ref{fig:snr_vs_crys}. It is seen that the SNR reached in the simulations compare well with the measured SNR values, especially for micron-sized grains (Table\,\ref{IDfeat}). The diagonal-striped areas correspond to crystalline mass fractions where the continuum over C23/C28 feature ratios could hardly match the median observed values, or which totally fail to reproduce them. This means that we are not able to discuss the detectability of features for crystalline mass fractions smaller than 5 and 15\% for grains sizes of 0.1 and 1.5\,$\mu$m, respectively. In the particular case of the 6.0\,$\mu$m grains, as explained above, we cannot compute SNR for C23 and C28 complexes as opacities do not match typical observations.

We see in Fig.\,\ref{fig:snr_vs_crys} that the SNR of the crystalline
features drops with increasing grain size (especially the $9.2\,\mu$m,
$11.3\,\mu$m features) and with decreasing crystalline mass fraction,
as expected. A crystalline fraction of $\sim 5$\% or larger seems to
be enough such that all four of the studied crystalline features are
observable if the grains are sub-micron in size. For the
$1.5\,\mu$m-sized grains, crystalline mass fractions of about 15\% are
required, (1)~to obtain spectra consistent with observations, and
(2)~to get the all the features studied detected with SNR$>$20. For
higher crystalline mass fractions, they all present very similar
behaviors. For $6.0\,\mu$m-sized grains, it becomes more difficult to
trace the detectability of the C23 and C28 features as opacities
become much less consistent with typical observations. For 9.2\,$\mu$m
and 11.3\,$\mu$m, crystalline fractions larger than 20\% and 15\%,
respectively, are required in order to detect these features with
SNR$>$20.

Since the observed features in our sample are mainly produced by
$\mu$m-sized grains (Sec.\,\ref{sec:gsize}), which in particular
produce a shoulder at the location of the $11.3\,\mu$m crystalline
feature, we conclude that the physical processes that preferentially
place $\mu$m-sized grains in the disk atmospheres do not contribute
much to the measured differences between the number of detections of the $11.3\,\mu$m feature compared to the number of detections of the C23 complex (Figs.\,\ref{fig:tt_feat} and \ref{fig:clouds}).

To summarize, we find an apparent {\it crystallinity paradox}, namely
a noticeable difference in detection statistics for the $11.3\,\mu$m
and C23 features, with more than 3 times more detections for C23
  compared to the number of detections of the 11.3\,$\mu$m forsterite
  feature. This seems counterintuitive as shorter wavelengths are
expected to probe warmer disk regions where grains can more
efficiently crystallize. We investigated the possibility that
crystalline features could be hidden by amorphous features. The effect
seems insufficient to explain the observations, but we cannot yet
firmly predict the real impact of this effect until the Spitzer
observations are confronted by a more detailed compositional analysis.
This is deferred to a future paper (Olofsson et al. 2009b, in prep.).

\section{Discussion}
\label{sec:discussion}

\subsection{On the lack of Fe-rich silicates}
Because of the large number of objects analysed in this paper, our
results raise challenging questions on the dynamics and (chemical)
evolution of planet forming disks regions that are in principle not
statistically biased. For instance, the lack of Fe-rich silicates in
our 96 star sample\footnote{except 5 marginal detections discussed in
  Sec.\,\ref{sec:ferich}} reinforces the result by \citet{Bouwman2008}
obtained on their 7-star sample. Silicates crystals studied in this
analysis may contain iron, as seen in Sec\,\ref{sec:ferich} but
magnesium is the most abundant element. Several scenarios can explain
this apparent lack of iron. First, \citet{Davoisne2006} found that a
reduction reaction during the thermal annealing of ferro-magnesian
amorphous silicates at temperatures below 1000\,K can produce pure
forsterite crystals plus spheroidal metallic particles. The iron is
therefore locked inside a metallic nanophase that cannot be
observed. Secondly, according to \citet{Nuth2006}, simple thermal
annealing in the warm inner regions of the disk cannot produce Fe-rich
silicate crystals. As the annealing timescale for iron silicates is
larger than for magnesium silicates (\citealp{Hallenbeck2000}), the
lifetime of such Fe-rich silicates against the evaporation timescale
is expected to be too short for these grains to crystallize and then
be transported outward in the disk. Still, thermal annealing induced
by shocks may be able to produce both Fe-rich and Mg-rich crystals at
the same time.

\subsection{On the need for turbulent diffusion and grain-grain
  fragmentation\label{sec:frag}}
\begin{figure}
  \centering
  \hspace*{-0.5cm}\includegraphics[angle=0,width=1.\columnwidth,origin=bl]{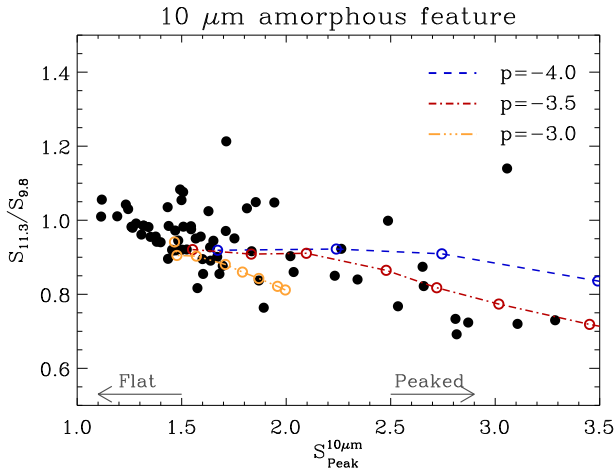}
  \caption{\label{fig:gs}Correlation between shape and strength for
    the amorphous 10\,$\mu$m feature compared to models assuming a
    power law grain size distributions with indexes between -4 and -3.
    For each colored curve, from right to left, open circles
    correspond to mimimum grain sizes $a_{\mathrm{min}}$ equal to 0.1,
    0.5, 0.75, 1.0, 1.5, 2.0 and 4.0\,$\mu$m. For blue points only the
    last four mininum grain sizes are represented, for red points the
    0.1\,$\mu$m point is not shown, as their
    $S_{\mathrm{Peak}}^{10\,\mu m}$ is larger than $3.5$.}
\end{figure}
The new grain size proxy introduced here for cold crystalline
silicates (via the C23 complex) indicates that the upper layers of the disks
are preferentially populated by $\mu$m-sized grains. This result also
holds for the warm disk region using the classical size proxy for the
carriers of the 10\,$\mu$m amorphous feature
(Fig.\,\ref{kurt_sp}). Adopting a purely sequential evolutionary
approach for the solid constituents in disks, where interstellar-like
grains coagulate to form micron-sized dust aggregates, then pebbles
and planetesimals, a straighforward conclusion from these results
would be that grain agglomeration occured in the first few AU and that
we are witnessing an intermediate, regular step of the growth of solid
particles along their way to form planets. Nevertheless, this picture
is most certainly an oversimplified view of the processes at work in
planet forming regions around young stars as it does not fit with
basic dynamical considerations. Indeed, the short settling time-scale
of $\mu$m-sized grains in a laminar disk, and their expected fast
growth which will accelerate sedimentation
\citep[][]{Dullemond2005,Laibe2008}, both argue in favor of some
processes that sustain the disk atmosphere with grains that have
mid-infared spectroscopic signatures (grains smaller than about
10$\,\mu$m) over a few Myr. Therefore, our results support the
replenishment of the disk upper regions by the vertical transport of
dust particles \citep[e.g. turbulent diffusion,][]{Fromang2009},
likely combined with grain-grain fragmentation to balance the expected
efficient growth of solid particulates \citep[that accelerates
sedimentation,][]{Dullemond2005, Laibe2008}. In this picture, the
grains observed at mid-IR wavelengths would then result from the
destruction of much larger solid particles \citep[an assumption
consistent with the presence of millimeter-sized grains in most TTauri
disks, e.g. review by][]{Natta2007}, rather than from the direct
growth from interstellar-like grains.

\subsection{On the depletion of submicron-sized
  grains}\label{sec:depl}
\begin{figure}
  \centering
 \hspace*{-0.5cm}\includegraphics[angle=0,width=1.\columnwidth,origin=bl]{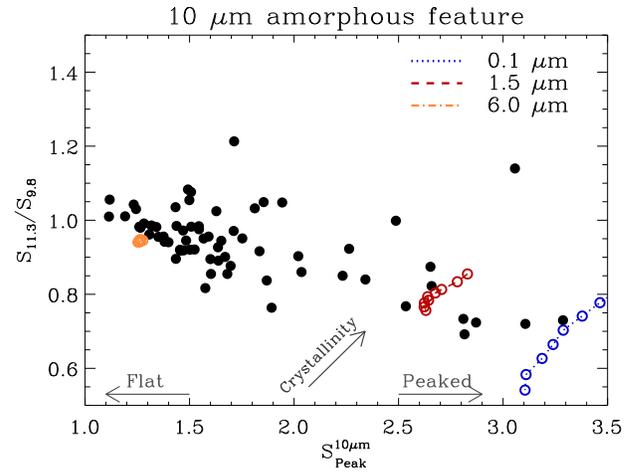}
 \caption{\label{fig:S11_cr}Correlation between shape and strength for
   the amorphous 10\,$\mu$m feature compared to models assuming a
   variation in crystallinity fraction, for three different grain sizes. For each grain size, increasing crystallinity goes from bottom left to top right, from 0\% to 50\% crystallinity (0, 5, 10, 15, 20, 25, 30, 40, 50\%)}
\end{figure}
A consequence of the above conclusion is that one would have to
explain the general apparent depletion of submicron-sized grains in
the disk zones probed by mid-IR spectroscopy. The fact that we see a
flat, boxy 10\,$\mu$m feature profile for most TTs, indeed indicates
that there cannot be many submicron-sized grains, because their
emission would overwhelm the few large grains. Grain-grain collisions
are nevertheless expected to produce fragments with a range of sizes
that are likely to extend to the submicron domain (i.e. the size of
the monomers that form $\mu$m-sized aggregates). This effect might be
enhanced by the large velocity fluctuations (a fraction of the local
sound speed) found at high altitudes above the disk midplane in disk
simulations with MHD turbulence \citep{Fromang2009}. Qualitatively,
submicron-sized grains, lifted in the disk atmospheres by turbulent
diffusion or produced locally, are therefore expected to be abundant,
which contradicts the IRS observations.

\subsubsection{Impact of the size distribution}
The above conclusion is based on our single, characteristic grain size
approach (Secs.\,\ref{sec:gsize} and \ref{sec:gsize2}) and one may
question whether this assumption affects our interpretation of the
observations. To answer that question, we calculate mean
cross-sections as follows:
\begin{eqnarray*}
  \sigma_{\mathrm{abs}} = \int_{a\mathrm{_{min}}}^{a\mathrm{_{max}}} \pi a^2 
  Q_{\mathrm{abs}}(a)\,\mathrm{d}n(a) / \int_{a\mathrm{_{min}}}^{a\mathrm{_{max}}}\mathrm{d}n(a),
\end{eqnarray*}
where $a$ is the grain radius, $Q_{\mathrm{abs}}$ is the dimensionless
Mie emission/absorption coefficient for a 50:50 mixture of amorphous
pyroxene and olivine, and $\mathrm{d}n(a)\propto a^p \mathrm{d}a$ is
the differential grain size distribution. We arbitrarily choose
$a\mathrm{_{max}}\,=\,100\,\mu$m (large enough such that it does not
affect the calculations for MRN-like size distributions), and we vary
the minimum grain size $a\mathrm{_{min}}$ between 0.1 and
4.0\,$\mu$m. Using a classical MRN ($p=-3.5$) grain size distribution,
we find that the diversity of $S_{11.3}/S_{9.8}$ and
$S_{\mathrm{Peak}}^{10\mu m}$ values is well reproduced by adjusting
the minimum grain size $a_{\mathrm{min}}$, and that the bulk of the
observations is reproduced for $a_{\mathrm{min}}$ larger than
2\,$\mu$m (Fig.\,\ref{fig:gs}), a result consistent with our single
grain size analysis. Nevertheless, another way to avoid having peaked
10\,$\mu$m amorphous features typical of pristine grains is to use a
flatter slope for the size distribution. For example, $p$ values
larger than $-3$ reduce the contribution of submicron-sized grains to
the mean absorption cross-section (see Fig.\,\ref{fig:gs} with
$p=-3$), creating an amorphous $10\,\mu$m feature that is large and
boxy despite the presence of some submicron-sized grains. Therefore, the majority of the points is located at the left side of Fig.\,\ref{fig:gs}, which either probes $\mu$m-sized grains as the small end
of the size distribution, or much flatter grain size distributions
compared to the MRN, meaning that the amount of small grains is at
least severely diminished in the regions probed by the Spitzer/IRS.

\subsubsection{Impact of the crystallinity}
Several studies (e.g. \citealp{Bouwman2001}, \citealp{Honda2006} or \citealp{2005}) mention the possible impact of the crystallinity on the shape versus strength correlation for the 10\,$\mu$m feature. According to these studies, the presence of crystalline grains can mimic the flattening of the feature.  \citet{Min2008} recently investigated in detail this effect by considering inhomogeneous aggregates. Their Fig.\,2 shows that $S_{\mathrm{Peak}}^{\mathrm{10\,\mu m}}$ remains approximately constant for crystalline fractions up to 20\% (the maximum fraction considered in their study) while $S_{11.3}/S_{9.8}$ rises, suggesting a shift along the $y$--axis of Fig.\,\ref{fig:gs} due to crystallinity rather than along the observed trend.

In order to further investigate the impact of crystallinity on this correlation, we use the synthetic spectra described in Sec.\,\ref{sec:contrast}. The only difference is that we do not normalize opacities anymore to obtain a constant ratio between feature strength and continuum. Normalizing opacities would lead to a constant $S_{\mathrm{Peak}}^{\mathrm{10\,\mu m}}$, which is not what we want to characterize here. We therefore normalize the opacities, for each grain size, so that only the strongest feature reaches a ratio of 2 between the continuum and the feature strength. Other features are normalized with the same factor and obviously present a higher ratio between continuum and feature strength. This normalization is done in order to reproduce spectra similar to observations, and still conserve the diversity of feature strengths. Subsequently, values for $S_{9.8}$, $S_{11.3}$ and $S_{\mathrm{Peak}}^{\mathrm{10\,\mu m}}$ are derived in the same way as in Sec\,\ref{sec:gsize}. Fig.\,\ref{fig:S11_cr} shows the impact of crystallinity on the shape versus strength for the 10\,$\mu$m feature. We consider three grain sizes (0.1, 1.5 and 6.0\,$\mu$m) and the same composition as in Sec.\,\ref{sec:contrast}: 50:50 mixture of olivine and pyroxene for the amorphous content and 50:50 of enstatite and forsterite for the crystalline population. The crystallinity fraction varies from 0\% to 50\% (0, 5, 10, 15, 20, 25, 30, 40 and 50\%), and increasing crystallinity goes from bottom left to top right on the plot. We note that for 6.0\,$\mu$m-sized grains the effect is negligible, while it is more dramatic for 0.1\,$\mu$m-sized grains. The spread induced by an increase of crystallinity is almost orthogonal to the direction of the correlation. This result was also found by \citet{Kessler-Silacci2006} (Fig.\,10e). The direction of this deviation can simply be understood when considering opacities for amorphous and crystalline grains. For the same mass of dust, crystalline grain emission will be stronger compared to amorphous grains. Therefore adding some crystalline grains to a purely amorphous dust content will give a 10\,$\mu$m feature stronger {\it and} broader (because of crystalline features at, e.g., 11.3\,$\mu$m in the case of forsterite). This will thus produce a higher value of $S_{11.3}/S_{9.8}$ and at the same time a stronger $S_{\mathrm{Peak}}^{\mathrm{10\,\mu m}}$.

Crystallinity therefore does impact the correlation and may be responsible for a dispersion in the direction orthogonal to the correlation, but according to the opacities, it cannot solely explain the observed trend.

\subsubsection{Hypothesis for the depletion of submicron-sized grains}

Several solutions could be envisaged to explain the apparent depletion
of submicron-sized grains at high altitudes above the disk
midplane. The fragmentation time-scale of grains in disk atmospheres may
be longer than the grain growth timescale, thereby quickly eliminating the
smallest and most freshly produced submicronic grains. The high
velocity fluctuations observed in MHD simulations at high altitudes,
do not, in principle, work in favor of this scenario. An alternative
possibility may be that the submicron-sized grains are evacuated from
the upper layers of the disks by stellar winds or radiation pressure. The
impact of radiation pressure on grains in young disks has been
explored by \citet{Takeuchi2003} and present several
advantages. First, it acts on short timescales as long as the grain
size reaches a threshold, usually referred to as the blowout size
limit, below which the radiation pressure force overcomes the
gravitational and gas drag forces. Second, the blowout size limit
falls in the right regime of grain sizes (close to a micrometer for
silicates) for solar-type stars in the presence of some gas.

It is interesting to apply the latter scenario to the case of more
luminous stars.  According to \citet{Kessler-Silacci2007}, HAeBe stars
tend to show a widest variety of 10$\,\mu$m silicate features,
including the peaked features attributed to submicron-sized
grains. Because of their higher temperature and luminosity, HAeBe
stars are expected to have upper layers that are even more ionized
than those of T\,Tauri stars, thereby enhancing the turbulence if it
is supported by disk magnetic fields, as well as velocity
fluctuations. This would therefore encourage the systematic production
of submicron dust fragments in disk atmospheres. On the other hand,
silicate grains of a few micrometers and smaller, are in principle
more prone to be eliminated by radiation pressure. Nevertheless, the
latter reasoning assumes an optically thin medium at the wavelengths
where grains absorb most, namely in the UV and visible. The optically
thin region at mid-IR wavelengths is likely to be essentially opaque
in the UV/visible, and this opacity will affect more strongly the most
luminous stars, thereby possibly reducing the blowout size limit.

Clearly, the depletion of submicron-sized grains in the disk
atmospheres of most T\,Tauri stars should be further investigated by
models to examine the likelihood of the scenarios discussed above
(competition between coagulation and fragmentation, stellar winds,
radiation pressure, ...)

\subsection{On radial mixing}

The coldest disk component probed by our IRS observations shows a
rather high crystallinity fraction. According to
Fig.\,\ref{fig:tt_feat}, the 23 and 28$\,\mu$m complexes are detected
in more than 50\% of the spectra. In a scenario where silicate
crystals form by thermal annealing, the production of crystalline
silicates can only occur on reasonably short time-scales near the star
(T~$>$~900\,K), namely in regions much hotter than those emitting at
$\lambda > 20\,\mu$m. The observed cold crystals would thus have been
radially transported there from the inner disk zone, which according
to two-dimensional models can occur around the midplane of accretion
disks \citep[][and references
therein]{Ciesla2009}. \citet{Keller2004}, for example, find that $\sim
5$ to 30\% of the total mass that is transported inward at high disk
altitudes is transported outward near the disk midplane. For a solar
mass star with an accretion rate of $\dot{M} = 10^{-6}M_{\odot}$/yr,
\citet{Ciesla2009} estimates the crystalline fraction at 10\,AU to be
about $\sim 40$\% by the end of his simulations. As the accretion rate
becomes smaller, the crystalline fraction drops accordingly: $\sim
10$\% and $\sim 1$\% for $\dot{M} = 10^{-7}M_{\odot}$/yr and
$10^{-8}M_{\odot}$/yr, respectively. However, \citet{Watson2009} find
only a weak correlation between the accretion rate and the crystalline
features arising at around 10$\,\mu$m, and no correlation with the
33$\,\mu$m feature, leading the authors to suggest that some
mechanisms may be erasing such trends. Alternatively, other
crystallisation processes at colder temperatures (e.g. via
  nebular shock, \citealp{Desch2002}) could also contribute to the
presence of crystals at significant distances from the central object
\citep[e.g.][]{Kimura2008}.

It is noteworthy that the warmer disk zone seems significantly less
crystalline than do the colder regions, although detailed
compositional analysis of the IRS spectra is required before firmly
concluding this must be so (Olofsson et al. 2009b, in prep.)
Interestingly, compositional fits to some individual IRS spectra by
\citet{Bouy2008} and \citet{Sargent2009a} do show the need for larger
amounts of silicate crystals in the outer regions compared to inner
regions (e.g. \object{2MASS J04442713+2512164} and
\object{F04147+2822}). If radial mixing leads to the large crystalline
fraction in the cold disk component, it leaves the disks with an
inhomogeneous chemical composition for what concerns the solid
phase. Some re-amorphisation processing of the crystalline grains, by
X-ray emission or cosmics rays for example, could also contribute to
the apparent lack of crystals in the inner regions of disks.

This result of a cold component being more crystalline compared to the
inner warm regions seems, at first glance, to be in contradiction with
the results by \citet{2004} who observed three HAeBe stars with the
mid-IR instrument MIDI for the VLTI interferometer. Using different
baselines, they found that the inner regions (1--2\,AU) of the disks
are more crystalline compared to the outer regions (2--20\,AU). Beside
the fact that their sample is limited to 3 intermediate mass objects
while we analyse a statistically significant TTs sample, direct
comparison of their observations with ours is not
straightforward. They do have spatial information, but only for the
10\,$\mu$m spectral region, while we cover a larger spectral
range. We, on the other hand, are observing the entire disk with
Spitzer/IRS, with no direct spatial information. An ideal and
challenging way to confront such kinds of interferometric data with
IRS would be to repeat similar observations for TTs and even better,
at wavelengths larger than 20\,$\mu$m. This will ultimately require
space-born interferometers because of the poor atmosphere transmission
in the mid-IR. With such observations, however, we would then be able
to constrain the crystalline fraction in the inner and outer regions
of the disk more directly.

\subsection{On the similarities with Solar System objects}

Studies of Solar Sytem asteroids and comets show strong similarities
with our results. First of all, laboratory studies of asteroidal and
cometary solids have show that they contain very little ISM-like
materials. The 81P/Wild 2 samples from the Stardust mission
\citet{McKeegan2006}, for example, show that this comet contains
high-temperature silicates and oxide minerals. These authors conclude
that such materials could not have been formed via annealing of
presolar amorphous silicates in the Kuiper belt, and also that they
could not have been formed from a single isotopic reservoir. Overall,
this suggests that comet 81P/Wild 2 sampled different regions of the
inner solar protoplanetary disk during its formation. One possible
transportation mechanism proposed is the combination of winds
associated with bipolar outflows. This hypothesis is also supported by
laboratory measurements from \citet{Toppani2006} on condensation under
high temperature, low-pressure conditions. Mg-rich silicates crystals
can rapidly ($\sim 1$h at $4 \times 10^{-3}$ bar) be produced from the
gas phase through condensation, confirming the fact that
crystalline grains can even be produced in bipolar outflows from
evolved stars. In still further work on the 81P/Wild 2 Stardust mission,
\citet{Zolensky2006}, found large abundances of crystalline
materials. Olivine is present in most of the studied particles, with
grain sizes ranging between submicron-sized to over 10\,$\mu$m, and it
is concluded that some materials in the comet have seen temperatures
possibly higher than 2000\,K. Similar to \citet{McKeegan2006},
large-scale radial transportation mechanisms inside the disk are
required to match the derived mineralogy of the olivine
samples. Similar results were found studying other solar system
bodies, for exemple, comet Hale-Bopp. Spectroscopic studies of this
object (\citealp{Wooden1999}, \citealp{Wooden2000}) revealed the
presence of crystalline olivine, crystalline ortho-pyroxene and
crystalline clino-pyroxene, in significant quantities.

All these results echo the {\it crystallinity paradox} we find in our
Spitzer observations, in the sense that the disk dust content is far
from being homogeneous, and even far from the simple picture with
crystalline grains close to the central object and amorphous content
in the outer regions.

\begin{figure*}
\begin{center}
\hspace*{+0.5cm}
\hbox to \textwidth
{
\parbox{0.6\textwidth}{\includegraphics[angle=0,width=0.6\textwidth,origin=bl]{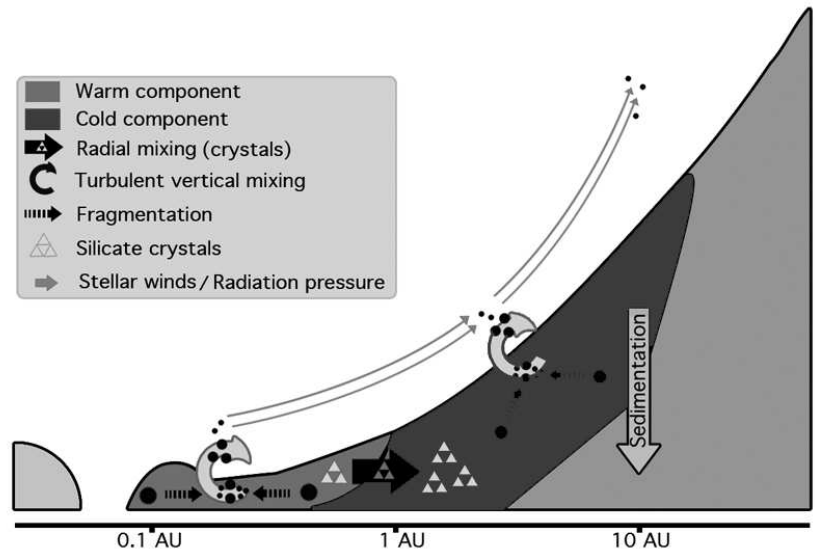}
}
\parbox{0.25\textwidth}{\caption{\label{fig:disksum}Schematic view of
    the TTauri disk regions probed by the Spitzer IRS spectroscopic
    observations. This illustrates some of the dynamical processes
    that are discussed in Sec.\,\ref{sec:discussion} and summarized in
    Sec.\,\ref{sec:summary}.}
}
}
\end{center}
\end{figure*}

\section{Summary and conclusion}
\label{sec:summary}

We have conducted in this paper a comprehensive statistical study of
crystalline silicates in proto-planetary (Class~II) disks which was
made possible thanks to the high sensitivity of the spectroscopic
instrument IRS (5--35$\,\mu$m) onboard the Spitzer Space
Telescope. This unprecedented work is among the first statistical
studies of crystalline silicates emission features over a large sample
of young solar-analog stars (96 objects). We find the following
results:
\begin{enumerate}

\item Crystallinity is not a marginal phenomenon in Class II disks
  around young solar analogs, as more than 3/4 of the objects show at
  least one crystalline feature in the IRS spectra. Crystalline
  features emitting at wavelengths larger than 20\,$\mu$m are in
  addition widely present: both the C23 and C28 complexes, produced by enstatite plus forsterite features, are present at a frequency larger than 50\%.

\item The crystalline silicates revealed by the IRS spectra are
  essentially Mg-rich silicates (forsterite, enstatite), with a small
  contribution of CaMg-rich silicates (diopside). We find no evidence
  of Fe-rich silicates (marginal evidence for fayalite) nor troilite
  (FeS) in our spectra.

\item We find that the amorphous 10\,$\mu$m feature is not correlated
  with the C23, C28 and 33.6\,$\mu$m crystalline features, neither on
  the basis of the frequencies at which they are detected nor on the
  fluxes they emit. Crystalline features emitted at wavelengths larger
  than 20\,$\mu$m are, on the other hand, correlated to each other
  (C23, C28 and 33.6\,$\mu$m), based either on detection frequencies
  and band fluxes.

\item From previous results, we conclude that within the spectral
  range of the Spitzer/IRS instrument, we are essentially probing two
  independent dust populations (see the illustration in
  Fig.\,\ref{fig:disksum}): a warm component, which is likely located
  close to the central object ($\leq 1$\,AU), emitting features in the
  10\,$\mu$m region, and a second component, located at larger radii
  ($\leq 10$\,AU) or deeper inside the disk; the latter cold component
  being responsible for the emission of both C23 and C28 complexes
  plus the 33.6\,$\mu$m forsterite feature.

\item For a large majority of objects, the upper layers of the disks are
  essentially populated by micron-sized grains, suggesting that
  submicron grains are largely depleted in disks atmospheres. For the
  amorphous 10\,$\mu$m feature, we find the known correlation between
  the shape and the strength of the feature used as a grain size proxy
  and showing that the grains are $\mu$m-sized or larger. We adapted
  this relationship to the C23 complex and find similar results for
  the crystalline grains producing this complex. We also noted that
  the grain size proxies for the 10$\,\mu$m and C23 features are
  uncorrelated.

\item We find a correlation between the shape of the SED
  ($F_{30}/F_{13}$) and the grain size proxy for the 10\,$\mu$m
  feature. We see rather small, warm amorphous silicates for flared
  disks and grains becoming larger as disks are flattening. This trend
  is almost reversed for the cold, C23 crystalline grains as small
  crystals are only found for flattened SEDs.

\item We identify an apparent {\it crystallinity paradox} since the
  cold crystalline features ($\lambda>20\,\mu$m) are much more
  frequently detected than the warm ones ($\lambda\sim10\,\mu$m). We
  show that crystalline features can be hidden by amorphous features,
  especially near $\lambda \simeq 10\,\mu$m, but this contrast effect
  is not sufficient to explain the difference regarding the apparition
  frequencies of the C23 complex and the 11.3\,$\mu$m forsterite
  feature, suggesting that the warm grain population is, on average,
  intrinsically more amorphous than the cold grain population.

\end{enumerate}

We argue that our Spitzer spectroscopic observations of disks around
young stars reveal ongoing dynamical processes in planet forming
regions of disks that future theoretical disk evolution and planet
formation models will have to explain:
\begin{itemize}

\item some mechanisms must supply the upper layers of the disks with
  micron-sized grains that should otherwise grow and settle
  rapidly. Vertical turbulent diffusion, accompanied by grain-grain
  fragmentation, would be natural candidates,

\item some mechanisms must act efficiently to remove the
  submicron-size grains from the regions which are optically thin at
  IRS wavelengths. Stellar winds and radiation pressure have been
  suggested as possible processes,

\item and, some mechanisms are responsible for rendering the colder
  disks regions much more crystalline than the warmer zones ({\it
    crystallinity paradox}). If this result has to do with dust radial
  mixing, it teaches us that the mixing does not result in a disk with
  a homogenous grain composition, and that it must transport crystals
  efficiently into the ``comet-forming regions''.

\end{itemize}


\begin{acknowledgements}
  The authors warmly thank Bram Acke for his help on the IDL routine
  to extract the crystalline features characteristics, J\'erome
  Al\'eon and Matthieu Gounelle for sharing with us their precious
  knowledge of small solid Solar System bodies, and Joel Green for the
  rich discussions we had. We also thank the anonymous referee for the
  very useful comments that helped improving this study. It is also a
  pleasure to acknowledge fruitful discussions with the Grenoble FOST
  team members, as well as the participants to the ANR (Agence
  Nationale de la Recherche of France) project ANR-07-BLAN-0221. This
  research is partly supported by the {\it Programme National de
    Physique Stellaire} (PNPS).
\end{acknowledgements}

\bibliography{biblio}

\onecolumn
\onecolumn
\begin{landscape}
\setlength\LTleft{0pt}
\setlength\LTright{0pt}
\begin{longtable}{@{\excs}lrrccccc}
\caption{Characteristics of the observations\label{obs}}\\
\hline \hline
Starname & RA (J2000) & Dec (J2000) & Class & SpT & AOR Key & Obs. date & Modules \\
\hline
\endfirsthead
\caption{continued.}\\
\hline\hline
Starname & RA (J2000) & Dec (J2000) & Class & SpT & AOR Key & Obs. date & Modules \\
\hline
\endhead
\hline
\endfoot
\endlastfoot
\multicolumn{6}{l}{Perseus}\\
\hline
\object{RNO\,15} & 03h27m47.68s & 30d12m04.3s & TTs (II) & - & 0005633280 & 2004-08-30 & SL, SH, LH\\
\object{LkH$\alpha$\,325} & 03h28m52.22s & 30d45m05.5s & CTTs (II) & K2$^a$ & 0015737600 & 2005-09-05 & SL, LL\\
\object{LkH$\alpha$\,270} & 03h29m17.66s & 31d22m45.1s & TTs (II) & K2.5-K7 & 0005634048 & 2005-02-08 & SL, SH, LH\\
\object{LkH$\alpha$\,271} & 03h29m21.87s & 31d15m36.3s & TTs (II) & K3-K5 & 0011827968 & 2005-02-10 & SL, SH, LH\\
\object{SSTc2d\,J033036.0+303024} & 03h30m35.94s & 30d30m24.4s & II & - & 0015737600 & 2005-09-05 & SL, LL\\
\object{SSTc2d\,J033037.0+303128} & 03h30m36.98s & 30d31m27.7s & II & - & 0015737600 & 2005-09-05 & SL, LL\\
\object{LkH$\alpha$\,326} & 03h30m44.05s & 30d32m46.7s & TTs (II) & G-M0 & 0005634304 & 2005-02-10 & SL, SH, LH\\
\object{SSTc2d\,J033052.5+305418} & 03h30m52.58s & 30d54m18.1s & II & - & 0015737088 & 2005-09-05 & SL, LL\\
\object{SSTc2d\,J033241.7+311046} & 03h32m41.76s & 31d10m46.6s & II & - & 0015737344 & 2005-09-05 & SL, LL\\
\object{LkH$\alpha$\,327} & 03h33m30.41s & 31d10m50.4s & TTs (II) & A9-K2 & 0005634560 & 2004-02-03 & SL, SH, LH\\
\object{SSTc2d\,J033341.3+311341} & 03h33m41.30s & 31d13m41.6s & II & - & 0015736832 & 2005-09-05 & SL, LL\\
\object{BD+31\,634} & 03h41m39.20s & 31d36m10.3s & HAeBe (II) & A8V & 0015736320 & 2005-09-05 & SL, LL\\
\object{SSTc2d\,J034219.3+314327} & 03h42m19.31s & 31d43m26.6s & II & - & 0015736064 & 2005-09-05 & SL, LL\\
\object{IRAS\,03406+3133} & 03h43m44.54s & 31d43m09.1s & - & - & 0015735296 & 2005-09-05 & SL, LL\\
\object{LkH$\alpha$\,330} & 03h45m48.29s & 32d24m11.8s & CTTs (II) & G3 & 0005634816 & 2004-09-02 & SL, SH, LH\\
\object{IRAS\,03446+3254} & 03h47m47.12s & 33d04m03.4s & TTs (II) & - & 0005635072 & 2004-09-29 & SL, SH, LH\\ \hline
\multicolumn{6}{l}{Taurus}\\
\hline
\object{LkCa\,8} & 04h24m57.08s & 27d11m56.5s & CTTs (II) & M0 & 0009832960 & 2005-02-08 & SH, LH\\
\object{IQ\,Tau} & 04h29m51.56s & 26d06m45.0s & WTTs (II) & M0-M0.5 & 0009832704 & 2005-02-09 & SH, LH\\
\object{FX\,Tau} & 04h30m29.62s & 24d26m45.1s & C+WTTs (II) & M1-M4 & 0009832448 & 2005-02-10 & SH, LH\\
\object{V710\,Tau} & 04h31m57.79s & 18d21m36.3s & C+WTTs (II) & M0.5-M3 & 0005636608 & 2004-09-29 & SH, LH\\
\object{RXJ0432.8+1735} & 04h32m53.23s & 17d35m33.7s & TTs (II) & M2$^b$ & 0015917824 & 2005-09-09 & SL, LL\\
\object{DN\,Tau} & 04h35m27.37s & 24d14m58.8s & CTTs (II) & M0 & 0009831936 & 2005-02-10 & SH, LH\\
\object{CoKu\,Tau\,3} & 04h35m40.94s & 24d11m08.6s & CTTs (II) & M1 & 0009831936 & 2005-02-10 & SH, LH\\
\object{CoKu\,Tau\,4} & 04h41m16.79s & 28d40m00.5s & CTTs (II) & M1.5 & 0005637888 & 2004-09-02 & SH, LH\\
\object{RR\,Tau} & 05h39m30.52s & 26d22m27.0s & HAeBe (II) & B8-A5 & 0005638400 & 2004-09-28 & SL, SH, LH\\ \hline
\multicolumn{6}{l}{Chamaleon}\\
\hline
\object{SX\,Cha} & 10h55m59.74s & -77d24m39.9s & TTs (II) & M0.5 & 0005639424 & 2004-08-31 & SH, LH\\
\object{SY\,Cha} & 10h56m30.47s & -77d11m39.4s & TTs (II) & M0 & 0005639424 & 2004-08-31 & SH, LH\\
\object{TW\,Cha} & 10h59m01.10s & -77d22m40.80s & TTs (II) & K0-M0 & 0005639680 & 2004-09-01 & SH, LH\\
\object{B35} & 11h07m21.49s & -77d22m11.8s & CTTs (II) & M2 & 0005639680 & 2004-09-01 & SH, LH\\
\object{VW\,Cha} & 11h08m01.50s & -77d42m28.70s & CTTs (II) & K2-K7 & 0005639680 & 2004-09-01 & SH, LH\\
\object{Hn\,9} & 11h09m18.16s & -76d30m29.2s & - & K4-M1$^c$ & 0009831168 & 2005-02-10 & SL, LL\\
\object{VZ\,Cha} & 11h09m23.80s & -76d23m20.7s & TTs (II) & K6-K7 & 0005640448 & 2004-09-02 & SH, LH\\
\object{WX\,Cha} & 11h09m58.75s & -77d37m08.9s & TTs (II) & K7-M0 & 0005640192 & 2004-09-01 & SH, LH\\
\object{ISO-Cha237} & 11h10m11.44s & -76d35m29.2s & TTs (II) & M0 & 0005640448 & 2004-09-02 & SL, LL\\
\object{C7-11} & 11h10m38.01s & -77d32m39.9s & TTs (II) & K3 & 0005640192 & 2004-09-01 & SH, LH\\
\object{HM\,27} & 11h10m49.62s & -77d17m51.7s & TTs (II) & K7 & 0005640192 & 2004-09-01 & SH, LH\\
\object{XX\,Cha} & 11h11m39.67s & -76d20m15.1s & TTs (II) & M1-M2 & 0005640448 & 2004-09-02 & SH, LH\\
\object{HD\,98922} & 11h22m31.67s & -53d22m11.4s & HAeBe (II) & B9 & 0005640704 & 2004-01-04 & SH, LH\\
\object{HD\,101412} & 11h39m44.46s & -60d10m27.7s & HAeBe (II) & B9.5 & 0005640960 & 2005-02-10 & SL, SH, LH\\
\object{T\,Cha} & 11h57m13.53s & -79d21m31.5s & TTs (II) & G2-K0 & 0005641216 & 2004-07-18 & SH, LH\\
\object{IRAS\,12535-7623} & 12h57m11.78s & -76d40m11.5s & TTs (II) & M0 & 0011827456 & 2004-08-31 & SH, LH\\
\object{ISO-ChaII\,13} & 12h58m06.70s & -77d09m09.5s & BD (II) & - & 0015918592 & 2005-09-13 & SL, LL\\
\object{RXJ1301.0-7654} & 13h00m53.23s & -76d54m15.2s & - & K1$^d$ & 0009830656 & 2005-03-12 & SL, LL\\
\object{Sz50} & 13h00m55.37s & -77d10m22.2s & TTs (II) & K7-M3 & 0011827456 & 2004-08-31 & SH, LH\\
\object{ISO-ChaII\,54} & 13h00m59.20s & -77d14m02.7s & TTs (II) & K3 & 0015735040 & 2005-08-13 & SL, SH, LL\\
\object{Sz52} & 13h04m24.90s & -77d52m30.3s & - & - & 0018215168 & 2006-07-03 & SL, LL \\
\object{Sz62} & 13h09m50.66s & -77d57m24.0s & - & M2$^e$ & 0009830400 & 2005-03-12 & SL, LL\\ \hline
\multicolumn{6}{l}{Lupus}\\
\hline
\object{HT\,Lup} & 15h45m12.87s & -34d17m30.6s & TTs (II) & K2 & 0005643264 & 2004-08-30 & SL, SH, LH\\
\object{GW\,Lup} & 15h46m44.68s & -34d30m35.4s & TTs (II) & M2-M4 & 0005643520 & 2004-08-30 & SL, SH, LH\\
\object{HM Lup} & 15h47m50.63s & -35d28m35.4s & TTs (II) & M4 & 0005643776 & 2004-08-30 & SL, LL\\
\object{Sz73} & 15h47m56.98s & -35d14m35.1s & TTs (II) & K2-M & 0005644032 & 2004-08-30 & SL, SH, LH\\
\object{GQ\,Lup} & 15h49m12.10s & -35d39m05.0s & TTs (II) & K7-M0 & 0005644032 & 2004-08-30 & SL, SH, LH\\
\object{Sz76} & 15h49m30.74s & -35d49m51.4s & - & M1$^f$ & 0015916288 & 2005-09-09 & SL, LL\\
\object{IM\,Lup} & 15h56m09.20s & -37d56m06.4s & TTs (II) & M0 & 0005644800 & 2004-08-30 & SL, SH, LH\\
\object{RU\,Lup} & 15h56m42.31s & -37d49m15.5s & CTTs (II) & K3-M0 & 0005644800 & 2004-08-30 & SL, SH, LH\\
\object{Sz84} & 15h58m02.50s & -37d36m02.8s & - & M5.5$^g$ & 0005644288 & 2004-03-25 & SL, LL\\
\object{RY\,Lup} & 15h59m28.39s & -40d21m51.2s & TTs (II) & K0-K4 & 0005644544 & 2004-08-30 & SL, SH, LH\\
\object{EX\,Lup} & 16h03m05.50s & -40d18m24.9s & TTs (II) & M0 & 0005645056 & 2004-08-30 & SL, SH, LH\\
\object{RXJ1603.2-3239} & 16h03m11.81s & -32d39m20.2s & TTs (II) & K7$^g$ & 0015917312 & 2005-09-09 & SL, LL\\
\object{Sz96} & 16h08m12.62s & -39d08m33.4s & - & M1.5$^f$ & 0016755200 & 2006-03-15 & SL, LL\\
\object{Sz102} & 16h08m29.70s & -39d03m11.2s & TTs (II) & M0 & 0009407488 & 2004-03-25 & SL, SH, LH\\
\object{SSTc2d\,J161159.8-382338} & 16h11m59.81s & -38d23m37.5s & II & - & 0015737856 & 2005-08-14 & SL, LL\\
\object{RXJ1615.3-3255} & 16h15m20.23s & -32d55m05.1s & TTs (II) & K5$^h$ & 0015916800 & 2005-09-09 & SL, LL\\ \hline
\multicolumn{6}{l}{Ophiuchus}\\
\hline
\object{AS\,205} & 16h11m31.35s & -18d38m26.1s & TTs (II) & K5 & 0005646080 & 2004-08-28 & SL, SH, LH\\
\object{Haro\,1-1} & 16h21m34.69s & -26d12m27.0s & CTTs (II) & K5-K7 & 0009833472 & 2005-03-12 & SL, SH, LH\\
\object{SSTc2d\,J162148.5-234027} & 16h21m48.49s & -23d40m27.4s & II & - & 0015920897 & 2006-03-15 & SL, LL\\
\object{SSTc2d\,J162221.0-230403} & 16h22m21.01s & -23d04m02.6s & I & - & 0015920897 & 2006-03-15 & SL, LL\\
\object{SSTc2d\,J162245.4-243124} & 16h22m45.40s & -24d31m23.9s & II & - & 0015920641 & 2006-03-15 & SL, LL\\
\object{SSTc2d\,J162332.8-225847} & 16h23m32.85s & -22d58m46.9s & II & - & 0015920641 & 2006-03-15 & SL, LL\\
\object{Haro\,1-4} & 16h25m10.51s & -23d19m14.5s & TTs (II) & K4-K6 & 0009833216 & 2005-03-12 & SH, LH\\
\object{VSSG1} & 16h26m18.86s & -24d28m19.7s & TTs (II) & - & 0005647616 & 2004-08-28 & SH, LH\\
\object{DoAr\,24E} & 16h26m23.38s & -24d21m00.1s & TTs (II) & K0-K1 & 0005647616 & 2004-08-28 & SH, LH\\
\object{DoAr\,25} & 16h26m23.68s & -24d43m14.00s & TTs (II) & K5$^i$ & 0012663808 & 2005-08-12 & SH, LH \\
\object{GY23} & 16h26m24.06s & -24d24m48.1s & TTs (II) & K5-M2$^j$ & 0005647616 & 2004-08-28 & SH, LH\\
\object{SR\,21} & 16h27m10.28s & -24d19m12.5s & TTs (II) & F4-G2.5 & 0005647616 & 2004-08-28 & SH, LH\\
\object{SSTc2d\,J162715.1-245139} & 16h27m15.14s & -24d51m38.9s & II & - & 0016754688 & 2006-03-15 & SL, LL\\
\object{SR\,9} & 16h27m40.27s & -24d22m04.0s & TTs (II) & K5-M2 & 0012027392 & 2004-09-02 & SH, LH\\
\object{SSTc2d\,J162816.7-240514} & 16h28m16.73s & -24d05m14.3s & II & - & 0016754944 & 2006-03-15 & SL, LL\\
\object{V853Oph} & 16h28m45.28s & -24d28m19.0s & TTs (II) & M1.5 & 0012408576 & 2005-03-12 & SH, LH\\
\object{ROX42C} & 16h31m15.75s & -24d34m02.1s & TTs (II) & K4-K6 & 0006369792 & 2005-03-12 & SH, LH\\
\object{ROX43A} & 16h31m20.13s & -24d30m05.1s & TTs (II) & G0 & 0015914496 & 2005-09-09 & SH, LH\\
\object{IRS60} & 16h31m30.89s & -24d24m39.7s & - & - & 0006370048 & 2005-09-09 & SH, LH\\
\object{Haro\,1-16} & 16h31m33.47s & -24d27m37.1s & TTs (II) & K2-K3 & 0012664064 & 2005-09-09 & SH, LH\\
\object{Haro\,1-17} & 16h32m21.94s & -24d42m14.7s & TTs (II) & M2.5 & 0011827712 & 2004-08-29 & SL, SH, LH\\
\object{RNO\,90} & 16h34m09.20s & -15d48m16.80s & TTs (II) & G5 & 0005650432 & 2004-08-28 & SL, SH, LH\\
\object{Wa\,Oph\,6} & 16h48m45.63s & -14d16m35.96s & TTs (II) & K7 & 0005650688 & 2006-03-15 & SL, SH, LH \\
\object{V1121\,Oph} & 16h49m15.31s & -14d22m08.6s & CTTs (II) & K5 & 0005650688 & 2006-03-15 & SL, SH, LH\\
\object{HD\,163296} & 17h56m21.29s & -21d57m21.9s & HAeBe (II) & A0-A2 & 0005650944 & 2004-08-28 & SH, LH\\ \hline
\multicolumn{6}{l}{Serpens}\\
\hline
\object{VV\,Ser} & 18h28m47.86s & 0d08m39.8s & HAeBe (II) & B1-A3 & 0005651200 & 2004-09-01 & SL, SH, LH\\
\object{SSTc2d\,J182850.2+00950} & 18h28m50.21s & 0d09m49.6s & II & M2$^k$ & 0013461505 & 2006-04-21 & SL, LL\\
\object{SSTc2d\,J182900.9+02931} & 18h29m00.90s & 0d29m31.5s & II & - & 0013210112 & 2005-04-17 & SL, SH, LL\\
\object{SSTc2d\,J182901.2+02933} & 18h29m01.20s & 0d29m33.2s & II & M0.5$^k$ & 0013461505 & 2006-04-21 & SL, LL\\
\object{SSTc2d\,J182901.8+02954} & 18h29m01.80s & 0d29m54.3s & II & K7$^k$ & 0013210112 & 2005-04-17 & SL, SH, LL\\
\object{SSTc2d\,J182909.8+03446} & 18h29m09.80s & 0d34m45.8s & - & M5 & 0013210624 & 2005-04-14 & SL, SH, LL\\
\object{SSTc2d\,J182928.2+02257} & 18h29m28.24s & -0d22m57.4s & II & - & 0013210368 & 2005-04-14 & SL, SH, LL\\
\object{EC69} & 18h29m54.35s & 1d15m01.8s & II & - & 0009407232 & 2004-03-27 & SL, LL\\
\object{EC82} & 18h29m56.89s & 1d14m46.5s & TTs (II) & M0 & 0009407232 & 2004-03-27 & SL, SH, LH\\
\object{EC90} & 18h29m57.75s & 1d14m05.9s & TTs (II) & - & 0009828352 & 2004-09-01 & SL, SH, LH\\
\object{EC92} & 18h29m57.88s & 1d12m51.6s & TTs (II) & K7-M2 & 0009407232 & 2004-03-27 & SL, SH, LH\\
\object{CK4} & 18h29m58.21s & 1d15m21.7s & TTs (II) & K3 & 0009407232 & 2004-03-27 & SL, SH, LH\\ 
\object{LkH$\alpha$\,348} & 18h34m12.65s & -0d26m21.7s & TTs (II) & F6 & 0009831424 & 2006-04-16 & SL,SH, LH \\ \hline
\object{BF\,Ori} & 05h37m13.26s & -6d35m00.6s & HAeBe (II) & A5-F6 & 0005638144 & 2004-10-03 & SL, SH, LH\\
\object{IRAS\,08267-3336} & 08h28m40.70s & -33d46m22.3s & TTs (II) & K2-K3 & 0005639168 & 2004-11-11 & SL, SH, LH\\
\object{HD\,135344} & 15h15m48.44s & -37d09m16.0s & HAeBe (II) & A0-F4 & 0005657088 & 2004-08-08 & SH, LH\\ \hline
\end{longtable}
Where reference for spectral type is not indicated, it comes from
\citet{Lahuis2007}, the third paper in the series of four c2d articles
on IRS observations of TTauri stars. a) \citet{Herbig1988}, b)
\citet{Wichmann1996}, c) \citet{Lawson1996}, d) \citet{Alcala1995}, e)
\citet{Hughes1992}, f) \citet{Chen1997}, g) \citet{Krautter1997}, h)
\citet{Padgett2006}, i) \citet{Andrews2007}, j)
\citet{Kessler-Silacci2006}, k) \citet{Oliveira2009}
\end{landscape}

\onecolumn
\begin{landscape}
\setlength\LTleft{0pt}
\setlength\LTright{0pt}
\begin{longtable}{@{\excs}lcccccccccc|cccc|c}
  \caption{\label{IDfeat}Signal-to-noise ratios for detections of
    crystalline silicate features,
    amorphous silicate 10 $\mu$m feature and silica.} \\
  \hline \hline
Starname & 9.2 & Amorphous 10.0 & 11.3 & 12.5 & 16.2 & 21.6 & C23 & 25.0 & C28 & 33.6 & $S_{11.3}/S_{9.8}$ & $S_{\mathrm{Peak}}^{10\mu m}$ & $S_{24}/S_{23}$ &$S_{\mathrm{Peak}}^{23\mu m}$ & $F_{30}/F_{13}$ \\
  \hline
\endfirsthead
\caption{continued.}\\
\hline \hline
Starname & 9.2 & Amorphous 10.0 & 11.3 & 12.5 & 16.2 & 21.6 & C23 & 25.0 & C28 & 33.6 & $S_{11.3}/S_{9.8}$ & $S_{\mathrm{Peak}}^{10\mu m}$ & $S_{24}/S_{23}$ &$S_{\mathrm{Peak}}^{23\mu m}$ & $F_{30}/F_{13}$ \\
\hline
\endhead
\hline
\endfoot
\endlastfoot
\multicolumn{11}{l}{Perseus}\\
\hline
RNO\,15 &- & 871 &- &- &- &- &- &- &- &- &0.92 &1.45 &- &- &1.13 \\
LkH$\alpha$\,325 &  39 & 599 &- &- &- &- &  35 &- &  39 &  40 &0.98 &1.51 &- &- &1.10 \\
LkH$\alpha$\,270 &  22 & 109 &- &- &- &- & 108 &- &- &- &1.01 &1.11 &0.51 &9.83 &1.59 \\
LkH$\alpha$\,271 &- & 106 &- &  13 &- &- &- &- &- &- &0.98 &1.26 &- &- &2.54 \\
SSTc2d\,J033036.0+303024 & 192 &1136 &- &  36 &  86 &- & 186 &- & 162 & 195 &1.05 &1.94 &0.90 &1.19 &1.22 \\
SSTc2d\,J033037.0+303128 &- & 330 &- &- &- &- &- &- &  33 &- &0.99 &1.32 &   - &- &0.97 \\
LkH$\alpha$\,326 &  43 & 242 &- &  39 &  94 &  47 & 101 &  90 & 140 &- &1.01 &1.19 &   1.02 &1.09 &2.88 \\
SSTc2d\,J033052.5+305418 &- & 307 &- &8 &- &- &  14 &  10 &  22 &- &0.97 &   1.71 &0.85 &1.18 &1.60 \\
SSTc2d\,J033241.7+311046 &- & 302 &- &- &- &  62 &  36 &- &  81 &- &0.94 &   1.65 &1.30 &1.30 &1.55 \\
LkH$\alpha$\,327 &- & 105 &- &  19 &- &- & 141 &- &- & 198 &1.06 &1.12 &0.22 &   6.60 &1.23 \\
SSTc2d\,J033341.3+311341 &- &- &  45 &  25 &- &- &  35 &- &  60 &- &- &- &   0.98 &1.06 &2.01 \\
BD+31\,634 &- & 421 &- &- &  93 &- &  69 &- &- &- &- &- &- &- &2.82 \\
SSTc2d\,J034219.3+314327 &- & 171 &- &8 &- &- &  21 &- &  36 &- &1.05 &   1.50 &0.93 &1.12 &1.44 \\
IRAS\,03406+3133 &- & 301 &- &  34 &- &- &  87 &  55 &  72 &  88 &1.08 &1.49 &   0.77 &1.33 &1.36 \\
LkH$\alpha$\,330 &- & 247 &  $57^{\mathrm{PAHs}}$ &- &- &- &- &- &- &- &0.99 &1.32 &- &- &  11.74 \\
IRAS\,03446+3254 &- &- &- &- &  34 &- &- &5 &- &- &- &- &- &- &   3.21 \\
\hline
\multicolumn{11}{l}{Taurus}\\
\hline
LkCa\,8 &- &- &- &- &- &- & 154 &  64 &- &- &- &- &0.92 &1.73 &2.28 \\
IQ\,Tau &- &- &- &- &- &- &- &- &  97 &- &- &- &- &- &1.63 \\
FX\,Tau &- &- &- &- &  75 &- & 116 &- &- &- &- &- &0.46 &6.67 &1.87 \\
V710\,Tau &  58 & 293 &- &- &  28 &  35 &  23 &- &  27 &- &0.96 &1.35 &0.99 &1.05 &1.41 \\
RXJ0432.8+1735 &- &- &- &- &- &- &4 &- &- &- &- &- &0.95 &1.23 &   1.57 \\
DN\,Tau &- &- &- &  79 &- &- &- &- &- &- &- &- &- &- &1.77 \\
CoKu\,Tau\,3 &- &- &- &- &  50 &- &- &- &- &- &- &- &- &- &1.05 \\
CoKu\,Tau\,4 &- &  72 &- &- &- &- &- &- &- & 164 &0.98 &1.26 &- &- &  16.96 \\
RR\,Tau &- &- & 175 &  63 &- &- &  82 &- &  56 &  99 &- &- &- &- &1.69 \\
\hline
\multicolumn{11}{l}{Chamaleon}\\
\hline
SX\,Cha &- & 683 &- &- &- &- & 227 &- & 122 &- &0.86 &1.68 &0.85 &2.41 &   1.72 \\
SY\,Cha &- &- & 230 &- &- &- & 216 &- & 228 &- &- &- &0.50 &6.51 &   2.16 \\
TW\,Cha &- & 349 &- &- &- &- &  18 &  17 &  17 &- &0.73 &2.81 &0.69 &   1.56 &2.07 \\
B35 &- &  96 &8 &  10 &- &- &  15 &- &- &  15 &0.92 &1.52 &1.00 &   1.06 &2.39 \\
VW\,Cha &- & 511 &- &  20 &  42 &  82 &  47 &  34 &  75 &  88 &0.96 &1.59 &   0.94 &1.07 &2.08 \\
Hn\,9 &- & 395 &- &- &- &- &  26 &  10 &  22 &  31 &0.95 &1.57 &0.94 &   1.09 &1.54 \\
VZ\,Cha &- & 259 &- &- &- &  21 &  16 &- &  18 &- &0.94 &1.40 &0.71 &   1.52 &1.03 \\
WX\,Cha &- & 444 &- &- &  21 &- &  27 &  24 &  14 &- &0.93 &1.64 &0.70 &   1.55 &0.99 \\
ISO-Cha237 &- & 123 &  12 &- &- &- &- &- &- &- &0.94 &1.38 &- &- &   1.09 \\
C7-11 &  31 & 203 &- &7 &- &- &  21 &  14 &  25 &  58 &1.04 &1.43 &0.98 &1.04 &1.51 \\
HM\,27 &- & 339 &- &- &- &  41 & 101 &- &- &- &0.85 &1.60 &0.96 &1.14 &   2.03 \\
XX\,Cha &  45 & 230 &- &  16 &- &- &  26 &- &  27 &- &0.92 &1.46 &0.97 &   1.06 &1.70 \\
HD\,98922 &- &- & 191 &  46 &- &- & 260 &- &- & 103 &- &- &- &- &0.70 \\
HD\,101412 &  36 & 409 &- &- &- &- & 137 &- &- &- &- &- &- &- &0.92 \\
T\,Cha &- &- &  $27^{\mathrm{PAHs}}$ &- &- &- &- &- &- &- &- &- &- &- &8.26 \\
IRAS\,12535-7623 &- & 438 &- &- &  16 &- &  23 &  11 &  23 &  13 &0.95 &1.75 &   0.69 &1.47 &1.06 \\
ISO-ChaII\,13 &- &- &  43 &5 &- &- &  57 &- &  55 &  33 &- &- &0.98 &   1.20 &0.84 \\
RXJ1301.0-7654 &- & 806 &- &- &- &  70 &  66 &  48 &  25 &- &0.84 &1.87 &   0.89 &1.40 &1.62 \\
Sz50 &- & 108 &- &- &  24 &  57 &  44 &- &  27 &- &0.99 &1.28 &0.98 &   1.09 &2.31 \\
ISO-ChaII\,54 &- &- & 171 &- &  69 &- & 258 &- & 371 & 409 &- &- &0.94 &   1.21 &2.54 \\
Sz52 &- & 147 &- &  27 &- &- &  81 &  36 &  43 &  65 &0.97 &1.47 &0.83 &   1.28 &1.22 \\
Sz62 &- & 171 &- &  27 &- &  17 &  13 &  14 &  33 &- &1.08 &1.51 &0.93 &   1.14 &1.17 \\
\hline
\multicolumn{11}{l}{Lupus}\\
\hline
HT\,Lup &- & 699 &- &- &- &- &- &- &- &- &0.96 &1.38 &- &- &1.53 \\
GW\,Lup &- & 145 &- &4 &- &- &  13 &  11 &- &  26 &0.90 &1.43 &0.92 &   1.27 &1.56 \\
HM\,Lup &- & 313 &- &- &- &- &  16 &- &- &- &0.88 &1.70 &0.97 &1.04 &   1.78 \\
Sz73 &- & 443 &- &9 &  73 &- &  46 &- &  71 &- &0.98 &1.34 &0.97 &   1.04 &2.30 \\
GQ\,Lup &  31 & 360 &- &- &- &- &  20 &  23 &  34 &- &0.96 &1.31 &0.97 &   1.04 &2.04 \\
Sz76 &- &  83 &- &- &4 &- &4 &3 &8 &- &0.82 &1.58 &0.95 &   1.08 &2.31 \\
IM\,Lup &  65 & 803 &- &- &- &- &  34 &- &  42 &- &0.89 &1.64 &0.93 &   1.17 &1.85 \\
RU\,Lup &- & 446 &- &  25 &- &- &  95 &- & 157 &  94 &0.98 &1.54 &0.94 &   1.07 &1.94 \\
RY\,Lup &- &1690 &- &- &- &- &  52 &- &  78 &- &0.69 &2.82 &0.99 &1.09 &   3.92 \\
EX\,Lup &- & 856 &- &- &- &- &  79 &  55 &- &  59 &0.76 &1.89 &0.74 &   1.41 &1.56 \\
RXJ1603.2-3239 &- &- &- &- &- &- &  14 &- &- &  14 &- &- &0.42 &2.39 &   2.74 \\
Sz96 &- & 849 &- &- &  25 &  33 &  64 &- &  72 & 107 &0.84 &2.34 &0.93 &   1.13 &1.61 \\
Sz102 &- & 231 &- &  34 &- &- &- &- &- &- &0.90 &2.02 &- &- &3.95 \\
SSTc2d\,J161159.8-382338 &- & 187 &- &3 &  21 &- &  24 &- &  24 &  46 &0.92 &   2.26 &0.94 &1.15 &1.81 \\
RXJ1615.3-3255 &  16 & 226 &- &6 &9 &- &  13 &- &  18 &- &1.14 &3.06 &   1.03 &1.08 &6.52 \\
\hline
\multicolumn{11}{l}{Ophiuchus}\\
\hline
AS\,205 &- & 796 &- &- &- &- &- &- &- &- &0.90 &1.60 &- &- &1.76 \\
Haro\,1-1 &- & 551 &- &- &- &- &  20 &- &  39 &- &0.72 &2.87 &1.00 &1.02 &4.70 \\
SSTc2d\,J162148.5-234027 &- & 253 &  37 &  18 &- &- &  62 &  26 &  52 & 189 &1.02 &1.63 &0.96 &1.10 &3.77 \\
SSTc2d\,J162221.0-230403 &- & 121 &- &1 &- &- &- &- &- &- &0.82 &2.66 &   - &- &5.69 \\
SSTc2d\,J162245.4-243124 &- & 408 &- &- &  87 &- & 103 &- &- &  83 &1.03 &1.81 &0.97 &1.06 &3.20 \\
SSTc2d\,J162332.8-225847 &- & 193 &- &  17 &  14 &- &  12 &- &- &- &1.00 &   2.49 &1.00 &1.19 &1.77 \\
Haro\,1-4 &- & 304 &- &8 &- &- &- &- &- & 505 &1.05 &1.85 &- &- &   2.44 \\
VSSG1 &- &- &- &  12 &- &- &  37 &  26 &  40 &- &- &- &0.83 &1.21 &   1.06 \\
DoAr\,24E &- &- & 100 &- &  23 &- &  20 &- &  37 &- &- &- &0.96 &1.09 &   1.92 \\
DoAr25 &- &- &  84 &- &- &- &  86 &- & 103 & 258 &- &- &0.93 &1.04 &   3.31 \\
GY23 &- &- &  81 &- &  41 &- &  23 &8 &  49 &- &- &- &0.84 &1.20 &   1.01 \\
SR\,21 &- &- & $103^{\mathrm{PAHs}}$ &  19 &- &- &- &- &- &- &- &- &- &- &  10.85 \\
SSTc2d\,J162715.1-245139 &- & 172 &  22 &- &- &- &  39 &- &  34 &- &0.86 &   2.04 &0.96 &1.08 &3.98 \\
SR\,9 &- & 715 &- &- &- &- &  17 &- &  39 &  32 &0.77 &2.53 &0.84 &1.22 &2.66 \\
SSTc2d\,J162816.7-240514 &  19 &- &  83 &- &  13 &- &  69 &- &  54 &  63 &- &- &0.93 &1.15 &1.19 \\
V853Oph &- &- &- &- &- &- &  78 &- &  80 &- &- &- &0.99 &1.06 &1.69 \\
ROX42C & 130 & 278 &- &  25 &  57 &- &  76 &- &  89 &- &0.95 &1.48 &0.97 &   1.07 &1.90 \\
ROX43A & 175 &2574 &- &- & 223 &- & 307 &- & 306 & 274 &0.87 &2.65 &0.92 &   1.16 &1.68 \\
IRS60 &- & 405 &- &- &- &- &  26 &3 &  35 &  86 &0.98 &1.44 &0.98 &   1.07 &1.69 \\
Haro\,1-16 &- &1248 &- &- &- &- &  35 &- &  63 &- &0.72 &3.11 &0.89 &   1.13 &4.30 \\
Haro\,1-17 &  55 & 374 &- &- &- &- &  21 &- &  20 &  25 &0.98 &1.55 &0.88 &   1.19 &1.96 \\
RNO\,90 &- & 520 &- &- &- &- &  39 &  40 &- &- &0.92 &1.47 &0.79 &1.31 &   1.75 \\
Wa\,Oph\,6 &  64 & 374 & 132 &  41 &- &- &  73 &- &  90 &- &1.04 &1.23 &0.98 &1.05 &1.66 \\
V1121\,Oph &- &2663 &- &- &- &- & 177 &- & 161 &- &0.85 &2.23 &1.02 &   1.20 &1.87 \\
HD\,163296 &- &- & 278 &  32 &- &- & 186 &- &- &  56 &- &- &- &- &1.56 \\
\hline
\multicolumn{11}{l}{Serpens}\\
\hline
VV\,Ser &- & 659 &- &- &- &- &- &- &- &- &- &- &- &- &0.76 \\
SSTc2d\,J182850.2+00950 &- & 379 &- &- &- &- &  41 &  54 &  50 &- &0.90 &   1.67 &0.86 &1.38 &1.85 \\
SSTc2d\,J182900.9+02931 &- & 313 &- &- &- &- &  25 &  16 &  39 &  34 &0.98 &   1.27 &0.89 &1.15 &1.75 \\
SSTc2d\,J182901.2+02933 &- &- &  26 &  15 &  19 &- &  10 &- &  30 &- &- &- &1.03 &1.07 &1.45 \\
SSTc2d\,J182901.8+02954 &- &- &  18 &- &- &  32 &  18 &9 &  43 &  40 &- &- &0.90 &1.11 &1.46 \\
SSTc2d\,J182909.8+03446 &- & 852 &- &- &- &- &- &- &- &- &0.92 &1.83 &- &- &0.78 \\
SSTc2d\,J182928.2+02257 & 141 &1136 &- &- &- &- &  54 &- & 108 &- &1.21 &   1.71 &1.01 &1.08 &1.14 \\
EC82 &- &2618 &- &- &- &- &- &- &- &- &0.73 &3.29 &- &- &3.33 \\
EC90 &  46 &- &- &- & 113 &- & 193 &- &- & 148 &- &- &0.97 &1.21 &   1.61 \\
CK4 &- & 305 &- &- &- &- &- &- &- &- &0.92 &1.50 &- &- &2.47 \\
LkH$\alpha$\,348 &- & 475 &- &- &- &- &- &- &- &- &1.03 &1.24 &- &- &0.66 \\
\hline
BF\,Ori &- &2113 &- &- &  57 &- &- &- &  22 &  26 &- &- &- &- &0.84 \\
IRAS\,08267-3336 &- & 598 &- &- &- &- &- &- &- &- &- &- &- &- &3.12 \\
HD\,135344 &- &- & 304 &- &- &- &- &- &- &- &- &- &- &- &9.88 \\
\hline
\end{longtable}
\end{landscape}


\end{document}